\documentclass[twocolumn]{aa} 

\usepackage{graphicx}
\usepackage[varg]{txfonts}
\usepackage[percent]{overpic}
\usepackage{color}
\usepackage{multirow}
\usepackage{natbib,twoopt} 
\usepackage[colorlinks=true,urlcolor=blue,linkcolor=red]{hyperref} 
\bibpunct{(}{)}{;}{a}{}{,} 

\makeatletter
  \newcommandtwoopt{\citeads}[3][][]{\href{http://adsabs.harvard.edu/abs/#3}%
    {\def\hyper@linkstart##1##2{}%
     \let\hyper@linkend\@empty\citealp[#1][#2]{#3}}}
  \newcommandtwoopt{\citepads}[3][][]{\href{http://adsabs.harvard.edu/abs/#3}%
    {\def\hyper@linkstart##1##2{}%
     \let\hyper@linkend\@empty\citep[#1][#2]{#3}}}
  \newcommandtwoopt{\citetads}[3][][]{\href{http://adsabs.harvard.edu/abs/#3}%
    {\def\hyper@linkstart##1##2{}%
     \let\hyper@linkend\@empty\citet[#1][#2]{#3}}}
  \newcommandtwoopt{\citeyearads}[3][][]%
    {\href{http://adsabs.harvard.edu/abs/#3}
    {\def\hyper@linkstart##1##2{}%
     \let\hyper@linkend\@empty\citeyear[#1][#2]{#3}}}
\makeatother

\begin{document}
   
   \title{Convective blueshifts in the solar atmosphere}
   \subtitle{{\textrm {III}.} High-accuracy observations of spectral lines in the visible}
   
   \author{J. L\"ohner-B\"ottcher\inst{\ref{inst_kis},\ref{inst_hao}} \and W. Schmidt\inst{\ref{inst_kis}} \and R. Schlichenmaier\inst{\ref{inst_kis}} \and T. Steinmetz\inst{\ref{inst_mpq},\ref{inst_menlo}} \and R. Holzwarth\inst{\ref{inst_mpq},\ref{inst_menlo}}}
    \institute{
        Kiepenheuer-Institut f\"ur Sonnenphysik, Sch\"oneckstr. 6, 79104 Freiburg, Germany; \email{jlb@leibniz-kis.de}\label{inst_kis} \and
        High Altitude Observatory, NCAR, 3080 Center Green Drive, Boulder, CO 80301, USA; \email{jlb@ucar.edu}\label{inst_hao} \and
        Max-Planck-Institut f\"ur Quantenoptik, Hans-Kopfermann-Strasse 1, 85748 Garching, Germany\label{inst_mpq} \and
        Menlo Systems GmbH, Am Klopferspitz 19, 82152 Martinsried, Germany\label{inst_menlo}}
   \date{Received 19 Dec 2018 / Accepted 22 Jan 2019}

  \abstract 
  {Convective motions in the solar atmosphere cause spectral lines to become asymmetric and shifted in wavelength. For photospheric lines, this differential Doppler shift varies from the solar disk center to the limb.}
  {Precise and comprehensive observations of the convective blueshift and its center-to-limb variation improve our understanding of the atmospheric hydrodynamics and ensuing line formation, and provide the basis to refine 3D models of the solar atmosphere.}
  {We performed systematical spectroscopic measurements of the convective blueshift of the quiet Sun with the Laser Absolute Reference Spectrograph (LARS) at the German Vacuum Tower Telescope. The spatial scanning of the solar disk covered 11 heliocentric positions each along four radial (meridional and equatorial) axes. The high-resolution spectra of 26 photospheric to chromospheric lines in the visible range were calibrated with a laser frequency comb to absolute wavelengths at the $\mathrm{1\,m\,s^{-1}}$ accuracy. Applying ephemeris and reference corrections, the bisector analysis provided line asymmetries and Doppler shifts with an uncertainty of only few $\mathrm{m\,s^{-1}}$. To allow for a comparison with other observations, we convolved the results to lower spectral resolutions.}
  {All spectral line {bisectors} exhibit a systematic center-to-limb variation. Typically, a blueshifted ``C''-shaped curve at disk center transforms into a less blueshifted ``\textbackslash''-shape toward the solar limb. The comparison of all lines reveals the systematic dependence of the convective blueshift {on the line depth}. The blueshift of the line minima describe a linear decrease with increasing line depths. The slope of the center-to-limb variation develops a reversal point at heliocentric positions between $\mu=0.7$ and 0.85, seen as the effect of horizontal granular flows in the mid photosphere. Line minima formed in the upper photosphere to chromosphere exhibit hardly any blueshift or even a slight redshift. Synthetic models yield considerable deviations from the observed center-to-limb variation.}
  {The obtained Doppler shifts of the quiet Sun can serve as an absolute reference for other observations, the relative calibration of Dopplergrams, and the necessary refinement of atmospheric models. Based on this, the development of high-precision models of stellar {surface convection} will advance the detection of (potentially habitable) exoplanets by radial velocity measurements.}

  \maketitle
  \titlerunning{Convective blueshifts in the solar atmosphere. III} 
  \authorrunning{L\"ohner-B\"ottcher et al.}

\section{Introduction}\label{sec_intro}

Spectroscopic observations of spatially unresolved quiet Sun regions typically yield photospheric lines to be Doppler shifted to slightly shorter wavelengths. Though, this convective blueshift \citep{Beckers1977} is not constant at all. It varies from line to line, depending on the line strength, excitation potential, wavelength region, and atmospheric formation layer \citep{1981A&A....96..345D,1984SoPh...93..219B}. The blueshift and characteristic line asymmetry of the line profile results from the superposition and lateral averaging of many different line profiles of the convective pattern of granulation and intergranular lanes \citep{1975A&A....43...45D,1981A&A....96..345D}. Due to their brightness, the hot, rising granules inducing blueshifted line profiles have a greater statistical contribution to the average profile, than the fainter redshifted profiles from the cooler intergranular lanes. Toward higher photospheric layers, the brightness pattern of the granulation reverses, whereas velocity gradients along the line of sight introduce differential Doppler shifts. In summary, the bisector representing the line shift at each depth along the average line profile reveals a typically ``C''-shaped line asymmetry.

Further, the line shift and asymmetry feature a significant center-to-limb variation. Also known as the limb effect \citep{1907AN....173..273H}, this variation involves a decrease in blueshift by a few hundred $\mathrm{m\,s^{-1}}$ and a transformation of the bisector from a convex ``C''-shape to a ``\textbackslash''-shape \citep{1976MNRAS.177..687A}. \citet{1978SoPh...58..243B} studied the systematic line-of-sight effects of granular motions and presented a theory which managed to {explain the} characteristics of the observed center-to-limb variation. By including line-of-sight variations of the granular intensity pattern, the vertical and horizontal flow pattern of adjacent granules, and opacity effects, they qualitatively reproduced: (1) the blueshift at disk center, (2) its slow decrease toward the solar limb, (3) the initial slight increase in blueshift of some spectral lines when moving away from disk center, and (4) even the slight redshift at the extreme solar limb, formerly known as ``supergravity'' redshift. Especially the consideration of the horizontal flow component of the overturning granular convection \citep{1985SoPh...99...31B} provided an explanation for the occasional first increase in blueshift of some lines \citep{1967ApJ...147.1100A} from disk center toward heliocentric positions around $\mu=\cos\theta=0.8$ (with $\theta$ being the heliocentric angle).

Since the onset of solar high-resolution spectroscopy in the 1960s, the convective blueshift and its center-to-limb variation has been frequently observed and extensively discussed \citep[for example, by][]{1981A&A....96..345D,1982SoPh...79....3B,1984SoPh...93..219B,1984cup..book.....B,1985A&A...150..256C,1986A&A...163..219C}. In this work, we present comprehensive and systematic observations of the convective blueshift, and the most accurate analysis of its center-to-limb variation thus far. After Paper I \citep{2018A&A...611A...4L} and Paper II \citep{2018arXiv181108685S} which were confined to the 6302\,\AA\ lines and to the 6173\,\AA\ line, respectively, this third article of the series ``Convective blueshifts in the solar atmosphere'' addresses the analysis of several frequently used spectral lines in the visible range of the solar spectrum. In Section\,\ref{sec_data}, we describe our systematic observations of the quiet Sun, and the calibration of absolute Doppler shifts at the unprecedented $\mathrm{m\,s^{-1}}$ accuracy. In Section\,\ref{sec_results}, we specify the differential line shift and its center-to-limb variation for the respective spectral lines, and contrast the results with existing synthetic models. In Section\,\ref{sec_discussion}, the direct comparison of all lines reveals the systematic behavior of the convective blueshift. In Section\,\ref{sec_conclusions}, we draw our conclusions on the convective blueshifts of the Sun. In Section\,\ref{sec_outlook}, we give an outlook to the field of exoplanet detection.

\section{Observations}\label{sec_data}

Solar observations were performed between May 7th 2016 and May 11th 2018 with the Laser Absolute Reference Spectrograph \citep[LARS,][]{Doerr2015,2017A&A...607A..12L} at the German Vacuum Tower Telescope (VTT) on Tenerife. Our measurements followed the same systematic procedure as initially presented in Paper I \citep{2018A&A...611A...4L} for the 6302\,\AA\ region. 

\begin{figure}[htbp]
\begin{center}
\includegraphics[trim=4.8cm 1.4cm 4.3cm 1.4cm,clip,width=\columnwidth]{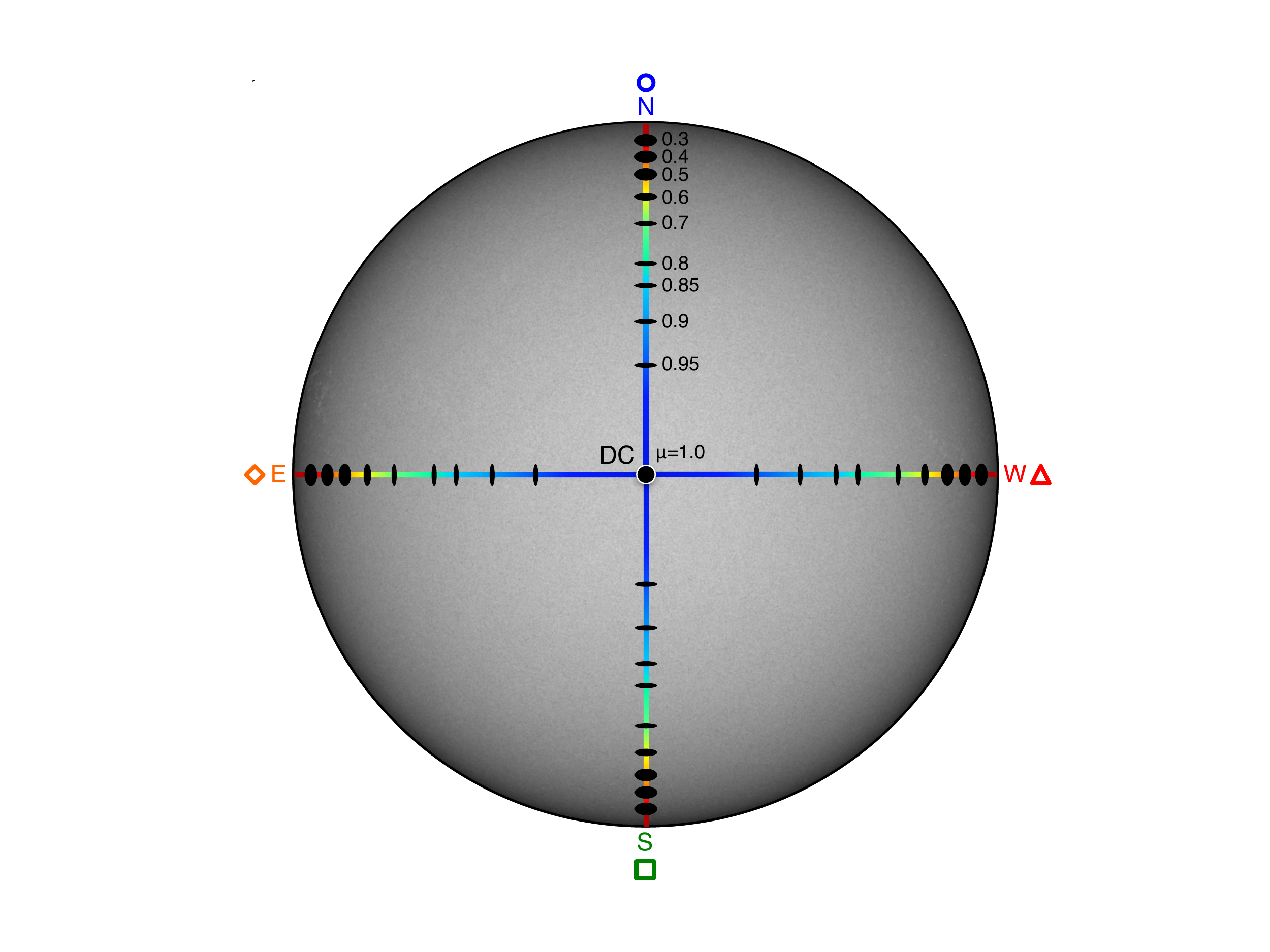}
\caption{Observation scheme. We systematically scanned the quiet Sun at up to eleven heliocentric positions from the disk center (DC, $\mu=1.0$) toward $\mu=0.3$, or $\mu=0.2$, close to the solar limb. The observations were performed along the meridional (N-S) and equatorial (E-W) radians: north (blue, circle), south (green, rectangle), east (orange, diamond), west (red, triangle). The black ellipses indicate the spatially integrated area (not to scale). Throughout this work, we maintain the color code (blue to red) of the heliocentric position along the axes, as well as the colors and symbols for the respective axes.}
\label{fig_sec2_observations_scheme}
\end{center}
\end{figure}

The observation scheme in Fig.\,\ref{fig_sec2_observations_scheme} illustrates the spatial sampling. To study the center-to-limb variation of the convective blueshift, we performed measurements at up to eleven heliocentric positions ($\mu=\cos\theta$, with $\theta$ being the heliocentric angle) along four radial axes (north, south, east, west). The sampling from $\mu=1.0$ (disk center) to $\mu=0.8$ was done in steps of $\Delta\mu=0.05$. From $\mu=0.8$ to $\mu=0.3$ (or $\mu=0.2$) close to the solar limb, the step size was set to $\Delta\mu=0.1$. 
To reduce the effect of acoustic oscillations and supergranular (predominantly horizontal) flows, we spatially integrated the sunlight over elliptical areas centered on the nominal positions. The pointing accuracy of the telescope was around 1\arcsec\ on the coordinate grid. The size and orientation of the covered elliptical area was adjusted according to the position on the solar disk and the local background effect of p-mode oscillations and supergranular flows (discussed in Paper I). In the quiet Sun, a size of up to 30--40\arcsec\ typically covers around 50 granules. A detailed description of the observation procedure is given in Paper I (and Table\,1 therein). Using G-band context images of LARS and full-disk magnetograms of the Helioseismic Magnetic Imager \citep[HMI,][]{2012SoPh..275..229S}, we verified that only quiet Sun regions were scanned. This guaranteed the consistency of our study, since strong magnetic fields would lead to a modification of the convective energy transport and reduction of the convective blueshift.

To perform the most accurate and precise spectroscopic analysis of the convective blueshifts so far, our observations had to meet a number of requirements. LARS was ideally suited since it combines the VTT's high-resolution echelle spectrograph with a laser frequency comb for an absolute wavelength calibration. The high spectral resolution ($\lambda/\Delta\lambda > 700\,000$ in the visible range) provides the required precision. The accuracy of the comb implies the unrestricted repeatability of the measurement. Since solar p-modes and supergranular flows superimpose the convective blueshift, identical instrumental conditions are needed to perform repetitive observations in order to reduce this "solar noise". As described in detail in \citet{2017A&A...607A..12L} and \citet{2018A&A...611A...4L}, we obtain a wavelength accuracy of around 0.02\,m\AA\ (or $\mathrm{1\,m\,s^{-1}}$) for the recorded single spectra. At this level, an identical illumination of the spectrograph and the pixels on the charge-coupled device (CCD) chip of the attached camera is crucial. Therefore, the light from the Sun and the laser frequency comb (and all other light sources) is spatially integrated and guided by optical single-mode fibers via a fiber switch device to the spectrograph.

\begin{table}[htbp]
\caption{Spectral regions and important spectral lines.}
\label{table1}
\centering
\tabcolsep=0.19cm
\begin{tabular}{c c c l c c}
\hline\hline
Spectral&Number of&\multicolumn{2}{c}{Spectral line}&$g_\mathrm{eff}$&Height\\ 
region&data sets&Ion&$\quad\lambda_0$ [\AA]&&[km]\\ 
\hline\\[-0.35cm]
5250\,\AA&65&\ion{Fe}{I}&5250.2084$^\ast$&3.00&310\\
&&\ion{Fe}{I}&5250.6453$^\ast$&1.50&360\\
5381\,\AA&75&\ion{C}{I}&5380.3308$^+$&1.00&40\\
5434\,\AA&91&\ion{Fe}{I}&5434.5232$^\ast$&0.00&550\\
&&\ion{Fe}{I}&5432.9470$^\ast$&0.50&250\\
5576\,\AA&75&\ion{Fe}{I}&5576.0881$^\ast$&0.00&360\\
5896\,\AA&67&\ion{Na}{I}&5895.92424$^+$&1.33&800\\
6149\,\AA&76&\ion{Fe}{II}&6149.2460$^\times$&1.33&130\\
6173\,\AA&62&\ion{Fe}{I}&6173.3344$^\ast$&2.50&270\\
6302\,\AA&99&\ion{Fe}{I}&6301.5008$^{+}$&1.67&340\\
&&\ion{Fe}{I}&6302.4932$^{+}$&2.50&260\\
\hline
\end{tabular}
\tablefoot{Laboratory wavelength $\lambda_0$ in air: $^\ast$measured with LARS, $^+$observed wavelength from NIST, $^\times$Ritz wavelength from NIST.}
\end{table}

To infer the systematic convective blueshift of the solar photosphere, we had to obtain an adequate statistical sampling of observations. In total, we recorded 610 data sets in eight different wavelength regions (see Table\,\ref{table1}). Thus, we reached a total data volume of 203 hours of observation in which the frequency comb guaranteed identical conditions for direct comparison. With 800 (or 480) observation cycles (each consisting of one solar and one frequency comb spectrum) and a cycle time of 1.5\,s (or 2.5\,s), each data set described a 20\,min time sequence in order to trace p-modes and perform a temporal average over around four 5\,min-oscillations. The camera exposure time was set to 0.5\,s (or 1.0\,s), depending on the wavelength region.

To obtain the outstanding accuracy at the $\mathrm{m\,s^{-1}}$ level, we had to perform a valid and careful data calibration. This was done in the same manner as presented in Paper I and illustrated in \citet{2017A&A...607A..12L}, using the LARS data pipeline developed by \citet{Doerr2015}. The two crucial steps are the absolute wavelength calibration of the solar spectrum with the comb spectrum, and the precise reduction of all (non-convective) systematic relative motions between the telescope and the observed region on the Sun. The latter can sum up to systematic Doppler shifts of a few $\mathrm{km\,s^{-1}}$.

To calibrate the wavelength grid of the solar spectrum, the mode spectrum of the laser frequency comb served as an absolute ruler \citep{Steinmetz+etal2008}. The solar and comb spectra are exemplarily shown in Fig.\,\ref{fig_sec3_spectra_5250}. By the unambiguous determination of the mode numbers (with a spacing of 8.0\,GHz), we obtained the absolute frequency of each mode and, finally, the pixel-wise dispersion on the detector. The alternate recording of the comb and the solar spectrum enabled the calibration of each solar spectrum with an instrumental accuracy of $\mathrm{1\,m\,s^{-1}}$.

To obtain the convective blueshift of the Sun, we reduced all systematic orbital, radial and rotational motions of the Sun and Earth with respect to each other. By using the ephemeris code developed by \citet{Doerr2015}, which in turn is based on NASA's Spacecraft Planet Instrument C-matrix Events (SPICE) toolkit \citep{Acton1996}, we reach a model uncertainty of $\mathrm{0.1\,m\,s^{-1}}$ at solar disk center. For heliographic positions close to the solar limb, the uncertainty from the applied (spectroscopic) rotation model of \citet{1990ApJ...351..309S} increases to $\mathrm{4\,m\,s^{-1}}$. 

Finally, the calibrated solar spectra are only affected by the constant gravitational redshift of $\mathrm{+635\,m\,s^{-1}}$ (caused by the Sun and Earth according to the principle of equivalence and the general theory of relativity) and the local solar activity itself. Since the systematic convective blueshift of the solar atmosphere is superimposed by temporal variations of convective motions, acoustic oscillations, supergranulation, and large-scale flows, its analysis must be based on temporal and statistical averaging. Temporal averaging of each 20\,min sequence reduced the uncertainty of the mean Doppler shift to a few $\mathrm{m\,s^{-1}}$. According to error propagation of systematic and statistical errors, we yield a total uncertainty of around $\mathrm{5\,m\,s^{-1}}$ for the mean Doppler shift of each observation sequence (compare Fig.\,\ref{fig_sec3_clv_linecore_Fe52502}). 

To measure absolute Doppler shifts of spectral lines, the air wavelength $\lambda_0$ of the laboratory reference has to be provided at an accuracy level comparable to that of our solar measurements.. The National Institute of Standards and Technology Atomic Spectra Database \citep[NIST ASD,][]{NIST_ASD} lists the rest wavelength of spectral lines with an uncertainty of around $\mathrm{1\,m\AA}$. In the visible range around $\lambda_0=5000\,\AA$, this translates into a Doppler uncertainty of $\mathrm{60\,m\,s^{-1}}$. To increase the accuracy of our study, we measured the laboratory wavelengths of most iron lines with the hollow cathode lamp of LARS. The emission lines of the lamp (compare Fig.\,\ref{fig_sec3_spectra_5250}) were fitted with a symmetrical Voigt function. At the given instrumental accuracy, we obtained reference wavelength with an uncertainty of below $\mathrm{0.1\,m\AA}$ (or $\mathrm{2-4\,m\,s^{-1}}$). The Doppler velocity
\begin{equation}
\mathrm{v_{los}=c\cdot(\lambda-\lambda_0)/\lambda_0-v_{grs}}\label{eq1}
\end{equation}
results from the shift of the observed wavelength $\lambda$ with respect to the reference wavelength $\mathrm{\lambda_0}$, multiplied by the speed of light $\mathrm{c}$. The constant gravitational redshift $\mathrm{v_{grs}}$ was subtracted. 

The comprehensive analysis of the solar convective blueshift demands a well-considered sample of spectral lines. A selection of important lines within our final sample is listed in Table\,\ref{table1}. Since the convective blueshift is basically a photospheric phenomenon, we have focused our study on photospheric to lower chromospheric lines. In order to obtain a valid atmospheric sampling, we selected spectral lines with different line strength. The {average} formation height of the spectral line core above the quiet Sun optical depth unity at 5000\,\AA\ is given in the right column of Table\,\ref{table1}. {We note that the given values are only simple estimates in line with classical one-dimensional atmospheres} \citep{1991sopo.work.....N,1994A&A...285.1012G,1998A&A...329..721S,1988A&AS...72..473B,1991PhDT.......113F,1991A&A...243..244G,1998A&A...332.1069K}. Moreover, our analysis included only those parts of spectral lines which were not deformed or corrupted by atomic or molecular blends, or by telluric lines.

Another important aspect for the line selection (in Table\,\ref{table1}) was the application of the line for solar observations and theoretical modeling. One of our final goals was to provide precise {reference values for the convective blueshift with respect to} the spectral line and the heliocentric position of the observed target. Such an indirect calibration of {Dopplergrams} will become important for the investigation of small-scale flows in the solar atmosphere. With regard to the high spatial resolution of the new 4\,m-class Daniel K. Inouye Solar Telescope \citep[DKIST,][]{2012ASPC..463..377R}, we focus our attention on the spectral lines and regions which will be observed with the first-light instruments VTF \citep[Visible Tunable Filter,][]{2012SPIE.8446E..77K,2014SPIE.9147E..0ES}, ViSP \citep[Visible Spectro-Polarimeter,][]{2012SPIE.8446E..6XD}, and DL-NIRSP \citep[Diffraction Limited Near Infrared Spectropolarimeter,][]{2014SPIE.9147E..07E}. In addition, we included the 5250\,\AA\ and 6173\,\AA\ for the calibration of Dopplergrams from IMAX \citep[Imaging Magnetograph eXperiment,][]{2011SoPh..268...57M}, CRISP \citep[CRisp Imaging SpectroPolarimeter,][]{2006A&A...447.1111S} and HMI. Further, our line selection enabled a direct comparison with theoretical syntheses of the solar convective blueshift by \citet{2011A&A...528A.113D} and \citet{2018ApJ...866...55C}. With a spectral sampling of less than $\mathrm{3\,m\AA\,pixel^{-1}}$ on a 2048 pixel wide detector, each spectral region covered a range of around 5.6\,\AA. Thus, we were able to include 26 spectral lines (11 of which are listed in Table\,\ref{table1}) for the analysis of systematics regarding the convective blueshift.

Throughout this work, we adopt common practice in astrophysics and use air wavelengths when referring to spectral lines or observed wavelengths. Negative Doppler velocities indicate blueshifts, positive velocities refer to redshifts.

\section{Results}\label{sec_results}
Our analysis of the convective blueshift involved several basic aspects. Spectral line shifts and asymmetries contain the essential information on the atmospheric distribution of the convective blueshift. The observation of the solar spectrum at high spectral resolution enabled the application of a bisector analysis to infer the asymmetry of a spectral line. In the following, we present the systematic center-to-limb variation of line bisectors. Due to the changing line asymmetry, we performed a quantitative center-to-limb study of the Doppler shift of different line segments. Within this scope, we focused on the center-to-limb shift of the line core. To allow for a comparison with other observations and theoretical studies of the convective blueshift, we transformed our observations to the lower spectral resolution of other spectroscopic instruments. The alteration of the spectral resolution causes a reduction of the line asymmetry and a change of the measured convective blueshift. In Section\,\ref{sec_results_FeI52502}, we will exemplarily present the complete set of analysis steps for the \ion{Fe}{I}\,5250.2\,\AA\ line. In the further course of this section, we apply these steps to all lines and present only the final outcome. Sections\,\ref{sec_results_5250} to \ref{sec_results_6302} correspond to the spectral regions as listed in Table\,\ref{table1}, ordered by wavelength.

\subsection{Lines around 5250\,\AA}\label{sec_results_5250}

\begin{figure}[htbp]
\begin{center}
\includegraphics[width=\columnwidth]{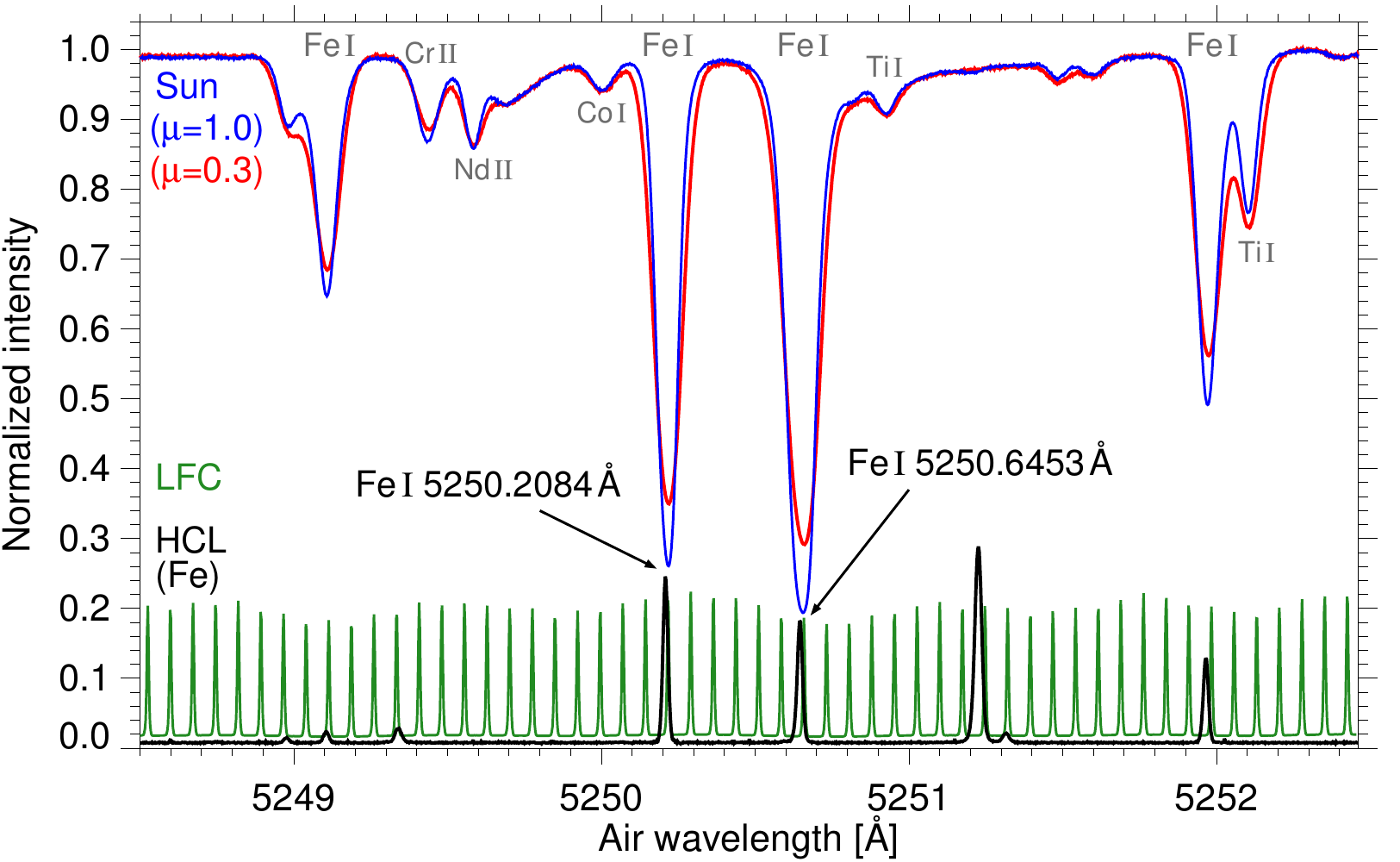}
\caption{Observed spectra around 5250\,\AA, with the quiet Sun absorption spectra at disk center ($\mu=1.0$, blue curve) and close to the solar limb ($\mu=0.3$, red curve). The continuum was normalized to 1 in both cases. The {atomic species} are stated in gray. The emission spectra of the laser frequency comb (LFC, green curve) and the iron hollow cathode lamp (HCL, black curve) are drawn at an arbitrary intensity scale.}
\label{fig_sec3_spectra_5250}
\end{center}
\end{figure}

We observed 65 sequences of the solar spectrum around 5250\,\AA\ shown in Fig.\,\ref{fig_sec3_spectra_5250}. The four strongest lines are all formed by neutral iron. The \ion{Fe}{I}\,5250.2\,\AA\ and \ion{Fe}{I}\,5250.6\,\AA\ lines are well-studied lines in solar physics. With Land\'e-factors of $g_\mathrm{eff}=3.0$ and $g_\mathrm{eff}=1.5$, both lines are Zeeman-sensitive and thus often employed for spectro-polarimetric observations of the lower to middle photosphere. To name an example, the IMAX instrument \citep{2011SoPh..268...57M} aboard the balloon-borne solar telescope Sunrise \citep{2011SoPh..268....1B} used the \ion{Fe}{I}\,5250.2\,\AA\ line to perform measurements of the dynamic photosphere at high spatial resolution. To infer accurate Doppler shifts, we measured the laboratory air wavelength of both lines with the iron hollow cathode lamp of LARS. As indicated in Fig.\,\ref{fig_sec3_spectra_5250}, we yield reference wavelengths of 5250.2084\,\AA\ and 5250.6453\,\AA\ with an uncertainty below 0.1\,m\AA\ (or a corresponding velocity error of around $\mathrm{3\,m\,s^{-1}}$). The analysis of the convective blueshifts of both lines is presented in Sections\,\ref{sec_results_FeI52502} and \,\ref{sec_results_FeI52506}. The \ion{Fe}{I} lines at 5249.1\,\AA\ and 5251.9\,\AA\ are blended by other spectral lines in the left, respectively right, line wing. We measured the reference wavelengths for both iron lines to 5249.1050\,\AA\ and 5251.9652\,\AA. Due to the blended line wings, the analysis of {the convective blueshift} was limited to the lower half of their line profiles. The results are displayed in Figs.\,\ref{fig_A1} and \ref{fig_A2}. Moreover, the solar spectrum (Fig.\,\ref{fig_sec3_spectra_5250}) exhibits a number of other weak spectral lines. Due to the lack of information on these lines, we analyzed only the \ion{Cr}{II}\,5249.4\,\AA\ line (as shown in Figs.\,\ref{fig_A1} and \ref{fig_A2}). Its reference wavelength of 5249.4346\,\AA\ was taken from the NIST ASD, which lists the observed wavelength with an uncertainty of 0.8\,m\AA.

\subsubsection{\ion{Fe}{I}\,5250.2\,\AA}\label{sec_results_FeI52502}
In this section, we exemplify the complete set of analysis steps to infer the convective blueshift of the \ion{Fe}{I}\,5250.2\,\AA\ line. This description serves as a guideline for all other spectral lines. 

\paragraph{Bisector analysis:} To study the differential Doppler shift of the spectral line, we performed a bisector analysis. Each point of a bisector curve represents the center of the line profile at the respective depth or intensity. Thus, the bisector provides the detailed asymmetry of the line and height-dependent Doppler shift along the solar atmosphere. With decreasing intensity from the continuum toward the line minimum at a normalized intensity of around 0.3, the bisector of \ion{Fe}{I}\,5250.2\,\AA\ at disk center captures the evolution of the convective blueshifts from the solar surface up to mid-photospheric altitudes of around 310\,km above \citep{1991sopo.work.....N}. Bisectors were calculated for intensities ranging from the line minimum to a normalized upper threshold level of around 93\% of the continuum intensity. To avoid oversampling and the inclusion of statistical fluctuations, we determined 30 to 40 intensity positions with an equidistant sampling increment. The computation of the bisector was done for each measurement of the 20\,min sequence. To obtain the systematic center-to-limb variation of the line shift, we averaged all bisectors of a heliocentric position (without distinction of the observed radial axis). 

\begin{figure}[h!]
\begin{center}
\includegraphics[width=\columnwidth]{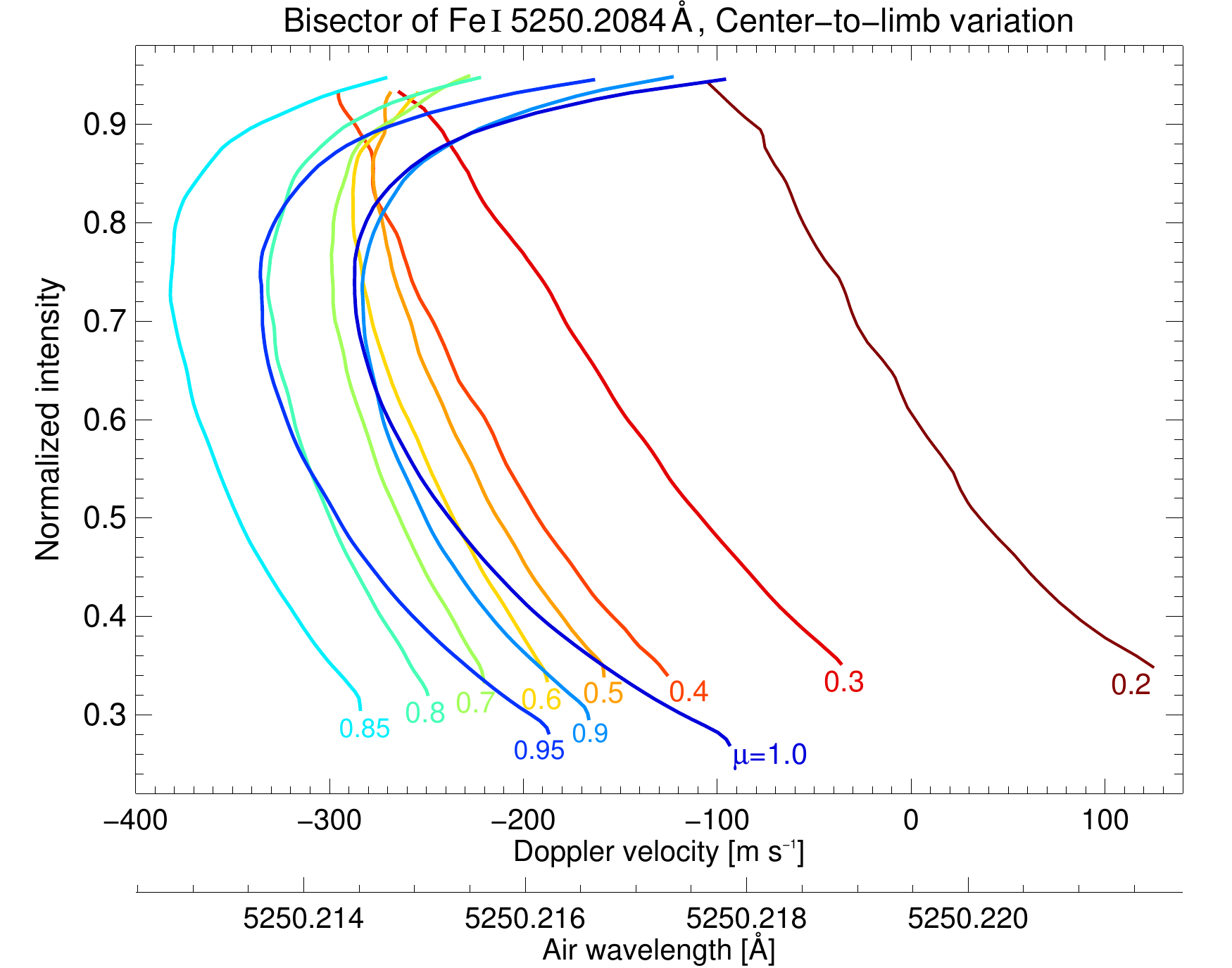}\\[0.2cm]
\caption{Variation of the \ion{Fe}{I}\,5250.2\,\AA\ line bisector from the solar disk center ($\mu=1.0$, blue curve) toward the limb ($\mu=0.2$, dark red curve). The normalized intensity is displayed against the absolute air wavelength and Doppler velocity. Each curve represents the average bisector for all measurements at the respective heliocentric position.}
\label{fig_sec3_bisectors_Fe52502}
\end{center}
\end{figure}

The significant change of the bisector from disk center ($\mu=1.0$) toward the solar limb ($\mu=0.2$) is shown in Fig.\,\ref{fig_sec3_bisectors_Fe52502}. The most apparent change is the transformation of the ``C''-shape at disk center into a ``\textbackslash''-shape when approaching the solar limb. At disk center, the bisector curve describes an initial increase in blueshift from around $\mathrm{-100\,m\,s^{-1}}$ at a normalized intensity of 0.94 toward a maximum blueshift of $\mathrm{-290\,m\,s^{-1}}$ at 0.76. Toward the line minimum at an intensity of 0.27, the blueshift decreases to around $\mathrm{-95\,m\,s^{-1}}$. From disk center to the heliocentric position $\mu=0.85$, the blueshift of the spectral line increases to a maximum of $\mathrm{-380\,m\,s^{-1}}$ at an intensity of 0.73, and $\mathrm{-285\,m\,s^{-1}}$ at the line minimum. Toward smaller $\mu$-values, the reversal of the blueshift in the upper third of the bisector vanishes. The impact of intergranular downflows has disappeared. Bisectors at $\mu\le0.4$ feature an almost straight decrease in blueshift from the continuum to the line core. Close to the solar limb at $\mu=0.2$, the blueshift even turns into a redshift of around $\mathrm{+120\,m\,s^{-1}}$. 

\begin{figure}[htbp]
\includegraphics[width=\columnwidth]{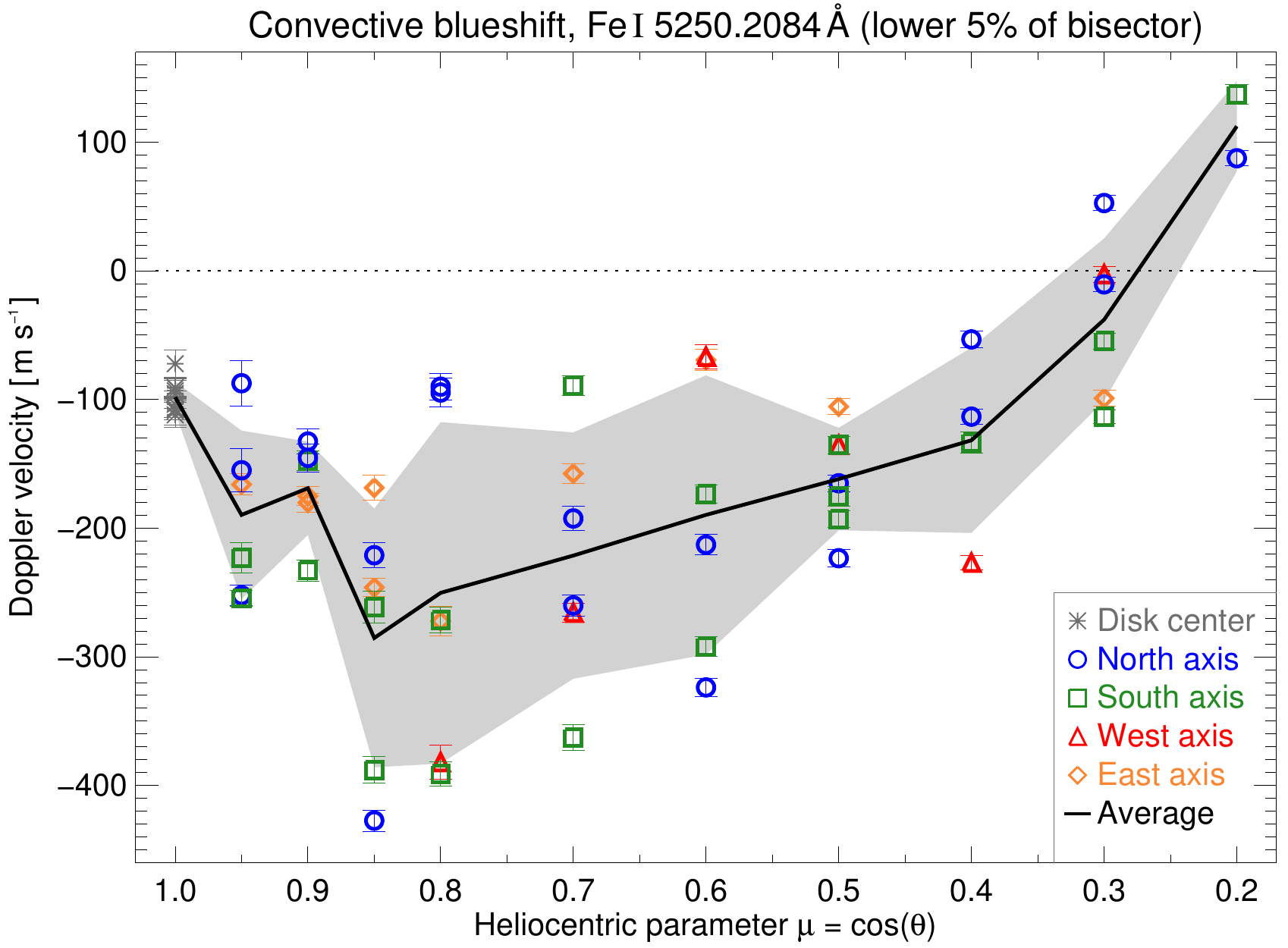}
\caption{Center-to-limb variation of the convective blueshift of the \ion{Fe}{I}\,5250.2\,\AA\ line core. Doppler velocities are plotted against the heliocentric position from the disk center ($\mu=1.0$) toward the solar limb. Each data point represents the mean velocity of the lower 5\,\% of the line bisector for the temporally averaged observation sequence. Error bars indicate the mean error. The four radial axes can be distinguished by the colors and symbols. The black solid line and the gray shaded area display the average center-to-limb variation and its standard deviation.}
\label{fig_sec3_clv_linecore_Fe52502}
\end{figure}

\paragraph{Convective blueshift of the line core:} To perform a more quantitative study of the convective blueshift and its center-to-limb variation, we scrutinize the evolution of the Doppler shift of the line core. The shift of the line core was defined as the average Doppler shift {within the (relative) lower 5\,\% intensity interval of each bisector which, in its entirety, extends from the line minimum (0\,\%) to the spectral continuum (100\,\%)}. Following, the mean convective blueshift of an observation sequence was calculated as the the temporal average over the individual measurements. The mean error of the average convective blueshift was of the order of $\mathrm{10\,m\,s^{-1}}$. Fig.\,\ref{fig_sec3_clv_linecore_Fe52502} displays the results for all observation sequences, plotted against their heliocentric position and sorted by radial axes. 
Across the solar disk, Doppler velocities are distributed in the range from blueshifts of up to $\mathrm{-430\,m\,s^{-1}}$ to redshifts of up to $\mathrm{+140\,m\,s^{-1}}$. The calculation of the mean center-to-limb curve highlights the distinct systematic trend. The convective blueshift increases from $\mathrm{-99\,m\,s^{-1}}$ at disk center ($\mu=1.0$) to a maximum of $\mathrm{-286\,m\,s^{-1}}$ at $\mu=0.85$. Toward the solar limb the blueshift decreases monotonically, and even turns into a redshift of $\mathrm{+111\,m\,s^{-1}}$ at $\mu=0.2$. The detailed numbers at all heliocentric positions are listed in Table\,\ref{table_sec3_resolution_comparison} (line section: Core; spectral resolution: 700\,000). The scatter around the mean center-to-limb curve varies with the heliocentric position. As indicated in Fig.\,\ref{fig_sec3_clv_linecore_Fe52502}, the standard deviation is smallest at disk center, with a minimum of $\mathrm{10\,m\,s^{-1}}$. We conclude that temporal averaging reduced the error caused by acoustic oscillations. Furthermore, given by the orthogonal line of sight, horizontal granular and supergranular flows do not affect the measurement of the systematic convective blueshift at disk center. However, these effects increase with increasing distance to the center of the solar disk. With a standard deviation of up to $\mathrm{130\,m\,s^{-1}}$, the scatter of the individual measurements is largest for heliocentric positions between $\mu=0.85$ and $\mu=0.6$. At the solar limb, we obtain a reduced scatter of the velocity distribution. Toward $\mu=0.2$, the standard deviation has decreased to around $\mathrm{30\,m\,s^{-1}}$. We infer that the spatial averaging successfully minimized the effect of supergranular flow fields. Moreover, the largely vertical p-mode oscillation do not impact the measurement due to the orthogonality with the line of sight.
 
\begin{figure}[htbp]
\textbf{a)}\\[-0.25cm]  \includegraphics[trim=0cm 0cm 0.5cm 0cm,clip,width=\columnwidth]{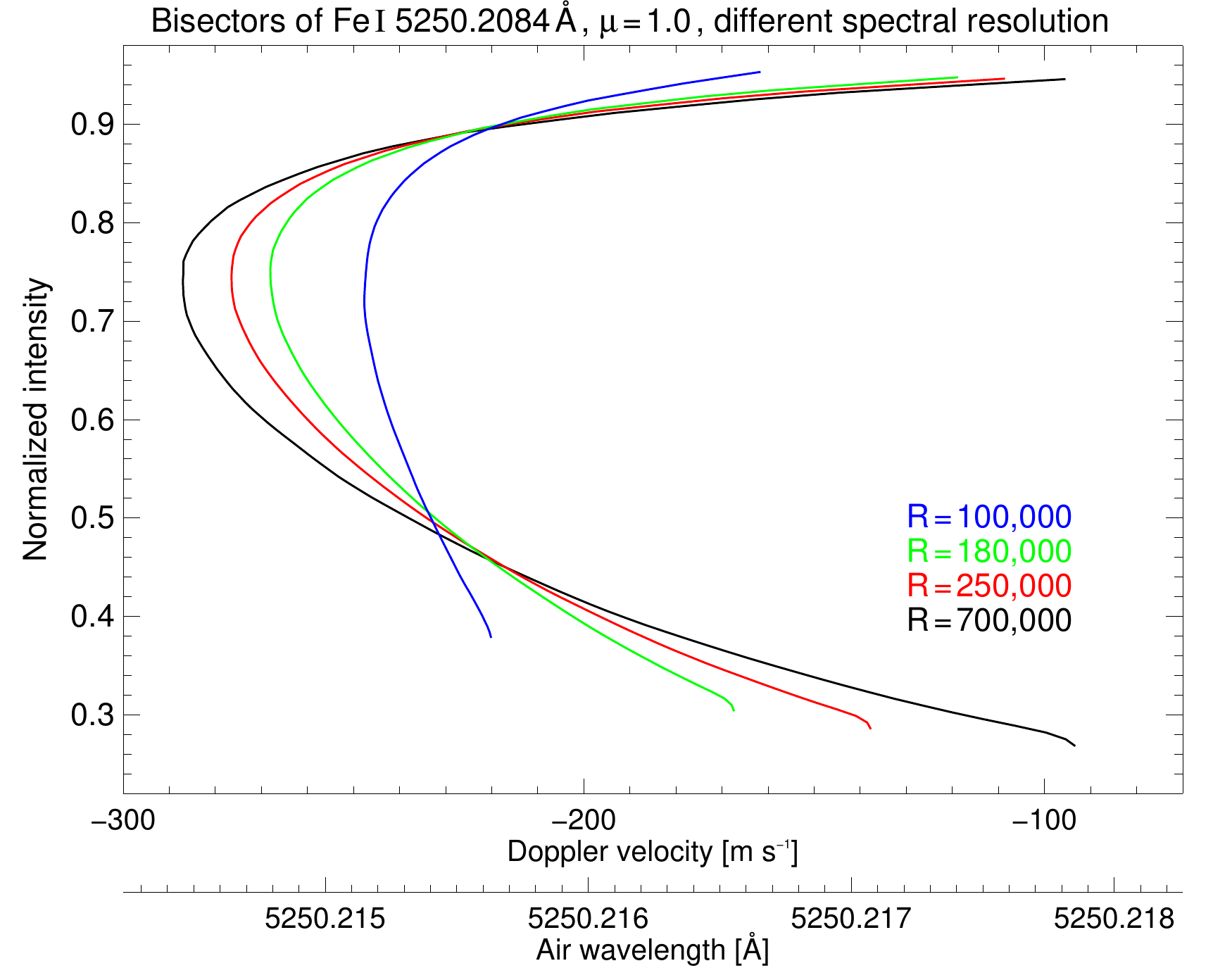}\\[0.1cm]
\textbf{b)}\\[-0.25cm]  \includegraphics[width=\columnwidth]{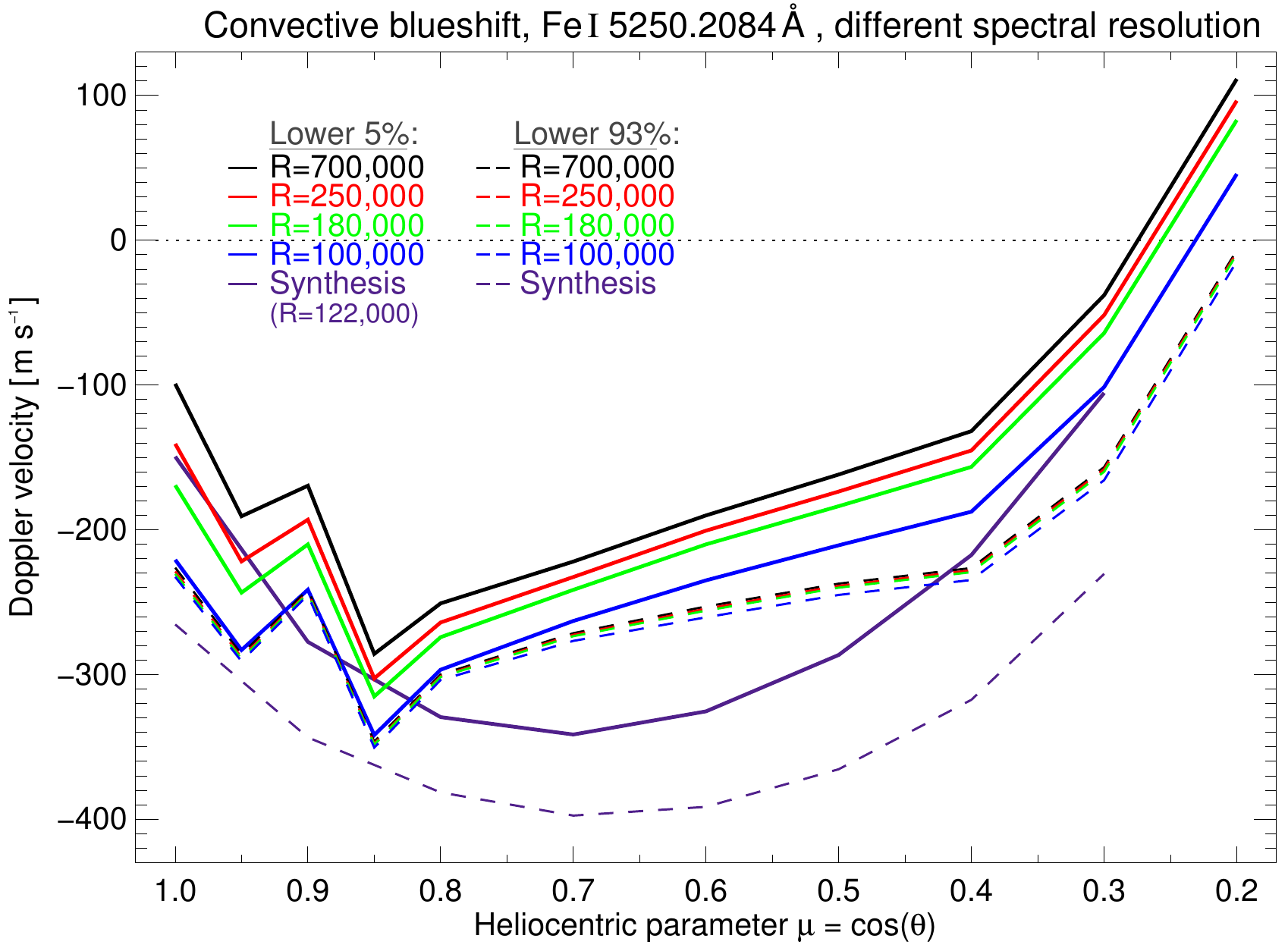}
\caption{Analysis of the convective blueshift of \ion{Fe}{I}\,5250.2\,\AA\ for different spectral resolutions $\mathrm{R}$, from 700\,000 (black) to 100\,000 (blue). {Panel a highlights the impact of the resolution on the line bisector (at $\mu=1.0$).} Panel b shows the center-to-limb variation of the convective blueshift, obtained for the the line core (lower 5\% of the bisector; solid lines) or the entire profile (lower 93\%; dashed lines). The syntheses of \citet{2011A&A...528A.113D} are compared as purple lines.}
\label{fig_sec3_clv_resolution_Fe52502}
\end{figure}

\paragraph{Convective blueshift of the entire line:} To examine the convective blueshift of the entire spectral line which provides a measure for the overall lower photosphere, we performed the same analysis as above but for the full {extent of the} bisector profile. {Compared to the case introduced above for the line core (averaged lower 5\,\% of the bisector), we extended the linear averaging along the bisector from the line minimum (at 0\,\%) to an upper threshold (here 93\,\%) which is close to the spectral continuum (at 100\,\%).} Depending on the disturbance of the spectral profile by line blends, this upper threshold can vary for other spectral lines. With regard to the bisectors shown in Fig.\,\ref{fig_sec3_bisectors_Fe52502}, it is obvious that the inclusion of the entire ``C''- or ``\textbackslash''-shape results in a stronger convective blueshift than in the case of the mere line core. The overall change of the mean center-to-limb variation is depicted in Fig.\,\ref{fig_sec3_clv_resolution_Fe52502} (panel b, curves at $\mathrm{R=700\,000}$). 
It manifests in a relative increase in blueshift accompanied by a change of the {scope of} the center-to-limb curve. {At the disk center, the relative increase in blueshift by more than $\mathrm{120\,m\,s^{-1}}$ is strongest}. At heliocentric positions around $\mu=0.8$, the {increase of around} $\mathrm{50\,m\,s^{-1}}$ is smallest. Toward the solar limb at $\mu=0.8$, the relative increase in blueshift is again $\mathrm{120\,m\,s^{-1}}${, which turns the average redshift of the line core into an overall slight blueshift.} The detailed numbers at all heliocentric positions are listed in Table\,\ref{table_sec3_resolution_comparison} (line section: Full; spectral resolution: 700\,000).

\paragraph{Changes by lowering the spectral resolution:} With a spectral resolution of more than 700\,000, LARS exceeds the spectral resolving capabilities of most instruments. To allow for a direct comparison with observations made by other instruments, we performed a numerical transformation of our measurements to lower spectral resolutions. For Dopplergrams which contain a quiet Sun region, this provides the opportunity of a relative velocity calibration with the absolute reference values obtained with LARS. The spectral resolution of an observation is coupled to the spectral point-spread function of the instrument. To compute spectra with a lower spectral resolution, we performed a convolution of the original LARS spectra with a Gaussian function, the width of which corresponds to the respective lower spectral resolution $\mathrm{R=\lambda/\Delta\lambda}$ of 250\,000, 180\,000, and 100\,000. We note that the spectral sampling was retained. The given spectral resolutions were selected with regard to the current and future prime solar spectrometers and spectrographs. Ordered in decreasing spectral resolution, these are ViSP \citep[$\mathrm{R\sim250\,000}$,][]{2012SPIE.8446E..6XD}, the GREGOR Fabry-P{\'e}rot Interferometer \citep[$\mathrm{R\sim250\,000}$,][]{2013OptEn..52h1606P}, the Interferometric BIdimensional Spectropolarimeter \citep[$\mathrm{R\sim250\,000}$,][]{2006SoPh..236..415C}, CRISP \citep[$\mathrm{R\sim122\,000}$,][]{2006A&A...447.1111S}, VTF \citep[$\mathrm{R\sim100\,000}$,][]{2012SPIE.8446E..77K}, HMI \citep[$\mathrm{R\sim81\,000}$,][]{2012SoPh..275..229S}, and IMAX \citep[$\mathrm{R\sim70\,000}$,][]{2011SoPh..268...57M}. 

Certainly, the reduction of the spectral resolution entails a broadening of the line profile accompanied by a decrease in line depth by about 15\%, from 0.73 ($\mathrm{1-I_{min}}$) at the original resolution $\mathrm{R=700\,000}$ to 0.62 at $\mathrm{R=100\,000}$. Next, we examined the asymmetry for the degraded line profiles. Fig.\,\ref{fig_sec3_clv_resolution_Fe52502} (panel a) displays the evolution of the line bisector by lowering the spectral resolution. In addition to the mentioned decrease in line depth, the steep gradient of the C-shaped bisector diminishes with decreasing spectral resolution. From $\mathrm{R=700\,000}$ to $\mathrm{R=100\,000}$, the maximum blueshift at normalized intensities around 0.74 decreases by $\mathrm{40\,m\,s^{-1}}$ to around $\mathrm{-250\,m\,s^{-1}}$. At the same time, the blueshift of the line core increases by more than $\mathrm{120\,m\,s^{-1}}$ to around $\mathrm{-220\,m\,s^{-1}}$. The strong effect of the spectral resolution on the line asymmetry reflects in the computation of the quantitative convective blueshift of the line core, defined as the average Doppler velocity of the lower 5\% of the bisector. All line core shifts are listed in Table\,\ref{table_sec3_resolution_comparison} (line section: Core) according to the respective heliocentric position and spectral resolution. To allow for an easier comparison, we display the center-to-limb variations of the line core in Fig.\,\ref{fig_sec3_clv_resolution_Fe52502} (panel b). All across the solar disk, the convective blueshift increases with decreasing spectral resolution. At disk center, we yield the maximal increase in blueshift. The difference between $\mathrm{R=700\,000}$ and $\mathrm{R=100\,000}$ is $\mathrm{122\,m\,s^{-1}}$. Toward the solar limb, the difference becomes smaller, reaching its minimum of $\mathrm{40\,m\,s^{-1}}$ at $\mu=0.7$. In the next step, we analyze the resolution-dependent change of the convective blueshift of the entire line (average over the lower 93\% of the bisector). The comparison is also shown in Fig.\,\ref{fig_sec3_clv_resolution_Fe52502} (panel b). Evidently, a lower spectral resolution hardly affects the mean line shift and its center-to-limb variation. The overall maximum difference is below $\mathrm{10\,m\,s^{-1}}$. Since this is the case for all spectral lines, we confine the listing in Table\,\ref{table_sec3_resolution_comparison} (line section: Full) to the convective shifts at $\mathrm{R=700\,000}$.
 
\paragraph{Comparison with theoretical models:} The final step of our analysis is the comparison our observations with the theoretical synthesis of the convective blueshift and its center-to-limb variation. \citet{2011A&A...528A.113D} carried out radiative transfer computations in LTE for \ion{Fe}{I}\,5250.2\,\AA\ and other spectral lines (\ion{Fe}{I}\,5250.6\,\AA, \ion{C}{I}\,5380.3\,\AA, \ion{Fe}{I}\,5576.1\,\AA, \ion{Fe}{I}\,6301.5\,\AA, and \ion{Fe}{I}\,6302.5\,\AA). The synthesized Doppler shifts of the line core are displayed in Fig.\,\ref{fig_sec3_clv_resolution_Fe52502} (panel b) for the spectral resolution of CRISP ($\mathrm{R=122\,000}$). From disk center at $\mu=1.0$ to heliocentric positions around $\mu=0.85$, the synthesized center-to-limb variation assorts well with the observations transformed to a spectral resolution of $\mathrm{R=180\,000}$. The error range of the synthesis was estimated to $\mathrm{50\,m\,s^{-1}}$. At the solar limb, the measured and synthetic values also agree well within one standard deviation. However, we find a major deviation of the synthesized center-to-limb curve from the observation for the range between $\mu=0.85$ and $\mu=0.5$. Whereas the observed curve describes an almost linear decrease in blueshift, the slope of the synthesized curve features a further increase in blueshift till $\mu=0.7$. The displaced maximum of the synthesized convective blueshift hints toward a slight overvaluation of horizontal flow speeds in the mid-photospheric granular convection pattern of the 3D hydrodynamical simulation. Moreover, we find a similar deviation for the synthesized convective blueshift of the entire spectral line, calculated as the average Doppler shift of the line bisector from the line minimum to an upper threshold of 95\% of the continuum intensity. The comparison of the measured and theoretical center-to-limb variation is overplotted in Fig.\,\ref{fig_sec3_clv_resolution_Fe52502} (panel b). The synthesis provides good results close to the solar disk center. However, the reproduction of the actual center-to-limb variation of the convective blueshift requires a refinement of the spectral line synthesis or the 3D hydrodynamical model itself.

\subsubsection{\ion{Fe}{I}\,5250.6\,\AA}\label{sec_results_FeI52506}

\begin{figure}[htbp]
\textbf{a)}\\[-0.3cm]  \includegraphics[width=\columnwidth]{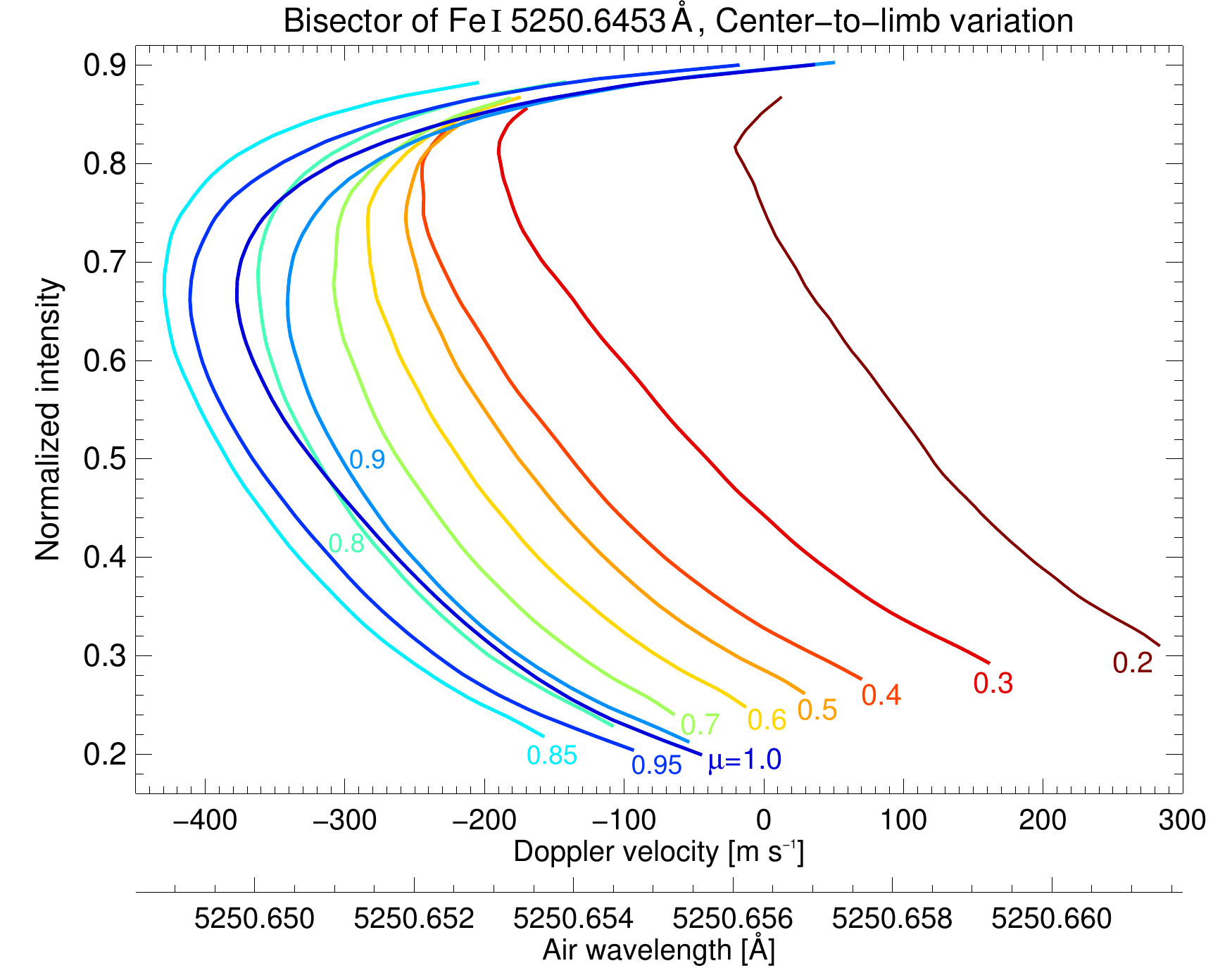}\\[0.2cm]
\textbf{b)}\\[-0.3cm]  \includegraphics[width=\columnwidth]{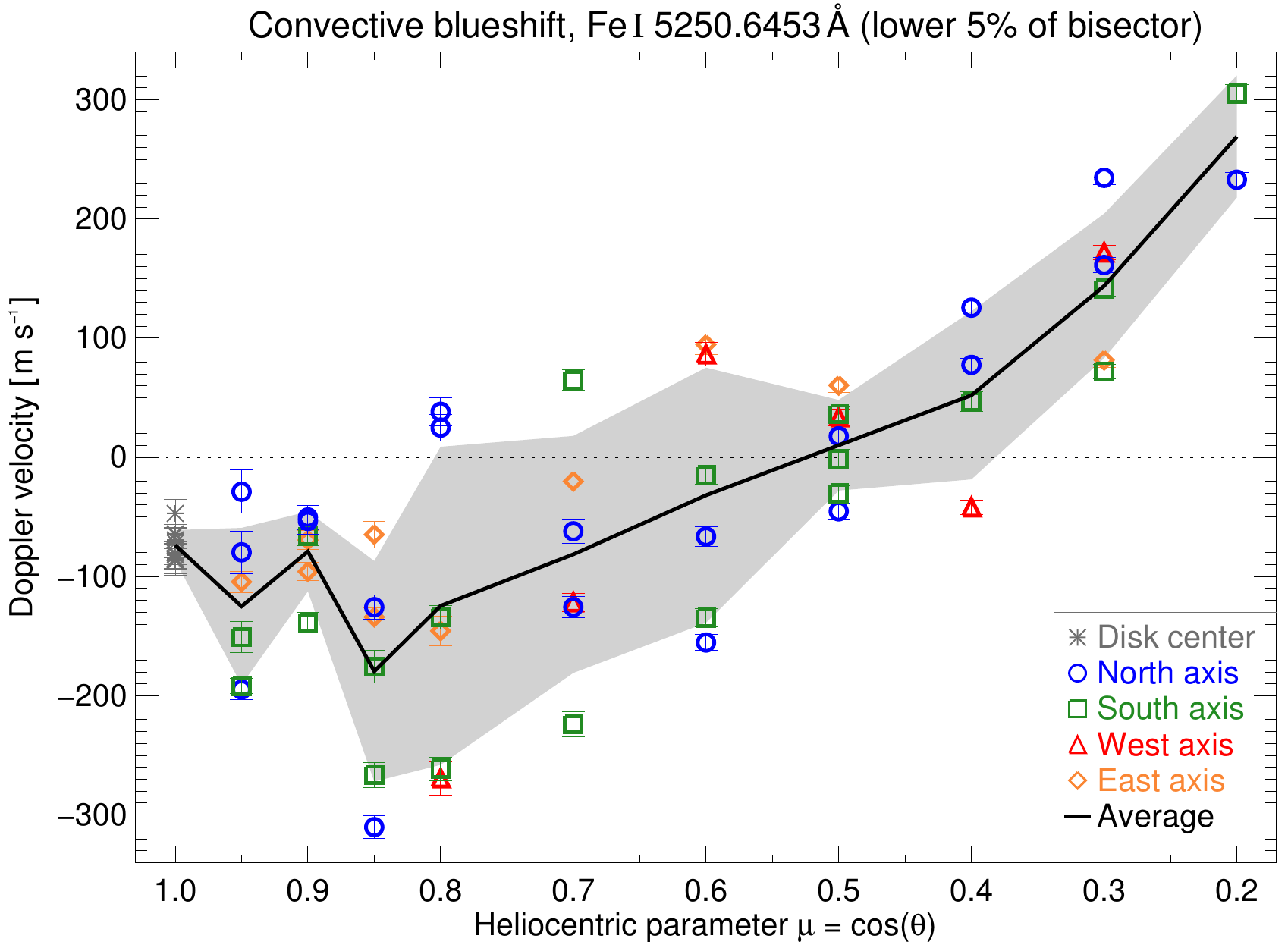}\\[0.2cm]
\textbf{c)}\\[-0.3cm]  \includegraphics[width=\columnwidth]{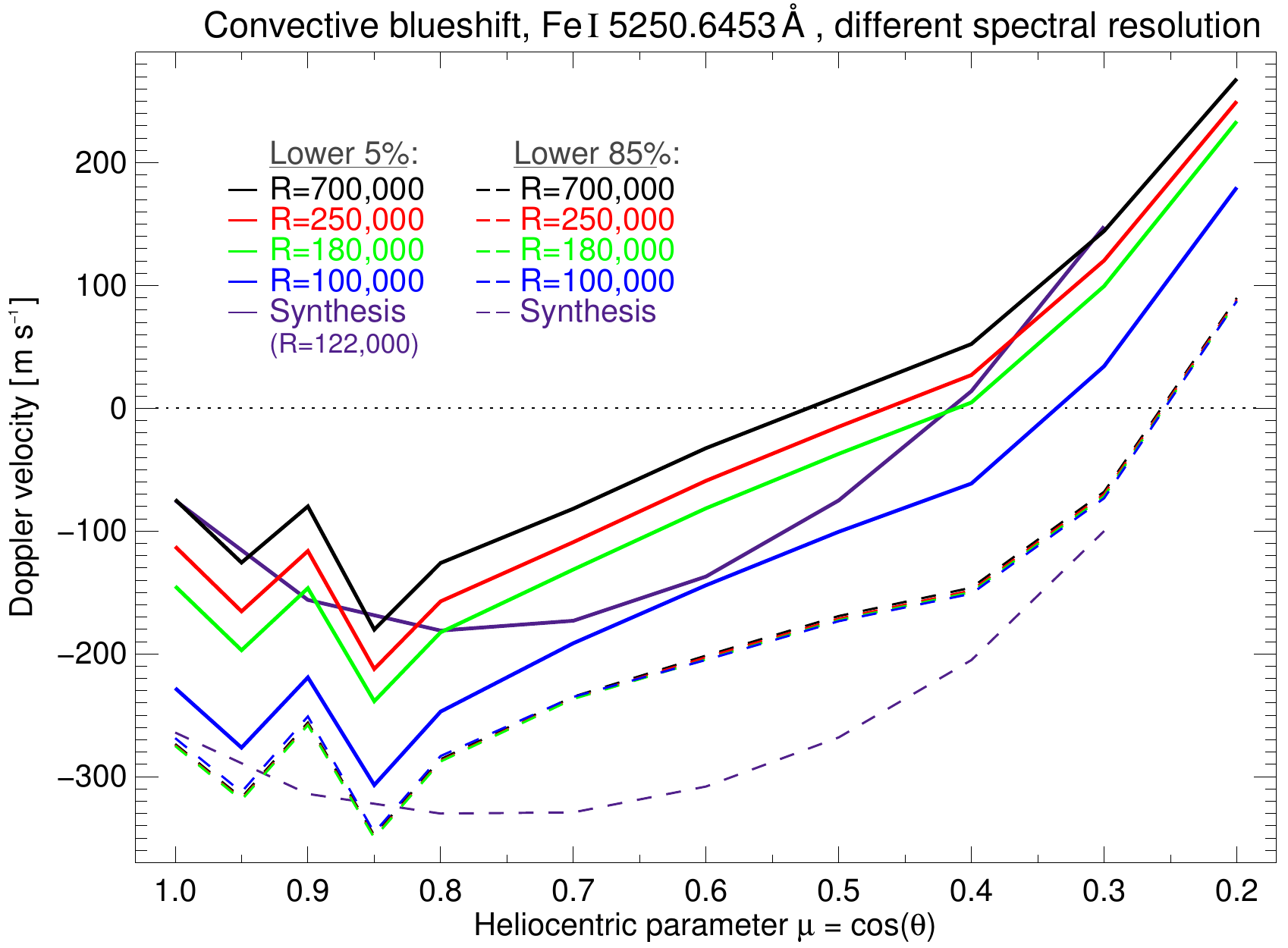}
\caption{Center-to-limb variation of the \ion{Fe}{I}\,5250.6\,\AA\ line. Panel a displays the average line bisectors from the disk center ($\mu=1.0$) toward the solar limb ($\mu=0.2$). Panel b shows the average convective blueshift of the line core (lower 5\,\% of the bisectors) for each observation. Colors and symbols indicate the axes. The average curve and its standard deviation are displayed as black solid line and gray shaded area. Panel c compares the observed convective blueshift for different spectral resolutions and line sections with the synthesis of \citet{2011A&A...528A.113D}. Dashed lines are close to each other or even overlay.}
\label{fig_sec3_analysis_Fe52506}
\end{figure}

The second important line of the 5250\,\AA\ region is the \ion{Fe}{I}\,5250.6\,\AA\ line. As shown in Fig.\,\ref{fig_sec3_spectra_5250}, the \ion{Fe}{I}\,5250.6\,\AA\ line is deeper than the neighboring \ion{Fe}{I}\,5250.2\,\AA\ line. Despite their proximity both line profiles do not affect each other. The \ion{Ti}{I} line at 5250.9\,\AA\ blends the outer wing of the \ion{Fe}{I}\,5250.6\,\AA\ line though. We thus had to limit the bisector analysis to normalized intensities below 0.9. The center-to-limb variation of the line asymmetry is displayed in Fig.\,\ref{fig_sec3_analysis_Fe52506} (panel a). 
The bisector at disk center manifests in a strong C-shape with a maximum blueshift of around $\mathrm{-380\,m\,s^{-1}}$ at an intensity (normalized to the continuum intensity) of 0.67, followed by a steep decrease in blueshift to around $\mathrm{-50\,m\,s^{-1}}$ at the line minimum at an intensity of 0.2. In comparison, the bisector of \ion{Fe}{I}\,5250.2\,\AA\ at disk center yielded a difference of only $\mathrm{200\,m\,s^{-1}}$ between the maximum blueshift and the line minimum. The center-to-limb variation of the \ion{Fe}{I}\,5250.6\,\AA\ bisector also exhibits the strongest overall blueshift for heliocentric positions around $\mu=0.85$. Toward the solar limb ($\mu=0.5-0.2$), the bisector transforms into the \textbackslash-shape with line core velocities indicating strong redshifts. The intensity of the line minimum increases from 0.2 at disk center to 0.31 at $\mu=0.2$. Former estimations of the average formation height of the line core to around 360\,km above the solar surface at $\tau_{5000\,\AA}=1$ suggest that the \ion{Fe}{I}\,5250.6\,\AA\ line samples slightly higher atmospheric layers than the \ion{Fe}{I}\,5250.2\,\AA\ line. 

In line with a higher formation layer, the line core of \ion{Fe}{I}\,5250.6\,\AA\ feature a weaker blueshift (or stronger redshifts at the limb) than the \ion{Fe}{I}\,5250.2\,\AA\ line. The distribution and center-to-limb variation of the line core velocities is displayed in Fig.\,\ref{fig_sec3_analysis_Fe52506} (panel b). At disk center, we obtain a mean blueshift of $\mathrm{-74\,m\,s^{-1}}$ with a standard deviation of below $\mathrm{15\,m\,s^{-1}}$. At $\mu=0.85$, the center-to-limb variation reaches its maximum blueshift of $\mathrm{-180\,m\,s^{-1}}$ with a standard deviation of $\mathrm{90\,m\,s^{-1}}$. Toward the solar limb, the blueshift decreases monotonically. It already turns into a redshift at $\mu=0.5$, and arrives at the maximum redshift of $\mathrm{+268\,m\,s^{-1}}$ at $\mu=0.2$. The values are listed in Table\,\ref{table_sec3_resolution_comparison} (line section: Core; $\mathrm{R}$: 700\,000). We register a change of the slope of the the center-to-limb variation in Fig.\,\ref{fig_sec3_analysis_Fe52506} (panel b). For \ion{Fe}{I}\,5250.6\,\AA, the increase in blueshift from $\mu=1.0$ to $\mu=0.85$ by around $\mathrm{100\,m\,s^{-1}}$ has halved, compared to the case of \ion{Fe}{I}\,5250.2\,\AA. Apparently, the \ion{Fe}{I}\,5250.6\,\AA\ and its higher line core formation capture the reversal point of the convective blueshift due to line-of-sight effects to a lesser extent.

The center-to-limb variation of the convective blueshift is displayed in Fig.\,\ref{fig_sec3_analysis_Fe52506} (panel c) for different spectral resolutions and bisector segments. All Doppler velocities are also listed in Table\,\ref{table_sec3_resolution_comparison}. Due to the strongly asymmetric shape of the bisectors, the lowering of the spectral resolution has a major effect on the line core velocity. At disk center, the convective blueshift increases by more than $\mathrm{150\,m\,s^{-1}}$ from $\mathrm{R=700\,000}$ to $\mathrm{R=100\,000}$. Toward the solar limb, the difference still amounts to around $\mathrm{100\,m\,s^{-1}}$. On the contrary, a change in spectral resolution hardly affects the average Doppler shift of the entire spectral line. Though the calculation of the mean line shift for the lower 85\% of the bisector yield significantly stronger blueshifts than obtained for the line core, the slope of the center-to-limb variation variation largely remains. 

The synthesized center-to-limb variation of \citet{2011A&A...528A.113D} is added in in Fig.\,\ref{fig_sec3_analysis_Fe52506} (panel c). For the line core, the synthesis (at $\mathrm{R=122\,000}$) provides blueshifts at disk center which are in line with our original observations (at $\mathrm{R=700\,000}$). However, considering the spectral resolution, the synthesis fails to reproduce the observations around disk center. For the average of the entire spectral line, the synthesis is in good agreement with our observations. Only at heliocentric positions between $\mu=0.7$ and $\mu=0.5$, the synthesis can not reproduce the measured center-to-limb variation.

\subsection{Lines around 5381\,\AA}\label{sec_results_5381}
\begin{figure}[htbp]
\begin{center}
\includegraphics[width=\columnwidth]{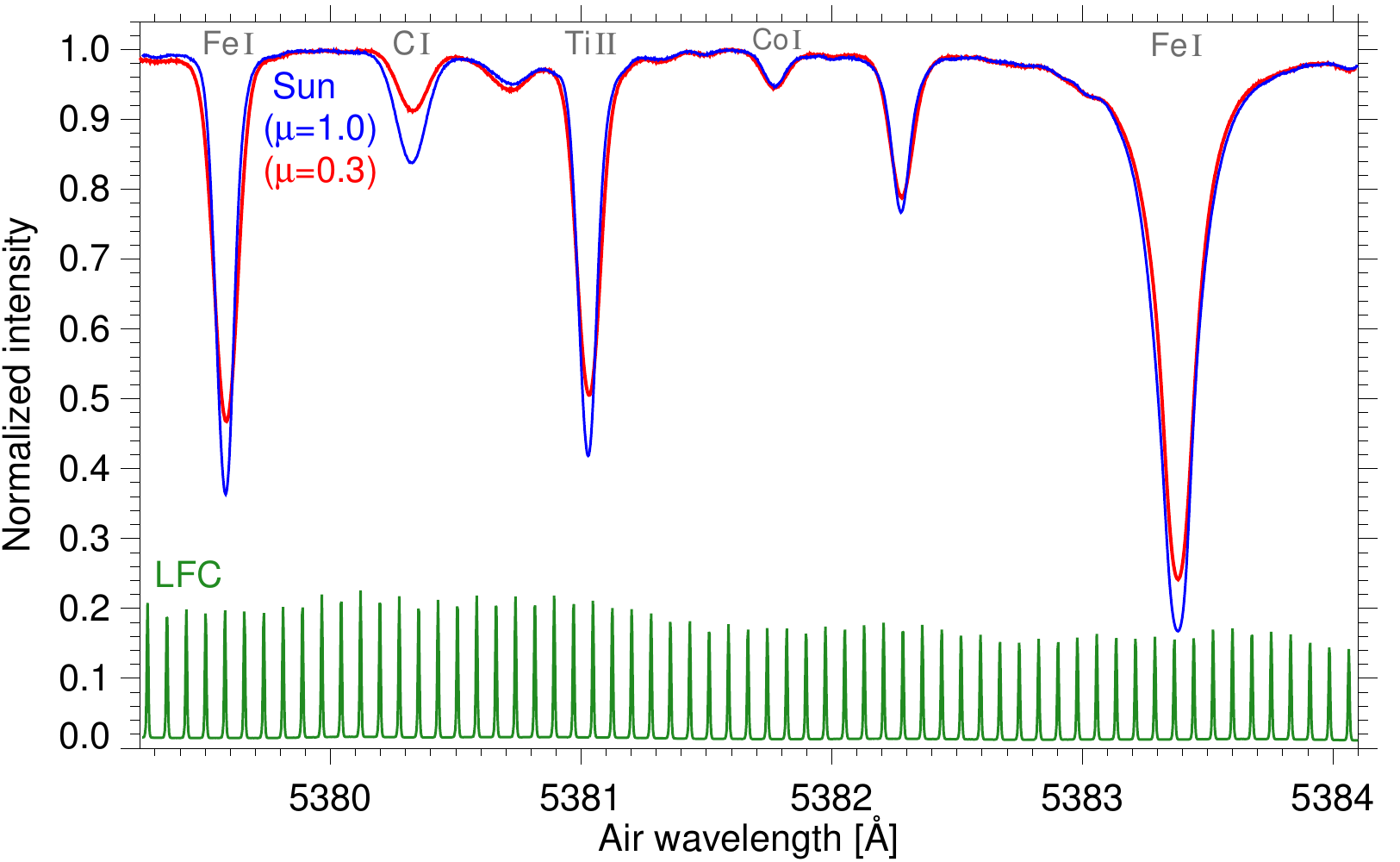}
\caption{Spectral region around 5381\,\AA, with the quiet Sun absorption spectra at the disk center ($\mu=1.0$, blue curve) and close to the solar limb ($\mu=0.3$, red curve). The {atomic species} are stated in gray. The spectrum of the laser frequency comb (LFC) is displayed as green curve.}
\label{fig_sec3_spectra_5381}
\end{center}
\end{figure}

Our observations of the 5381\,\AA\ region shown in Fig.\,\ref{fig_sec3_spectra_5381} covered 75 observation sequences from $\mu=1.0$ to $\mu=0.3$. We analyzed the convective blueshift of the spectral lines \ion{Fe}{I}\,5379.6\,\AA, \ion{C}{I}\,5380.3\,\AA, \ion{Ti}{II}\,5381.0\,\AA, and \ion{Fe}{I}\,5383.4\,\AA.

The reference wavelength for the calculation of Doppler shifts was taken from the NIST ASD. The observed air wavelengths (and their uncertainties) of the spectral lines amounted to \ion{Fe}{I}\,5379.5737\,\AA\ ($\pm$0.9\,m\AA), \ion{C}{I}\,5380.3308\,\AA\ ($\pm$0.3\,m\AA), \ion{Ti}{II}\,5381.0212\,\AA\ ($\pm$0.6\,m\AA), and \ion{Fe}{I}\,5383.3688\,\AA\ ($\pm$0.9\,m\AA). In the following section, we discuss the analysis of the convective blueshift for the \ion{C}{I}\,5380.3\,\AA\ line. The results for the other spectral lines are shown in Figs.\,\ref{fig_A3} and \ref{fig_A4}.

\subsubsection{\ion{C}{I}\,5380.3\,\AA}

\begin{figure}[htbp]
\textbf{a)}\\[-0.3cm]  \includegraphics[width=\columnwidth]{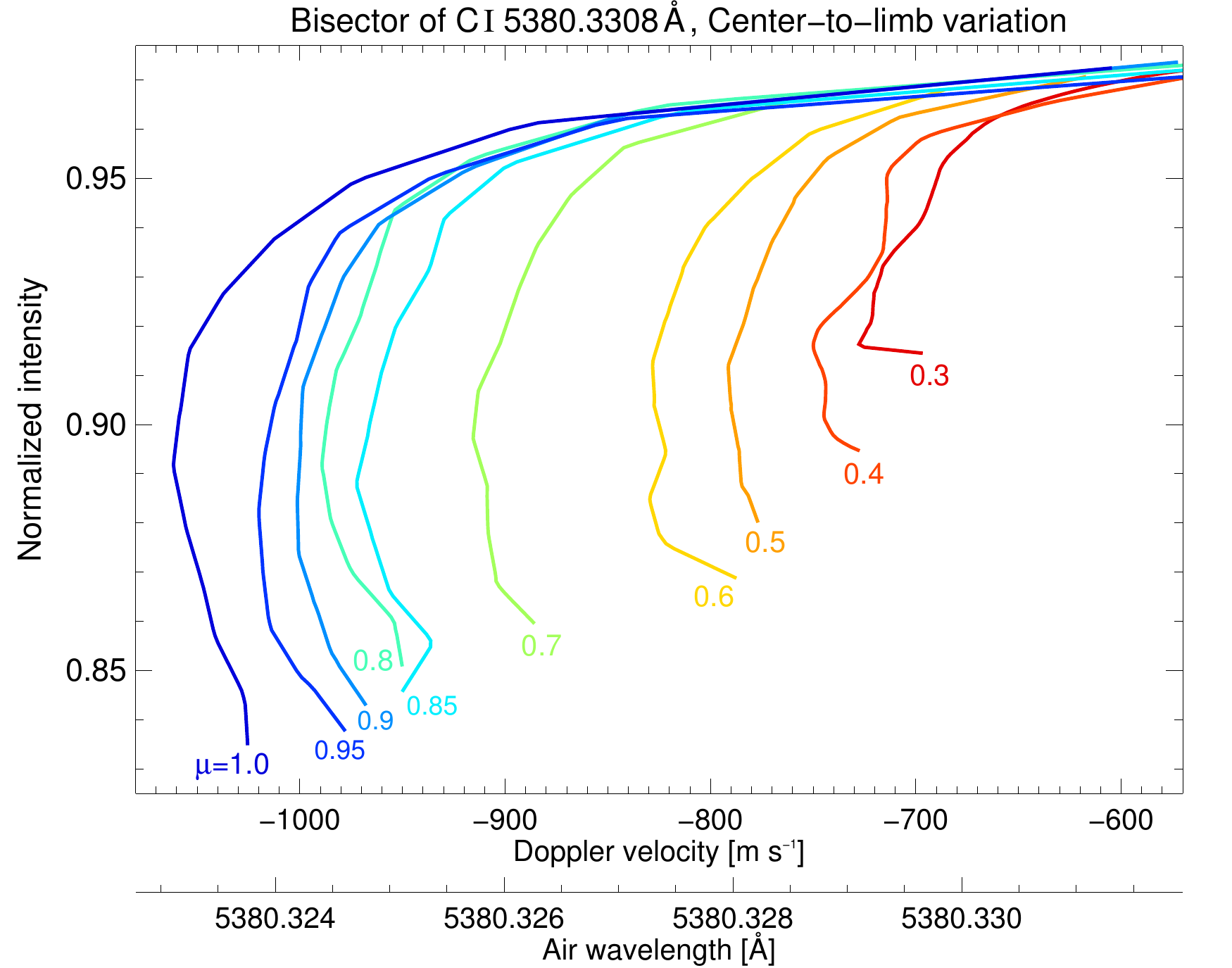}\\[0.2cm]
\textbf{b)}\\[-0.3cm]  \includegraphics[width=\columnwidth]{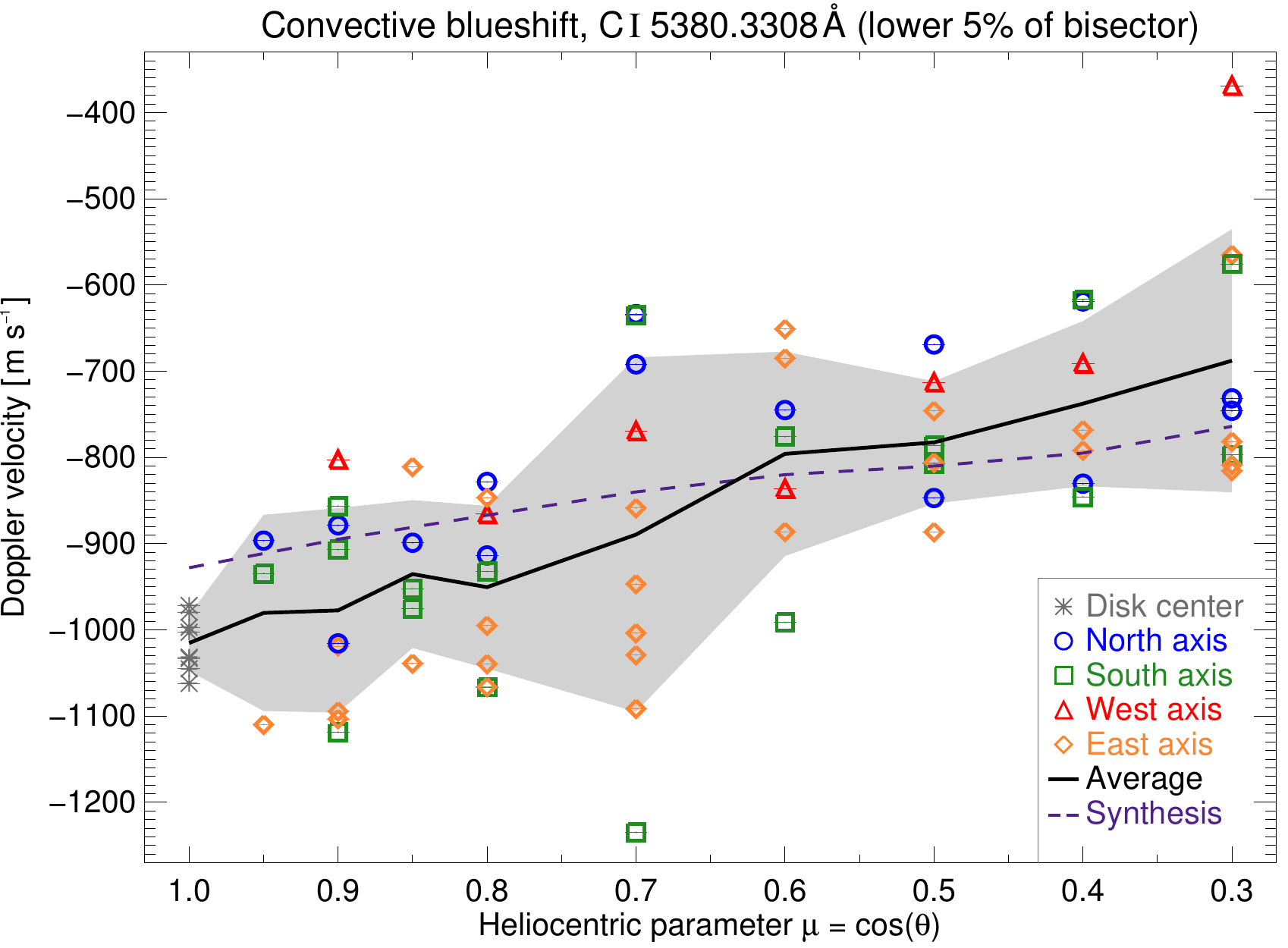}\\[0.2cm]
\textbf{c)}\\[-0.3cm]  \includegraphics[width=\columnwidth]{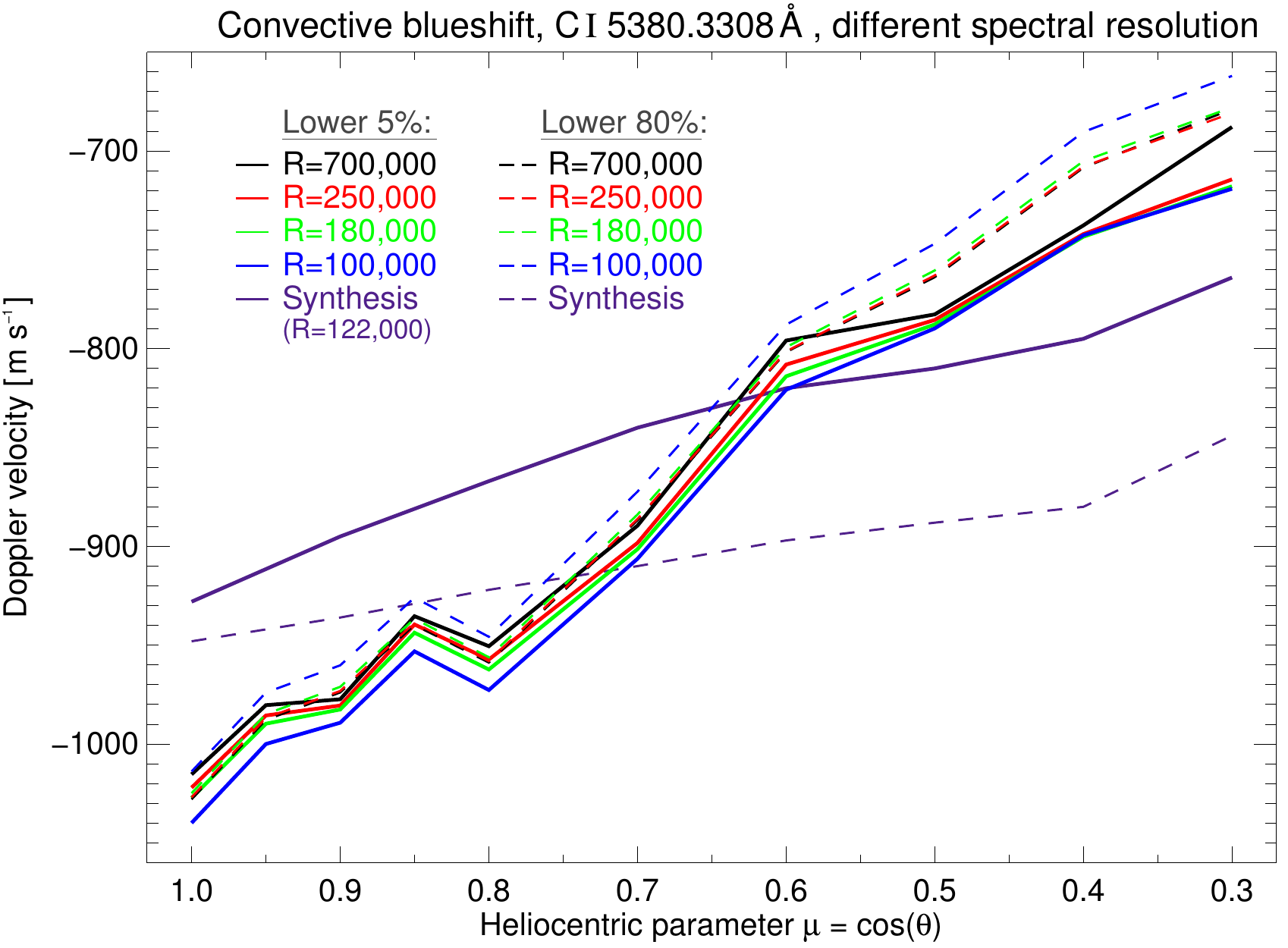}
\caption{Center-to-limb variation of the \ion{C}{I}\,5380.3\,\AA\ line. Panel a displays the average line bisectors from the disk center ($\mu=1.0$) toward the solar limb ($\mu=0.3$). Panel b shows the average convective blueshift of the line core (lower 5\,\% of the bisectors) for each observation. Colors and symbols indicate the axes. The average curve and its standard deviation are displayed as black solid line and gray shaded area. Panel c compares the observed convective blueshift for different spectral resolutions and line sections with the synthesis of \citet{2011A&A...528A.113D}.}
\label{fig_sec3_analysis_C53803}
\end{figure}

The core of the weak \ion{C}{I}\,5380.3\,\AA\ line is formed only around 40\,km above the solar surface at $\tau_{5000\,\AA}=1$ \citep{1991PhDT.......113F,1991A&A...243..244G}. The line samples the lowest layer of the photosphere and provides information about the near-surface convective blueshift. Thus, \ion{C}{I}\,5380.3\,\AA\ has been used for observations of convective flows in, for example, sunspot penumbrae \citep{1999A&A...349L..37S,2011Sci...333..316S,2011ApJ...734L..18J}. The line has also been of interest for measurements of intensity fluctuations of the Sun and other stars due to its temperature sensitivity \citep[e.g.,][]{1997ApJ...474..798G}. However, \citet{2012ASPC..463...99U} argued that Doppler measurements of regions with lower temperature are problematic. With decreasing temperature, the line disappears due to its high excitation potential and gets contaminated by MgH lines. Thus, the line should be almost non-existent in dark intergranular lanes. In conclusion, the contribution of the intergranular redshifts to the line profile should be small. 

The analysis of the convective blueshift shown in Fig.\,\ref{fig_sec3_analysis_C53803} confirms the prediction. We obtain a strong blueshift of the line core of up to $\mathrm{-1000\,m\,s^{-1}}$ at the disk center. It seems that only the strong vertical upflow of the bright granulation contributes to the line shift. Toward $\mu=0.3$ near the solar limb, the blueshift decreases monotonically to around $\mathrm{-700\,m\,s^{-1}}$. Despite the strong decrease in line depth, the shape of the bisector remains almost unaffected. As shown in panel c (and listed in Table\,\ref{table_sec3_resolution_comparison}), changes of the spectral resolution or the averaged bisector segment have minor impact on the convective blueshift. The synthesis of \citet{2011A&A...528A.113D} yields blueshifts of the same order, but with a significantly different gradient of the center-to-limb variation.

\subsection{Lines around 5434\,\AA}\label{sec_results_5434}

\begin{figure}[htbp]
\begin{center}
\includegraphics[width=\columnwidth]{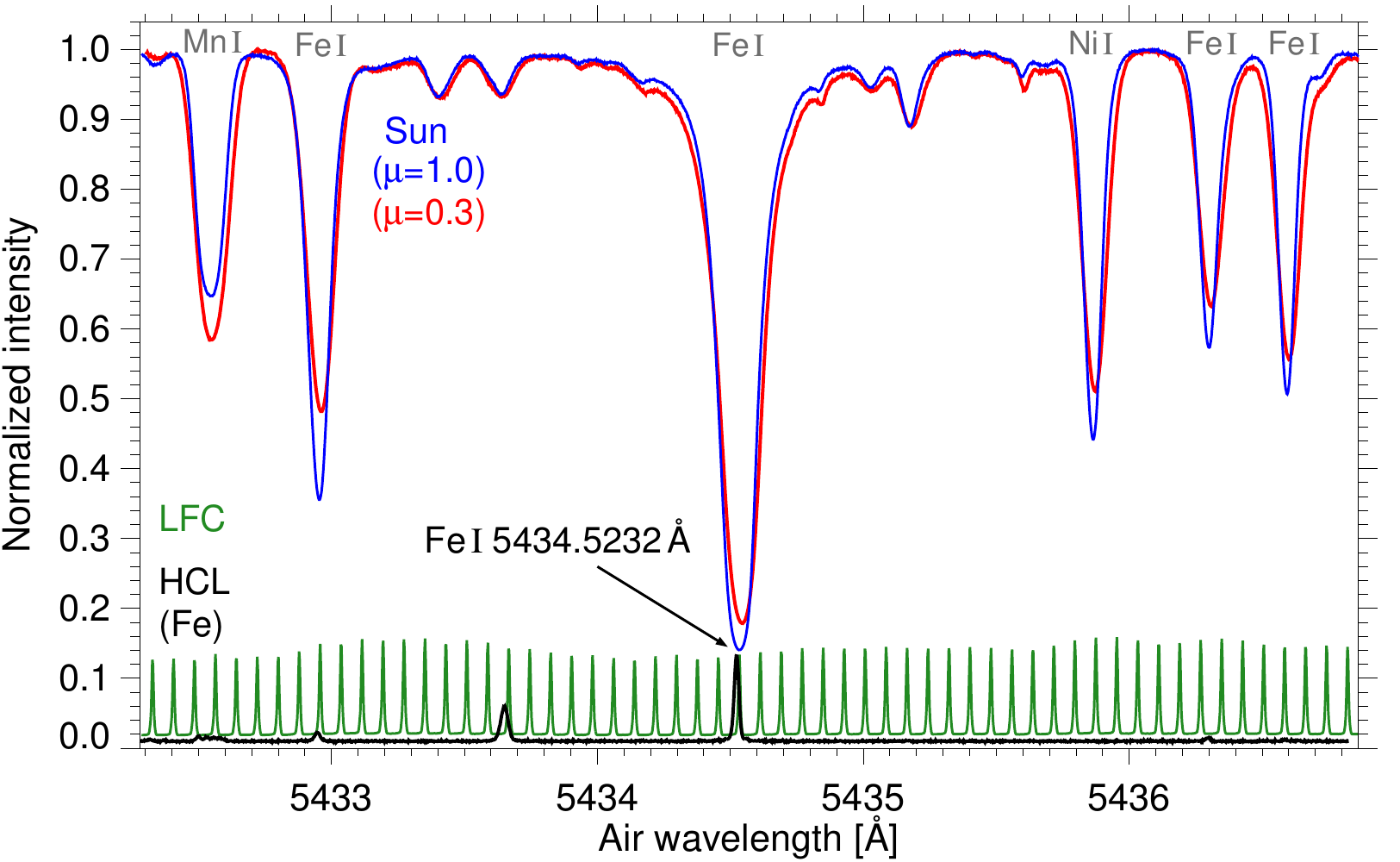}
\caption{Spectral region around 5434\,\AA, with the quiet Sun absorption spectra at the disk center ($\mu=1.0$, blue curve) and close to the solar limb ($\mu=0.3$, red curve). The {atomic species} are stated in gray. The emission spectra of the laser frequency comb (LFC, green curve) and the iron hollow cathode lamp (HCL, black curve) are displayed.}
\label{fig_sec3_spectra_5434}
\end{center}
\end{figure}

The observations of the 5434\,\AA\ region shown in Fig.\,\ref{fig_sec3_spectra_5434} comprised 91 observation sequences from $\mu=1.0$ to $\mu=0.2$. We analyzed the convective blueshift of the spectral lines \ion{Mn}{I}\,5432.5\,\AA, \ion{Fe}{I}\,5432.9\,\AA, \ion{Fe}{I}\,5434.5\,\AA, \ion{Ni}{I}\,5435.9\,\AA, \ion{Fe}{I}\,5436.3\,\AA, and \ion{Fe}{I}\,5436.6\,\AA.

We measured the reference wavelengths of \ion{Fe}{I}\,5432.9470\,\AA, \ion{Fe}{I}\,5434.5232\,\AA, and \ion{Fe}{I}\,5436.2947\,\AA\ with the iron hollow cathode lamp of LARS. The uncertainty was below 0.1\,m\AA. For \ion{Mn}{I}\,5432.539\,\AA, \ion{Ni}{I}\,5435.858\,\AA, and \ion{Fe}{I}\,5436.588\,\AA, the observed air wavelength from the NIST ASD was not accurate enough for the calculation of Doppler velocities. Matching the line profiles with similar well-known spectral lines allowed us to refine the reference wavelength to an uncertainty of 1\,m\AA. 

The spectral region was selected because of the \ion{Fe}{I}\,5434.5232\,\AA\ line. With a formation height of the line core of around 550\,km \citep{1991sopo.work.....N,1998A&A...332.1069K} above $\tau_{5000\,\AA}=1$, the line provides information about the entire photosphere from the solar surface up to the chromospheric transition. Since the line is not sensitive  ($g_\mathrm{eff}=0$) to magnetic fields, it is popular for photospheric Doppler velocity measurements \citep[e.g.,][]{2010A&A...522A..31B}. In Section \ref{sec_results_FeI5434}, we present the results of the analysis for the \ion{Fe}{I}\,5434.5\,\AA\ line. In Section \ref{sec_results_FeI54329}, we discuss the distinct center-to-limb variation of the convective blueshift of the \ion{Fe}{I}\,5432.9\,\AA\ line. Another interesting line is the \ion{Mn}{I}\,5432.5\,\AA\ line. Due to its extraordinary line behavior and hyperfine structure, the spectral line has been frequently used in theoretical studies of the manganese abundance and line formation \citep{1972SoPh...27..294M,2001A&A...369L..13D,2009A&A...499..301V}.
The analyses of the convective blueshift of \ion{Mn}{I}\,5432.5\,\AA, \ion{Ni}{I}\,5435.9\,\AA, \ion{Fe}{I}\,5436.3\,\AA, and \ion{Fe}{I}\,5436.6\,\AA\ are displayed in Figs.\,\ref{fig_A5} and \ref{fig_A6}, and the upper panels of Figs.\,\ref{fig_A7} and \ref{fig_A8}.

\begin{figure}[htbp]
\textbf{a)}\\[-0.3cm]  \includegraphics[width=\columnwidth]{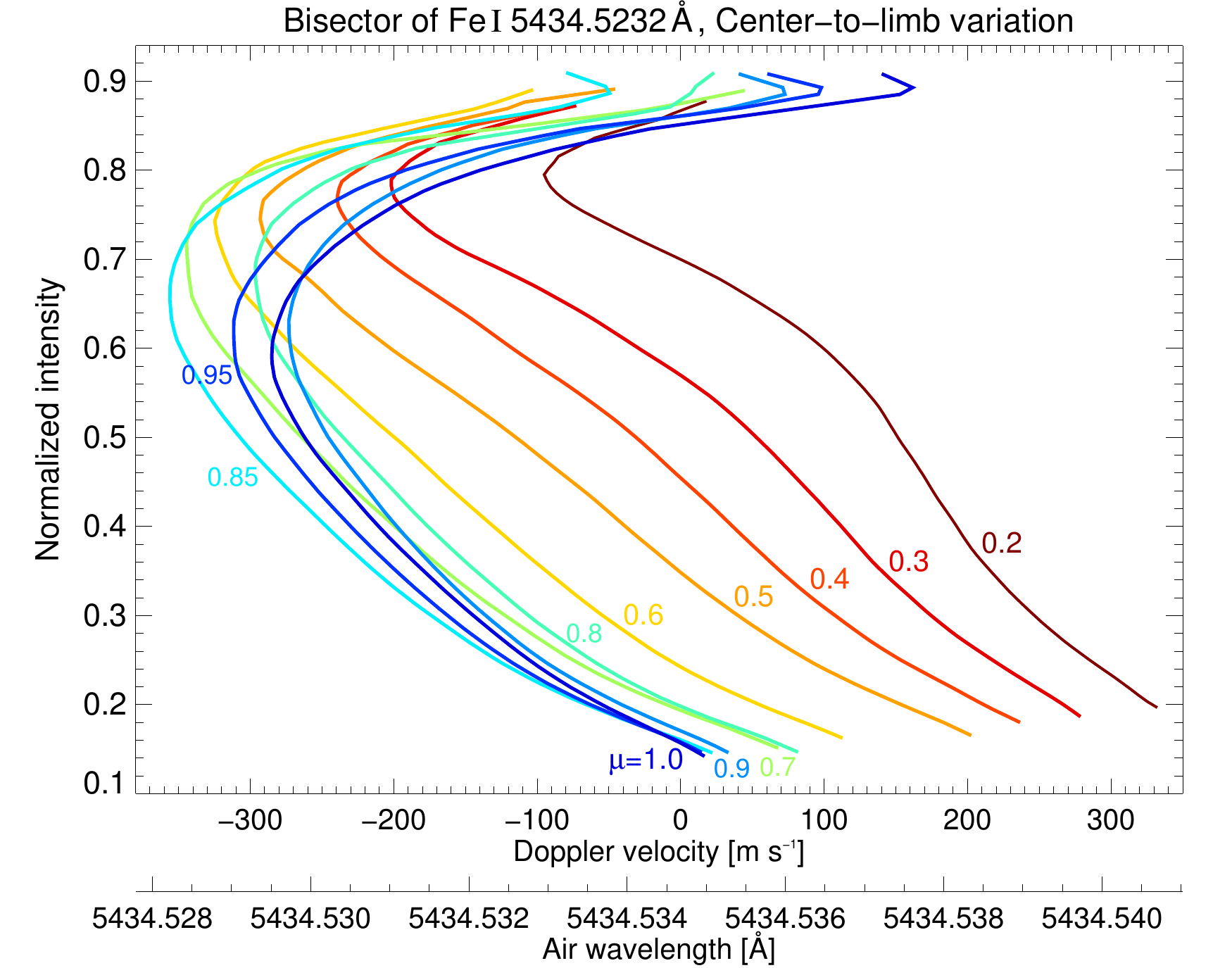}\\[0.2cm]
\textbf{b)}\\[-0.3cm]  \includegraphics[width=\columnwidth]{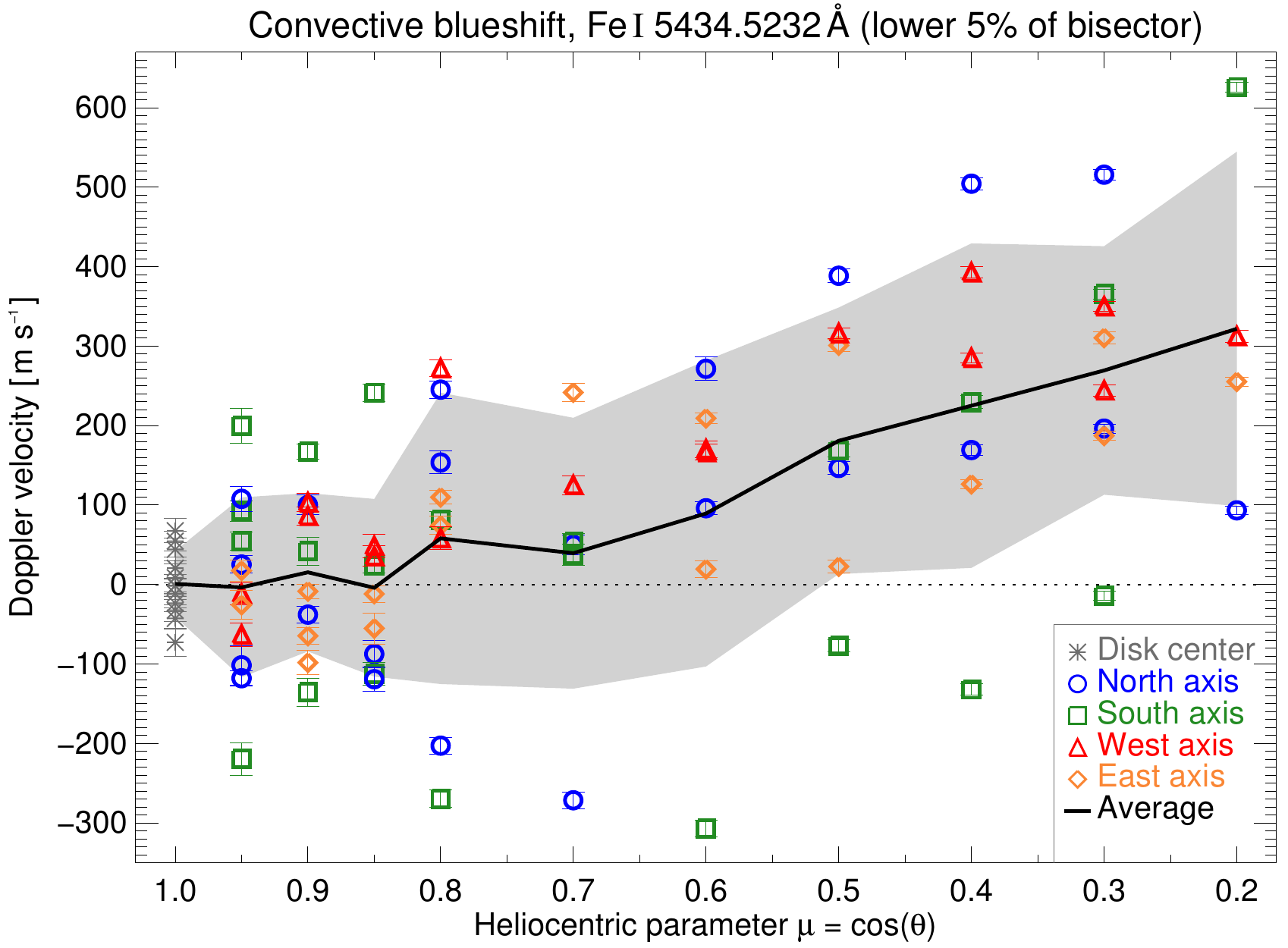}\\[0.2cm]
\textbf{c)}\\[-0.3cm]  \includegraphics[width=\columnwidth]{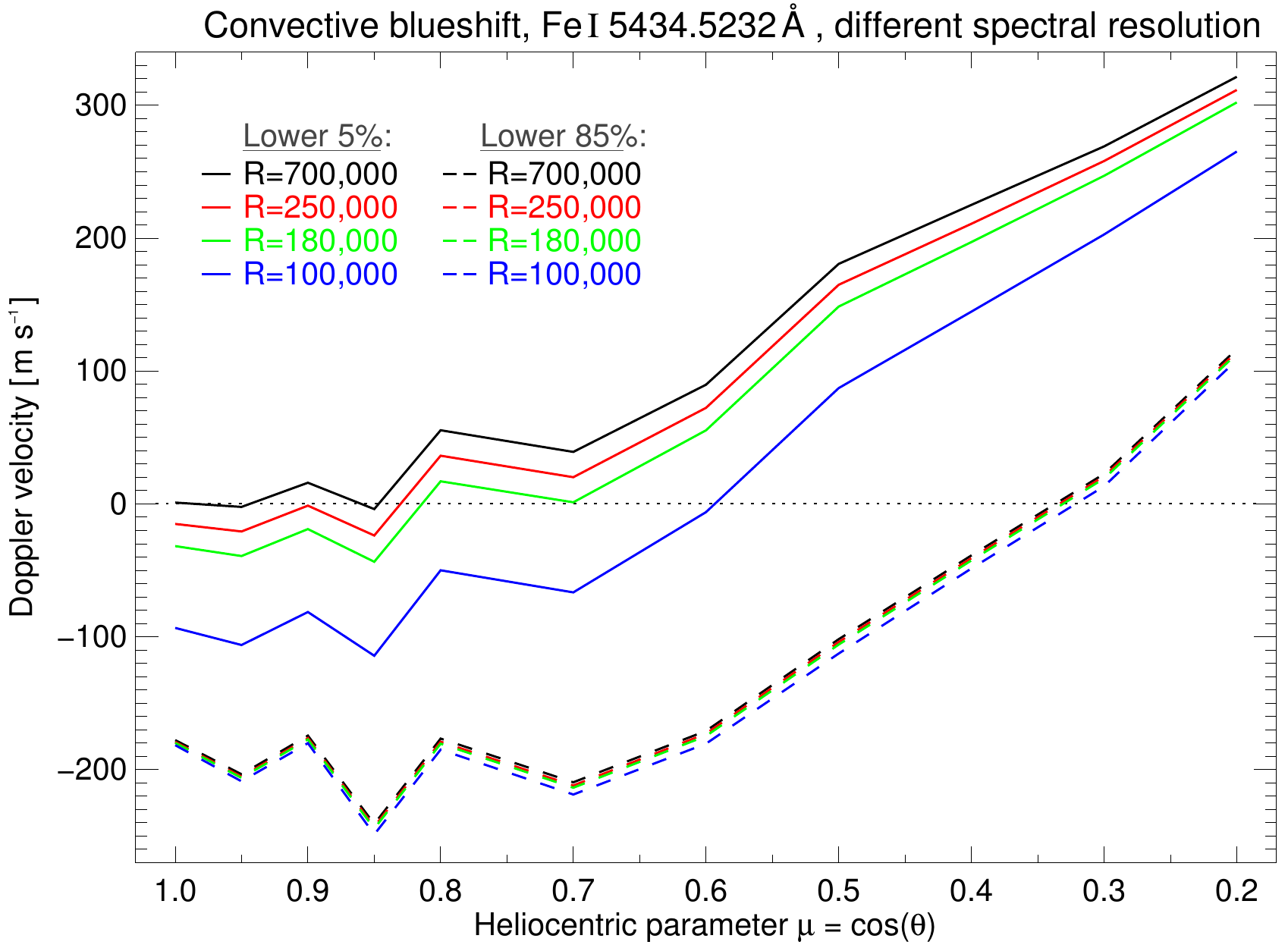}
\caption{Center-to-limb variation of the \ion{Fe}{I}\,5434.5\,\AA\ line. Panel a displays the average line bisectors from the disk center ($\mu=1.0$) toward the solar limb ($\mu=0.2$). Panel b shows the average convective blueshift of the line core (lower 5\,\% of the bisectors) for each observation. Colors and symbols indicate the axes. The average curve and its standard deviation are displayed as black solid line and gray shaded area. Panel c compares the observed convective blueshift for different spectral resolutions and line sections. Dashed lines are close to each other or even overlay.}
\label{fig_sec3_analysis_Fe5434}
\end{figure}

\subsubsection{\ion{Fe}{I}\,5434.5\,\AA}\label{sec_results_FeI5434}
The \ion{Fe}{I}\,5434.5\,\AA\ line is the broadest and deepest line of the spectral region in Fig.\,\ref{fig_sec3_spectra_5434}. At disk center, the normalized line minimum intensity is 0.14. Toward $\mu=0.2$ near the solar limb, the minimum intensity increases slightly to 0.2. The line bisector and its center-to-limb variation are shown in Fig.\,\ref{fig_sec3_analysis_Fe5434} (panel a). At heliocentric positions between $\mu=1.0$ and $\mu=0.7$, the bisector describes a pronounced C-shape with maximum blueshifts around $\mathrm{-300\,m\,s^{-1}}$ at normalized intensities between 0.6 and 0.7. The Doppler shift of the core indicates slight redshifts of around $\mathrm{+30\,m\,s^{-1}}$, even at disk center. We conclude that the line core forms hight enough in the atmosphere to exceed the zone of convective blueshifts. Toward the solar limb, the bisector further shifts to the red and develops a hump of the curvature at intensities around 0.6. We regard this turnover point as the transition into the chromosphere. 

The center-to-limb variation of the line core velocity in Fig.\,\ref{fig_sec3_analysis_Fe5434} (panel b) highlights the trend of increasing redshifts from around $\mathrm{0\,m\,s^{-1}}$ at the disk center to around $\mathrm{+300\,m\,s^{-1}}$ at $\mu=0.2$. We understand the limb redshift as the summed line-of-sight effect of the horizontally granular flow and the inverse granulation. 

As displayed Fig.\,\ref{fig_sec3_analysis_Fe5434} (panel c), a change of the spectral resolution from $\mathrm{R=700\,000}$ to $\mathrm{R=100\,000}$ induces a shift of the line core by up to $\mathrm{-100\,m\,s^{-1}}$. The calculation of the average line shift for the lower 85\% of the bisector results in an additional blueshift to around $\mathrm{-200\,m\,s^{-1}}$ near the disk center. Lowering the spectral resolution insignificantly affects the Doppler shift of the entire spectral line. Detailed values are listed in Table\,\ref{table_sec3_resolution_comparison}.

\subsubsection{\ion{Fe}{I}\,5432.9\,\AA}\label{sec_results_FeI54329}

\begin{figure}[htbp]
\textbf{a)}\\[-0.3cm]  \includegraphics[width=\columnwidth]{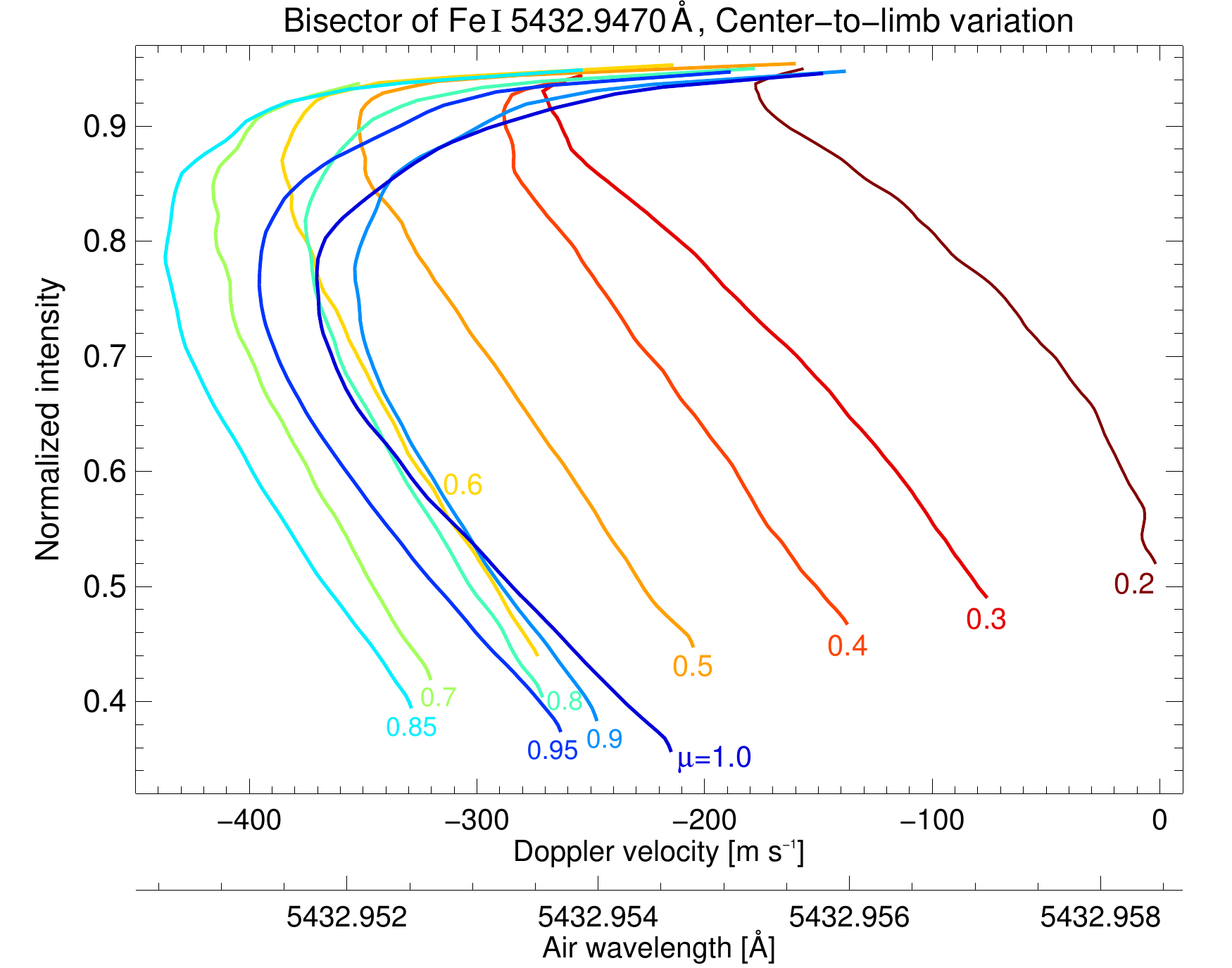}\\[0.2cm]
\textbf{b)}\\[-0.3cm]  \includegraphics[width=\columnwidth]{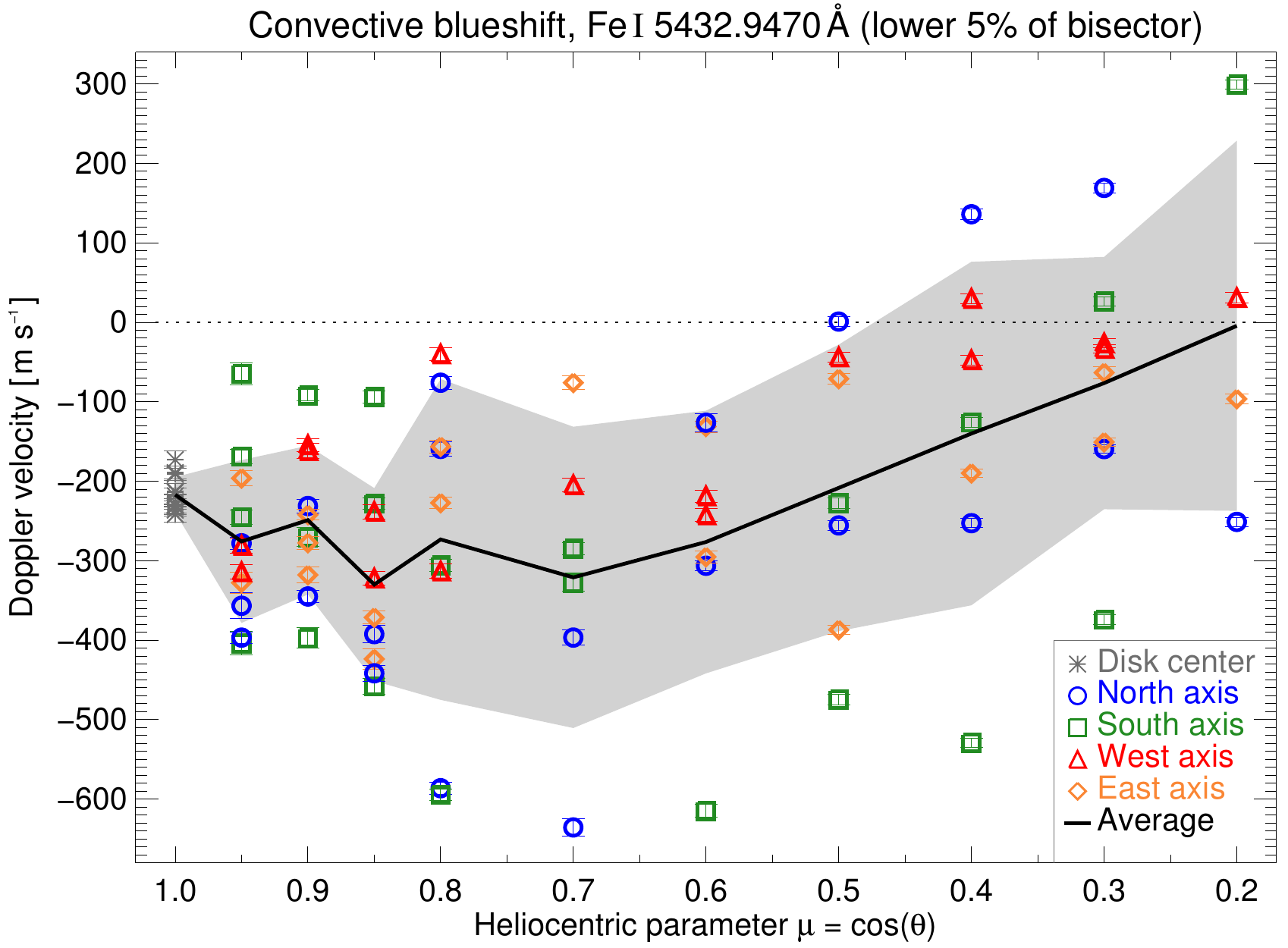}\\[0.2cm]
\textbf{c)}\\[-0.3cm]  \includegraphics[width=\columnwidth]{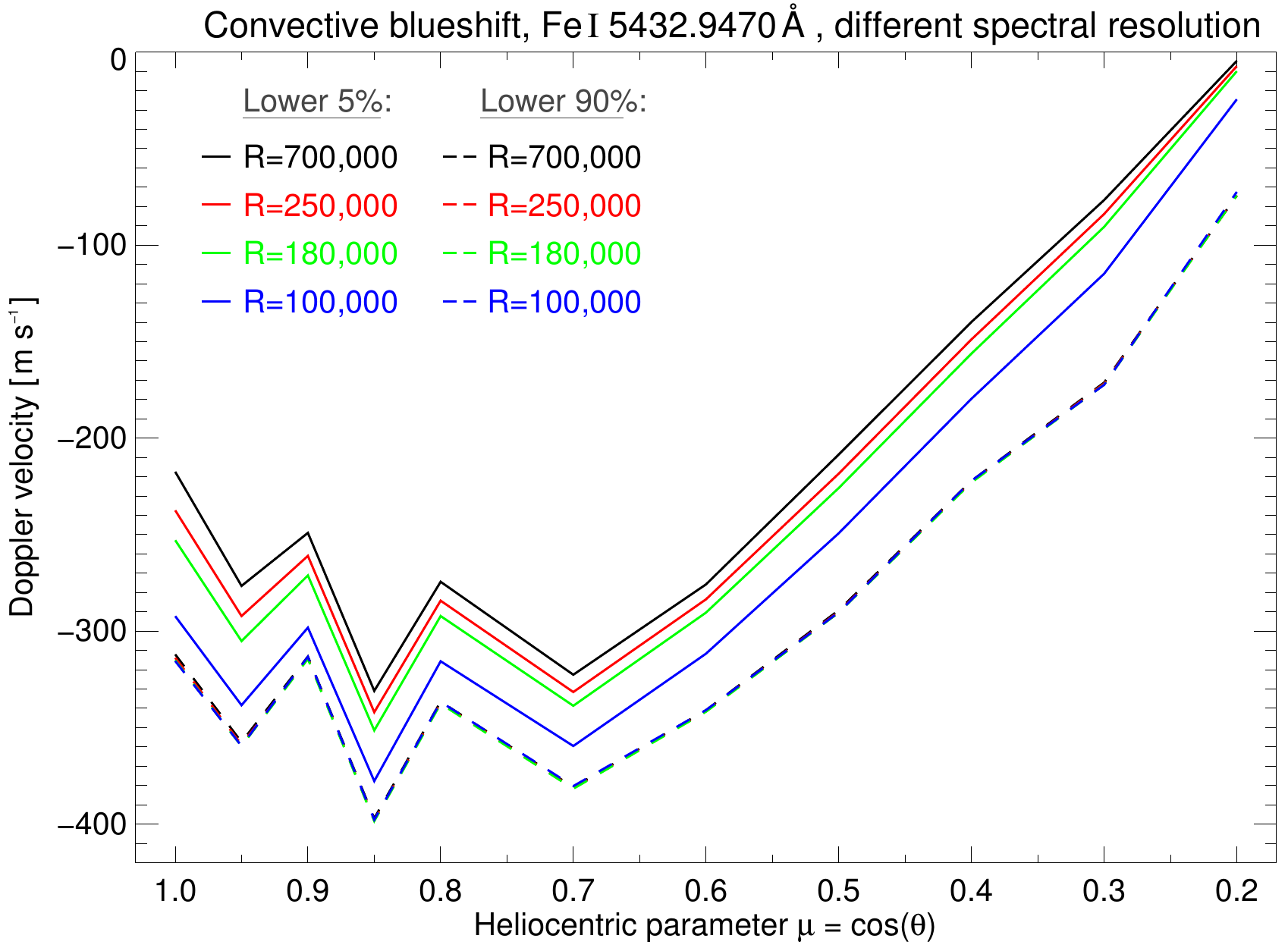}
\caption{Center-to-limb variation of the \ion{Fe}{I}\,5432.9\,\AA\ line. Panel a displays the average line bisectors from the disk center ($\mu=1.0$) toward the solar limb ($\mu=0.2$). Panel b shows the average convective blueshift of the line core (lower 5\,\% of the bisectors) for each observation. Colors and symbols indicate the axes. The average curve and its standard deviation are displayed as black solid line and gray shaded area. Panel c compares the observed convective blueshift for different spectral resolutions and line sections. Dashed lines are close to each other or even overlay.}
\label{fig_sec3_analysis_Fe54329}
\end{figure}

The \ion{Fe}{I}\,5432.9\,\AA\ line is a rather ordinary iron line covering the lower half of the photosphere. But exactly this is the reason for its exemplariness for the systematic convective blueshift and its center-to-limb variation. As displayed in Fig.\,\ref{fig_sec3_analysis_Fe54329} (panel a), the C-shape of the bisector becomes less pronounced from $\mu=1.0$ to $\mu=0.7$. At the same time, the blueshift of the bisector increases by up to $\mathrm{100\,m\,s^{-1}}$. At heliocentric positions between $\mu=0.6$ and $\mu=0.4$, the bisector describes a linear \textbackslash-shape for normalized intensities below 0.9. At the solar limb ($\mu\le0.3$), the curvature of the bisector reverses. The slope of the bisector indicates a saturation of the line core velocity at around $\mathrm{0\,m\,s^{-1}}$. The quantitative analysis of the line core velocity highlights the slope of the mean center-to-limb variation (panel b). The blueshift increases from $\mathrm{-217\,m\,s^{-1}}$ at the disk center to $\mathrm{-323\,m\,s^{-1}}$ at $\mu=0.7$, accompanied by an increase of the standard deviation by almost one order of magnitude. Toward the solar limb, the convective blueshift decreases monotonically to $\mathrm{-5\,m\,s^{-1}}$ at $\mu=0.2$. The lowering of the spectral resolution (panel c) leads to a stronger blueshift of the line core. Due to the distinct asymmetry of the line bisector, the effect is strongest at disk center. The overall stronger blueshift for the averaged bisector does not depend on the spectral resolution. Detailed velocities are listed in Table\,\ref{table_sec3_resolution_comparison}.

At this point, we anticipate that the \ion{Fe}{I}\,5432.9\,\AA\ line is one of the most suitable spectral lines to observe the reversal point of the center-to-limb variation of the convective blueshift. As we will later discuss in Section \ref{sec_discussion_depth}, the line depth (line minimum at around 0.4) and formation height (up to the mid photosphere) are ideally suited to capture the horizontally reversing flows of the granular convective motion. 

\subsection{Lines around 5576\,\AA}\label{sec_results_5576}
\begin{figure}[htbp]
\begin{center}
\includegraphics[width=\columnwidth]{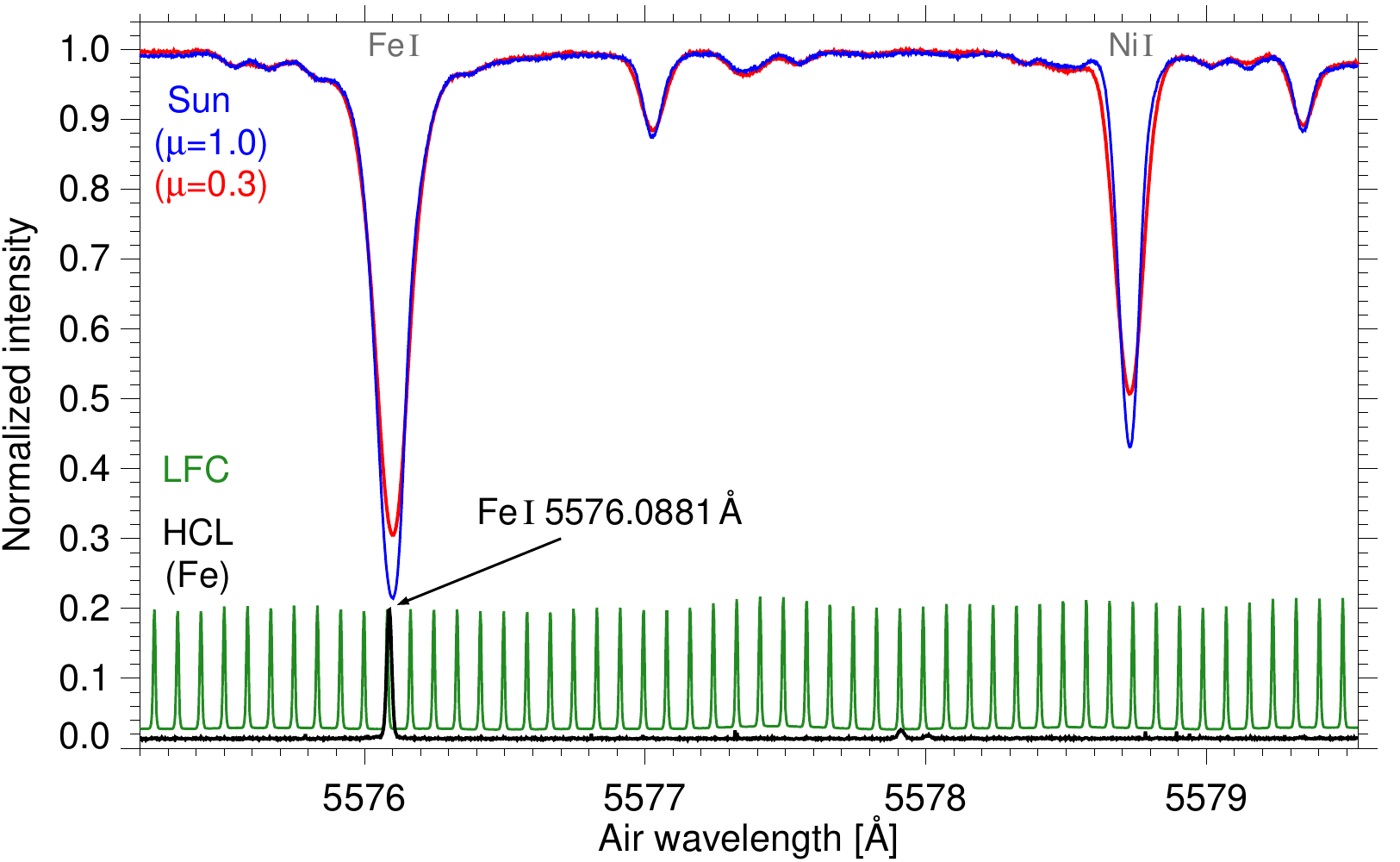}
\caption{Spectral region around 5576\,\AA, with the quiet Sun absorption spectra at the disk center ($\mu=1.0$, blue curve) and close to the solar limb ($\mu=0.3$, red curve). The {atomic species} are stated in gray. The emission spectra of the laser frequency comb (LFC, green curve) and the iron hollow cathode lamp (HCL, black curve) are displayed.}
\label{fig_sec3_spectra_5576}
\end{center}
\end{figure}

Our observations of the 5576\,\AA\ region shown in Fig.\,\ref{fig_sec3_spectra_5576} covered 75 observation sequences from $\mu=1.0$ to $\mu=0.2$. We analyzed the convective blueshift of the spectral lines \ion{Fe}{I}\,5576.1\,\AA\ and \ion{Ni}{I}\,5578.7\,\AA. 
Using the iron hollow cathode lamp of LARS, we measured the laboratory air wavelength of the \ion{Fe}{I} line to 5576.0881\,\AA\ with an uncertainty below 0.1\,m\AA. The reference wavelength of \ion{Ni}{I} line was refined to 5578.7204\,\AA\ by adaption with other spectral lines with an uncertainty of 1\,m\AA.

The main focus lies on the Zeeman-insensitive ($g_\mathrm{eff}=0$) \ion{Fe}{I}\,5576.1\,\AA\ line. The line core forms in the upper half of the photosphere around 310\,km \citep{1991sopo.work.....N,1994A&A...285.1012G,1998A&A...332.1069K} and 370\,km \citep{1975SoPh...43...33A,1988A&AS...72..473B} above the solar surface at $\tau_{5000\,\AA}=1$. Due to its velocity sensitivity, it is frequently used for high-resolution spectroscopy of photospheric convective line shift \citep{1982SoPh...79....3B,1984SoPh...94...49A}, acoustic waves \citep[e.g.,][]{2000A&A...363..306G,2009A&A...508..941B}, and sunspot flows \citep[e.g.,][]{1995A&A...298..260R,2004A&A...415..717T,2004A&A...415..731S}. In the following section, we present the yet most accurate measurements of the line shift and asymmetry. The results for the \ion{Ni}{I}\,5578.7\,\AA\ line are displayed in Figs.\,\ref{fig_A7} and \ref{fig_A8}.

\subsubsection{\ion{Fe}{I}\,5576.1\,\AA}

\begin{figure}[htbp]
\textbf{a)}\\[-0.3cm]  \includegraphics[width=\columnwidth]{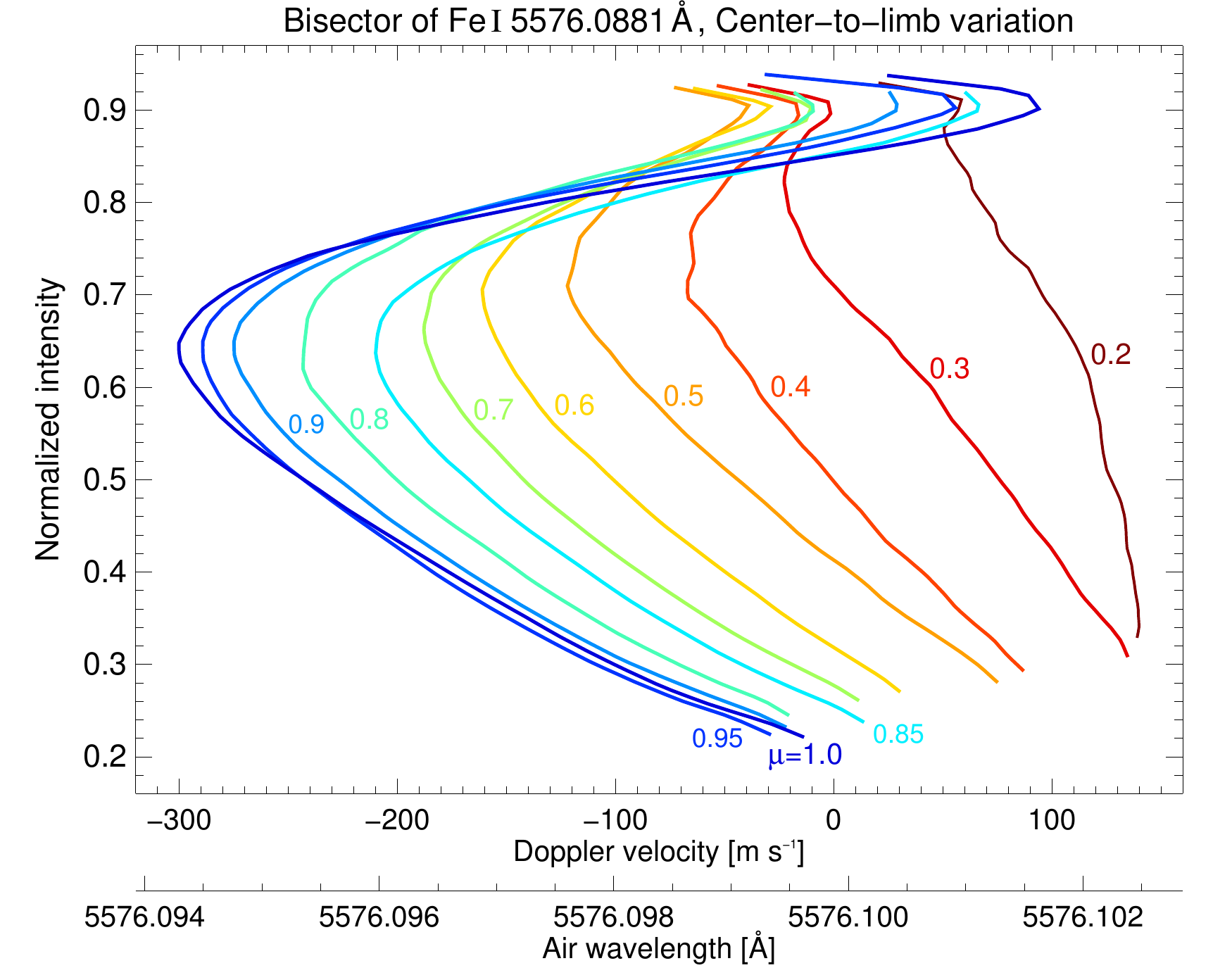}\\[0.2cm]
\textbf{b)}\\[-0.3cm]  \includegraphics[width=\columnwidth]{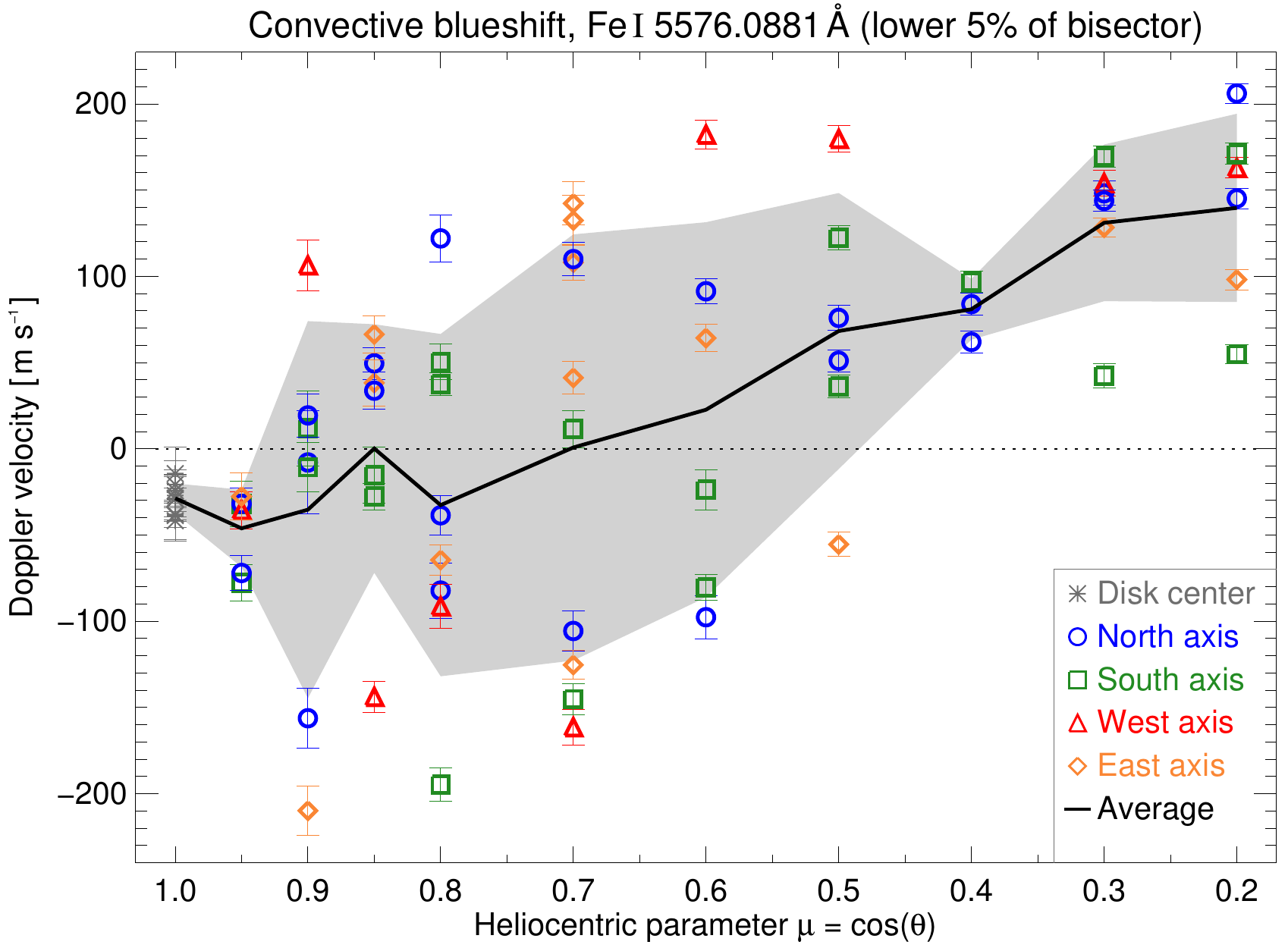}\\[0.2cm]
\textbf{c)}\\[-0.3cm]  \includegraphics[width=\columnwidth]{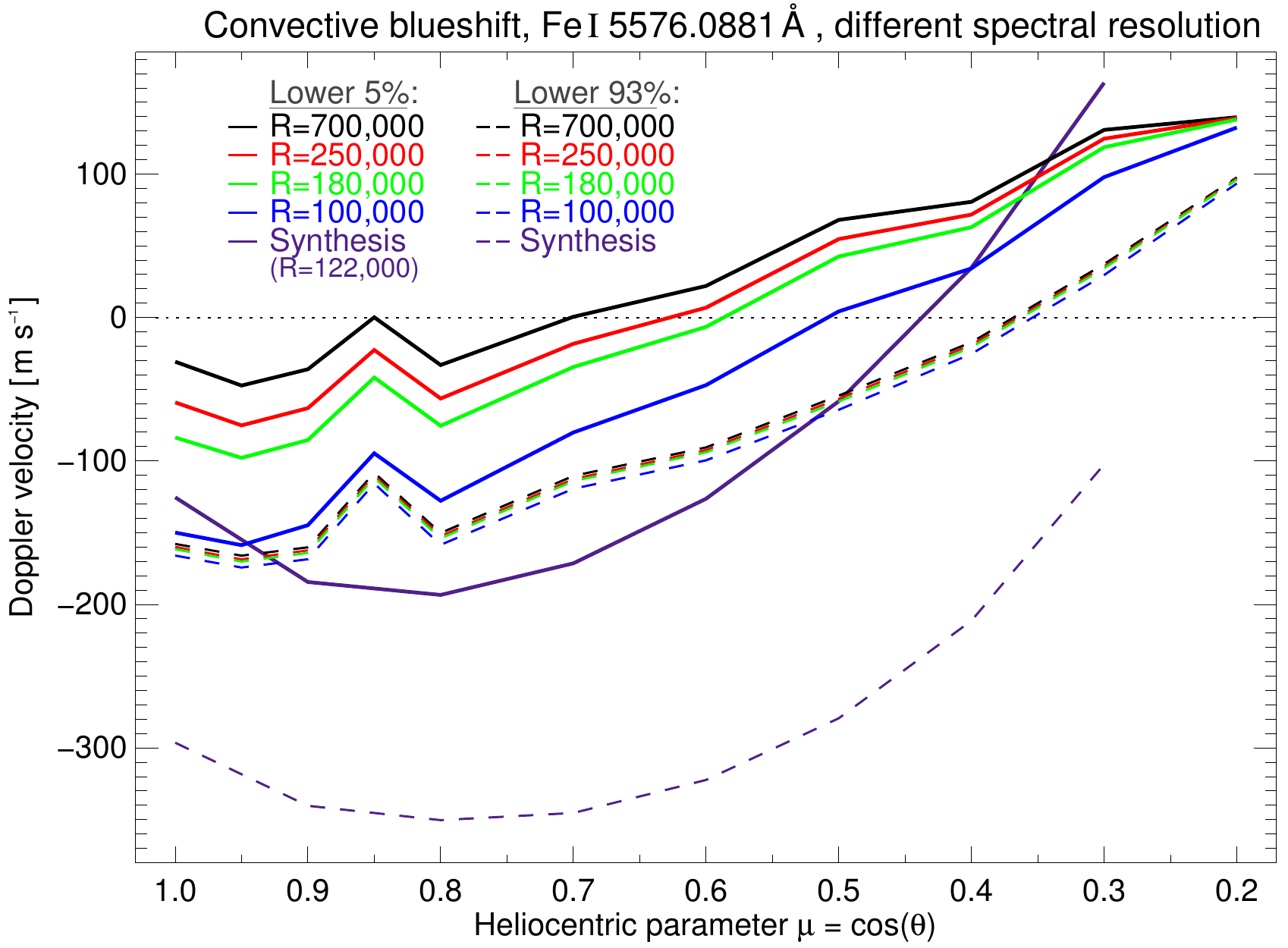}
\caption{Center-to-limb variation of the \ion{Fe}{I}\,5576.1\,\AA\ line. Panel a displays the average line bisectors from the disk center ($\mu=1.0$) toward the solar limb ($\mu=0.2$). Panel b shows the average convective blueshift of the line core (lower 5\,\% of the bisectors) for each observation. Colors and symbols indicate the axes. The average curve and its standard deviation are displayed as black solid line and gray shaded area. Panel c compares the observed convective blueshift for different spectral resolutions and line sections with the synthesis of \citet{2011A&A...528A.113D}. Dashed lines are close to each other or even overlay.}
\label{fig_sec3_analysis_Fe5576}
\end{figure}

The analysis of the Doppler shift and line asymmetry of \ion{Fe}{I}\,5576.1\,\AA\ is shown Fig.\,\ref{fig_sec3_analysis_Fe5576}. The bisector (panel a) features a distinct C-shape at disk center, ending in a weak blueshift of around $\mathrm{-20\,m\,s^{-1}}$ at the line minimum. Toward smaller $\mu$-values, the C-shape fades out, shifts toward longer wavelength, and reverses its curvature at the limb positions ($\mu\le0.3$). The bend of the bisector at normalized intensities above 0.9 results from a blend in the outer blue wing of \ion{Fe}{I}\,5576.1\,\AA. The center-to-limb variation of the line core velocity (panel b) yields an almost monotonic transition from weak blueshifts of $\mathrm{-30\,m\,s^{-1}}$ to redshifts of $\mathrm{+140\,m\,s^{-1}}$. At disk center, lowering the spectral resolution leads to considerably stronger shifts of the line core (panel c). Detailed velocities are listed in Table\,\ref{table_sec3_resolution_comparison}. Except for the shift of the line core at disk center, the synthesis of \citet{2011A&A...528A.113D} differs significantly from our observations. 


\subsection{Lines around 5896\,\AA}\label{sec_results_5896}
\begin{figure}[htbp]
\begin{center}
\includegraphics[width=\columnwidth]{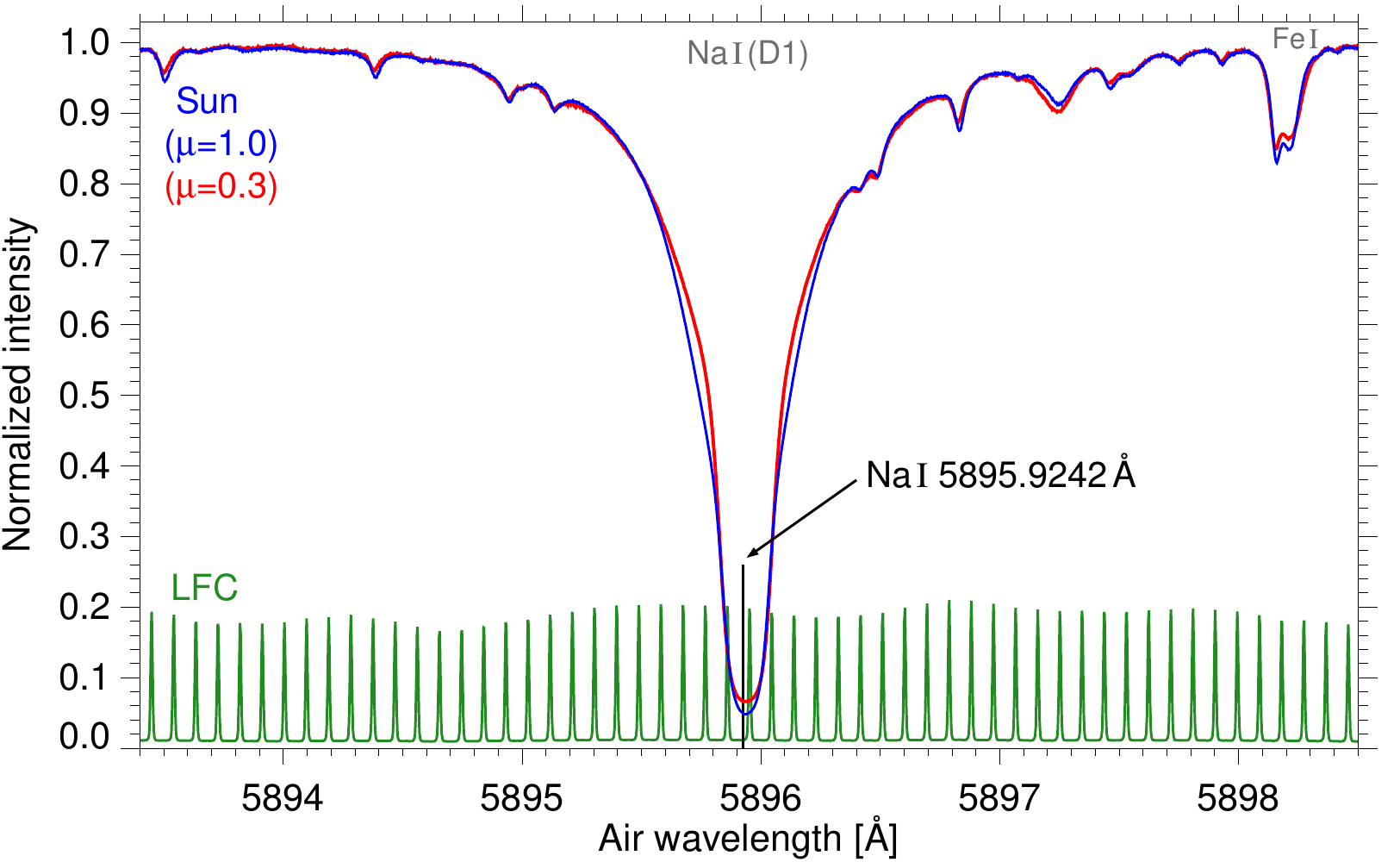}
\caption{Spectral region around 5896\,\AA, with the quiet Sun absorption spectra at the disk center ($\mu=1.0$, blue curve) and close to the solar limb ($\mu=0.3$, red curve). The {atomic species} are stated in gray. The emission spectra of the laser frequency comb (LFC) is displayed as green curve). The laboratory wavelength of the \ion{Na}{i} line is indicated.}
\label{fig_sec3_spectra_5896}
\end{center}
\end{figure}

The 5896\,\AA\ region shown in Fig.\,\ref{fig_sec3_spectra_5896} was observed within 67 sequences from $\mu=1.0$ to $\mu=0.2$. We analyzed the convective shift of the Fraunhofer line \ion{Na}{I}\,5895.9\,\AA, also known as \ion{Na}\,D$_1$.

The \ion{Na}{I}\,5895.9\,\AA\ line is one of the long-known and most-studied spectral lines of the Sun. The deep absorption line covers the entire photosphere {up to heights} around 800\,km \citep[e.g.,][]{1976PhDT.......336S,2001A&A...371.1128E,2010ApJ...709.1362L} above the solar surface at $\tau_{5000\,\AA}=1$. We have thus chosen to include \ion{Na}\,D$_1$ to obtain a comprehensive analysis of the photospheric convective blueshift and its transition into the chromosphere. The laboratory wavelength of the \ion{Na}{I}\,5895.9\,\AA\ line is well-known. The NIST ASD provides an observed air wavelength of $5895.92424\,\AA$ with an uncertainty of 0.03\,m\AA\ (or $\mathrm{1.5\,m\,s^{-1}}$).

\subsubsection{\ion{Na}{I}\,5895.9\,\AA}

\begin{figure}[htbp]
\textbf{a)}\\[-0.3cm]  \includegraphics[width=\columnwidth]{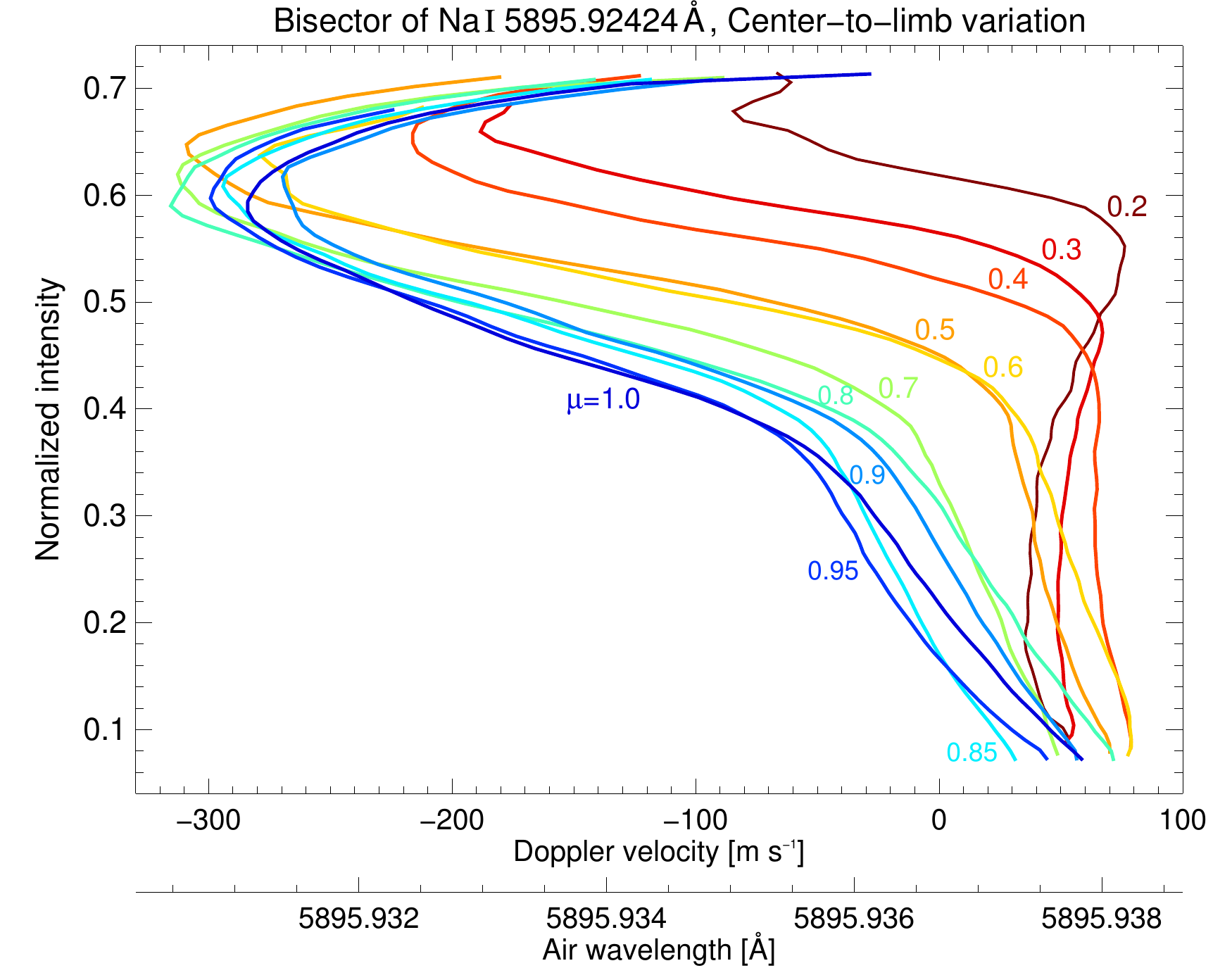}\\[0.2cm]
\textbf{b)}\\[-0.3cm]  \includegraphics[width=\columnwidth]{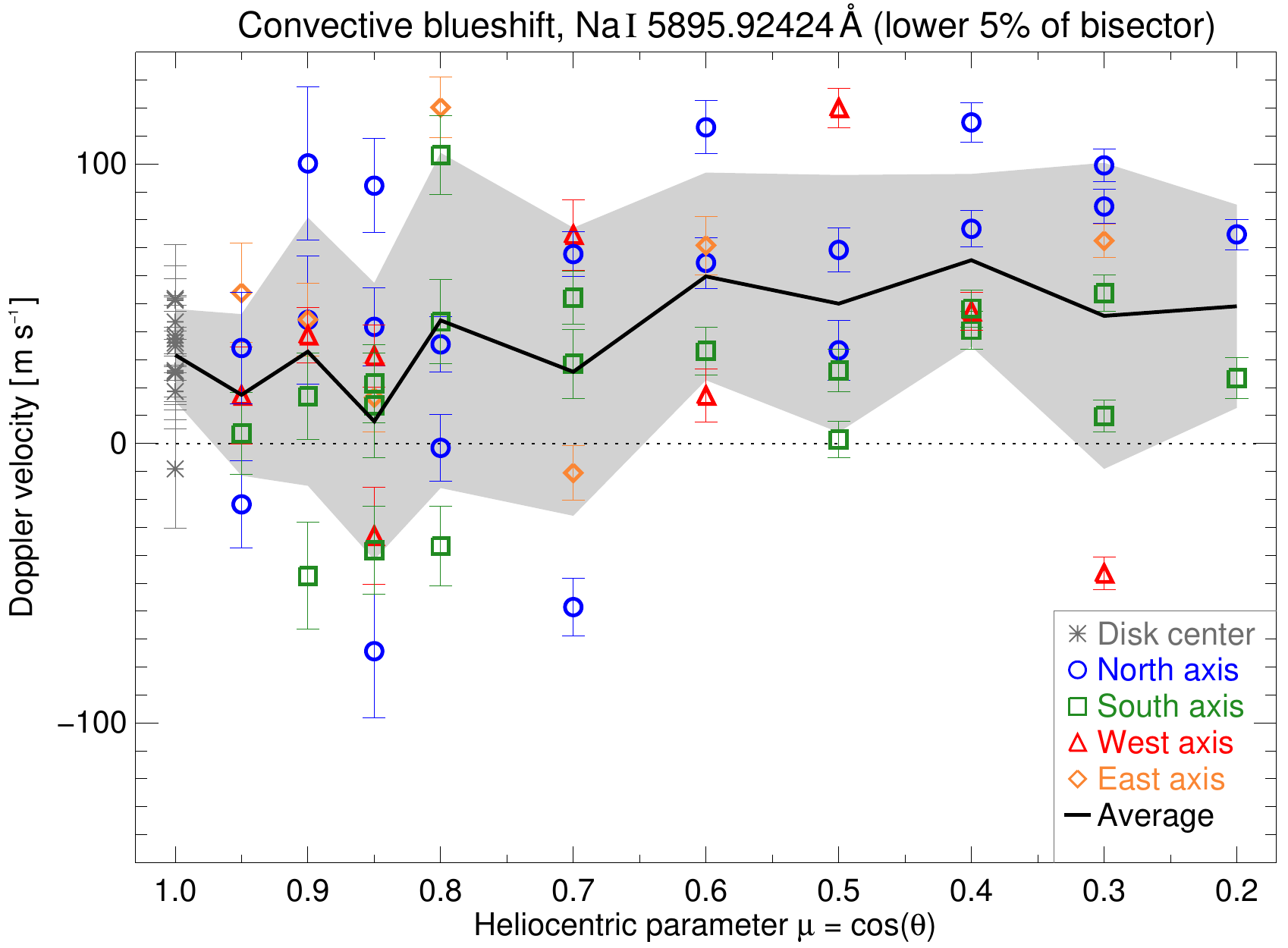}\\[0.2cm]
\textbf{c)}\\[-0.3cm]  \includegraphics[width=\columnwidth]{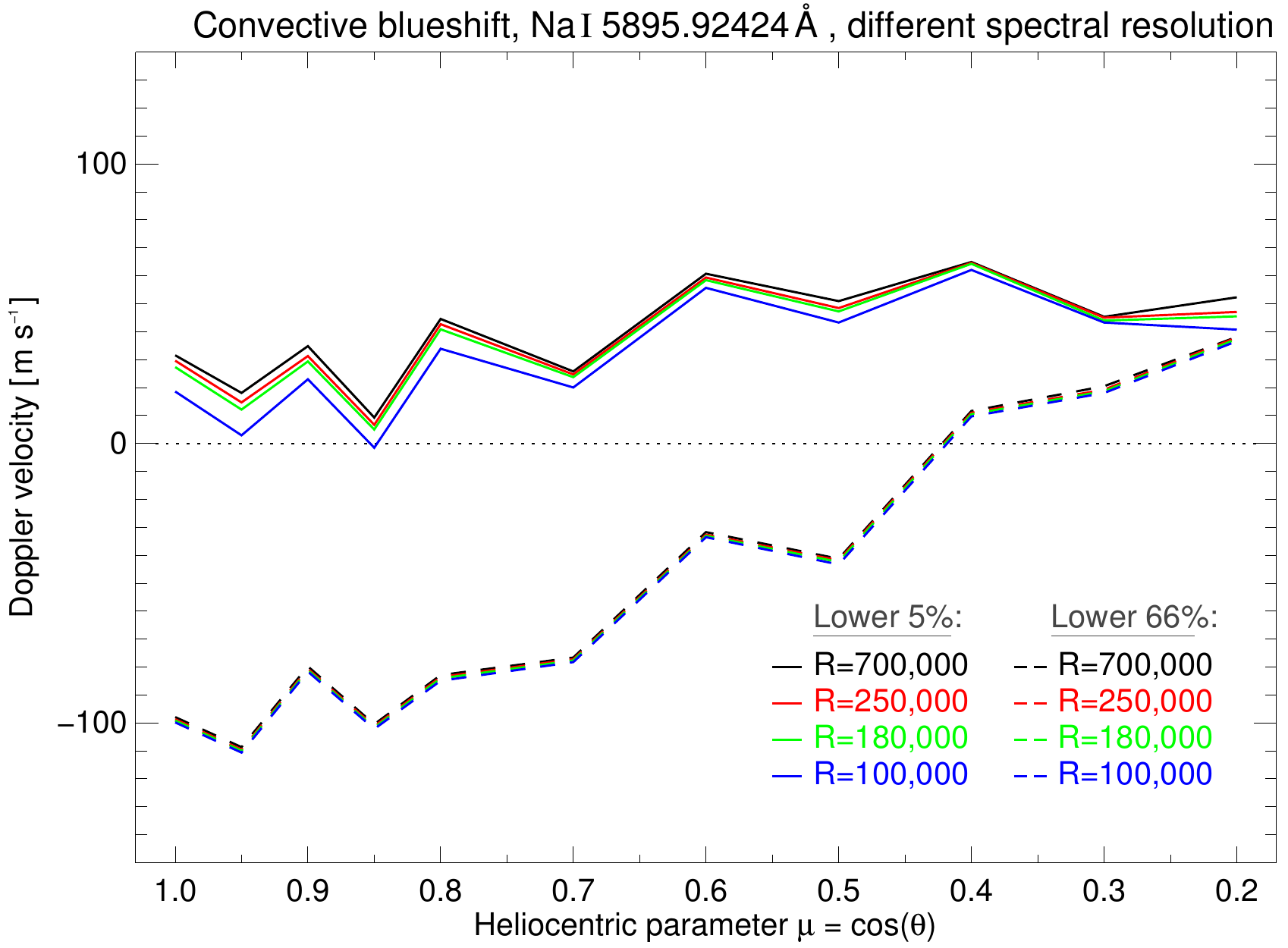}
\caption{Center-to-limb variation of the \ion{Na}{I}\,5895.9\,\AA\ line. Panel a displays the average line bisectors from the disk center ($\mu=1.0$) toward the solar limb ($\mu=0.2$). Panel b shows the average convective blueshift of the line core (lower 5\,\% of the bisectors) for each observation. Colors and symbols indicate the axes. The average curve and its standard deviation are displayed as black solid line and gray shaded area. Panel c compares the observed convective blueshift for different spectral resolutions and line sections. Dashed lines are close to each other or even overlay.}
\label{fig_sec3_analysis_Na5896}
\end{figure}

The line depth is virtually constant from disk center toward the limb. This could be explained either by the low temperature sensitivity of the line core or an isothermal formation layer. But as shown by the bisector in Fig.\,\ref{fig_sec3_analysis_Na5896} (panel a), the line asymmetry changes significantly. At disk center, the photospheric part of the bisector at normalized intensities above 0.4 is C-shaped. The bisector bends at an intensity of around 0.36, associated with the transition from the photosphere to the chromosphere. Toward the line minimum, the bisector describes an almost linear curve ending in a redshift of around $\mathrm{+50\,m\,s^{-1}}$. The center-to-limb variation of the bisector demonstrates how the line bend shifts toward longer wavelength and higher intensities. At $\mu\le0.3$, the curvature toward the line minimum even reverses. Nevertheless, the line minimum of each bisector features a redshift of $\mathrm{30-80\,m\,s^{-1}}$. The quantitative analysis of the line core yields a more or less constant redshift of around $\mathrm{+50\,m\,s^{-1}}$. {Due to the inverse granulation in the sampled uppermost photosphere, and its brighter descending elements, such a moderate redshift is thus plausible \citep{2007A&A...461.1163C}.}
Since \ion{Na}{I}\,5895.9\,\AA\ is a broad line, a decrease of the spectral resolution hardly affects the Doppler shift of the line core, and especially not the entire line (panel c). The detailed Doppler shifts are listed in Table\,\ref{table_sec3_resolution_comparison}.

\subsection{Lines around 6149\,\AA}\label{sec_results_6149}
\begin{figure}[htbp]
\begin{center}
\includegraphics[width=\columnwidth]{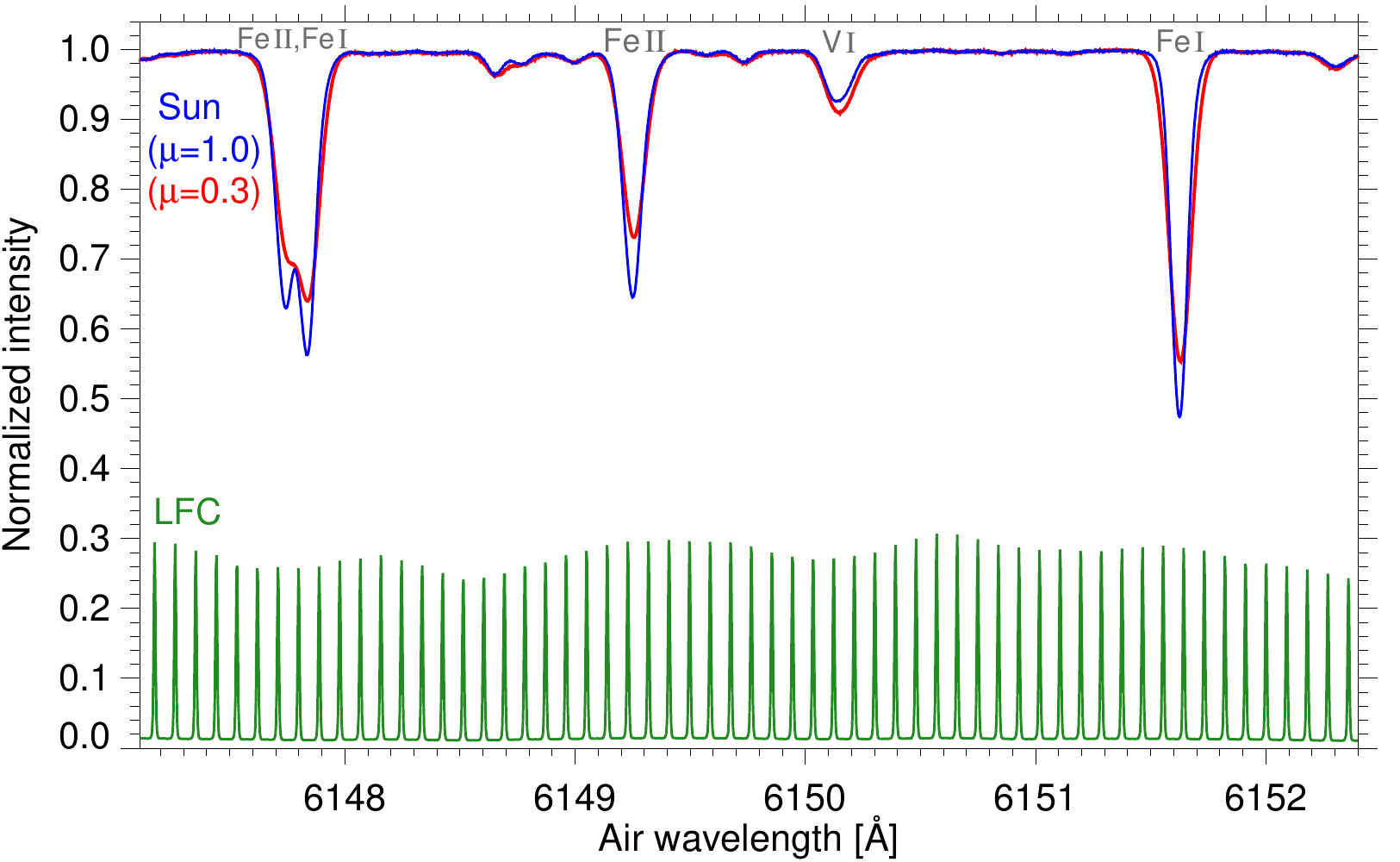}
\caption{Spectral region around 6149\,\AA, with the quiet Sun absorption spectra at the disk center ($\mu=1.0$, blue curve) and close to the solar limb ($\mu=0.3$, red curve). The {atomic species} are stated in gray. The spectrum of the laser frequency comb (LFC) is displayed as green curve.}
\label{fig_sec3_spectra_6149}
\end{center}
\end{figure}

The 6149\,\AA\ region is shown in Fig.\,\ref{fig_sec3_spectra_6149}. We performed 76 observation sequences from $\mu=1.0$ to $\mu=0.2$. The analysis of the convective blueshift included \ion{Fe}{II}\,6149.2\,\AA\ and \ion{Fe}{I}\,6151.6\,\AA.

Our main interest lies in the convective blueshift of the temperature-sensitive \ion{Fe}{II}\,6149.2\,\AA\ line. As a weak ionized iron line, it is formed in the hotter {parts} of the photosphere {which are predominantly assigned to the uprising and locally blueshifted granules. The formation height of the \ion{Fe}{II} line core} is estimated to not more than 150\,km above the solar surface at $\tau_{5000\,\AA}=1$. By the absence of linear polarization and instrumental Stokes Q/U to V crosstalk, the line possesses rare polarization properties \citep{1993SoPh..143..229L,1994A&AS..103..293V}, hence it is often applied for measurements of the magnetic field of sunspots \citep{1993A&A...279..243B,2005A&A...434..317B,2005A&A...443L...7B} and stars \citep{1997A&A...318..429R,2014A&A...568A..38B}. In our study, we applied the Ritz air wavelength of 6149.2460\,\AA\ taken from the NIST ASD as the Doppler reference with an uncertainty of $\pm$0.6\,m\AA. The observed air wavelength of \ion{Fe}{I}\,6151.6177\,\AA\ was given by the NIST ASD with an uncertainty of $\pm$1.1\,m\AA. The analysis results for the latter are displayed in Figs.\,\ref{fig_A7} and \ref{fig_A8} (lower panels).

\subsubsection{\ion{Fe}{II}\,6149.2\,\AA}

\begin{figure}[htbp]
\textbf{a)}\\[-0.3cm]  \includegraphics[width=\columnwidth]{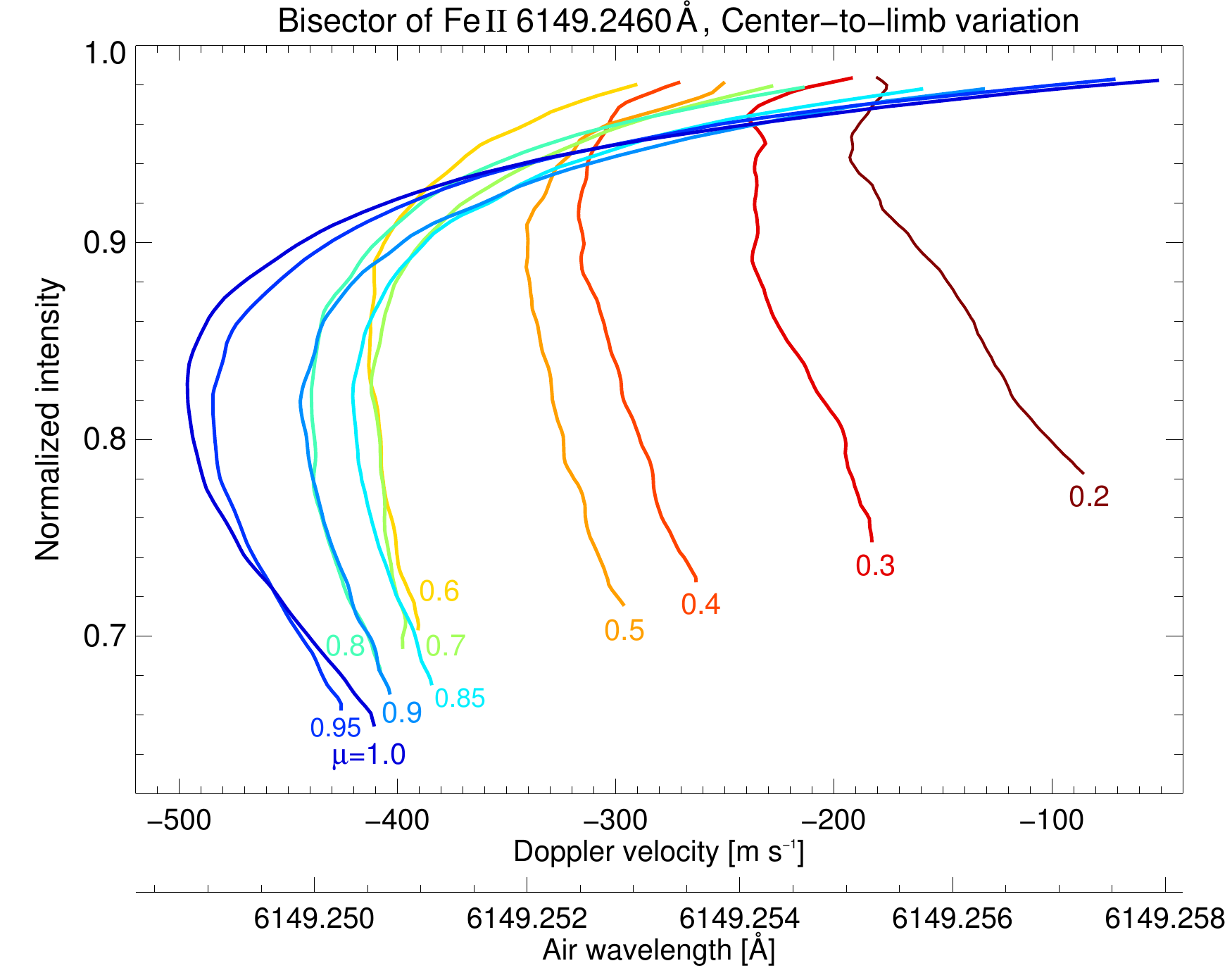}\\[0.2cm]
\textbf{b)}\\[-0.3cm]  \includegraphics[width=\columnwidth]{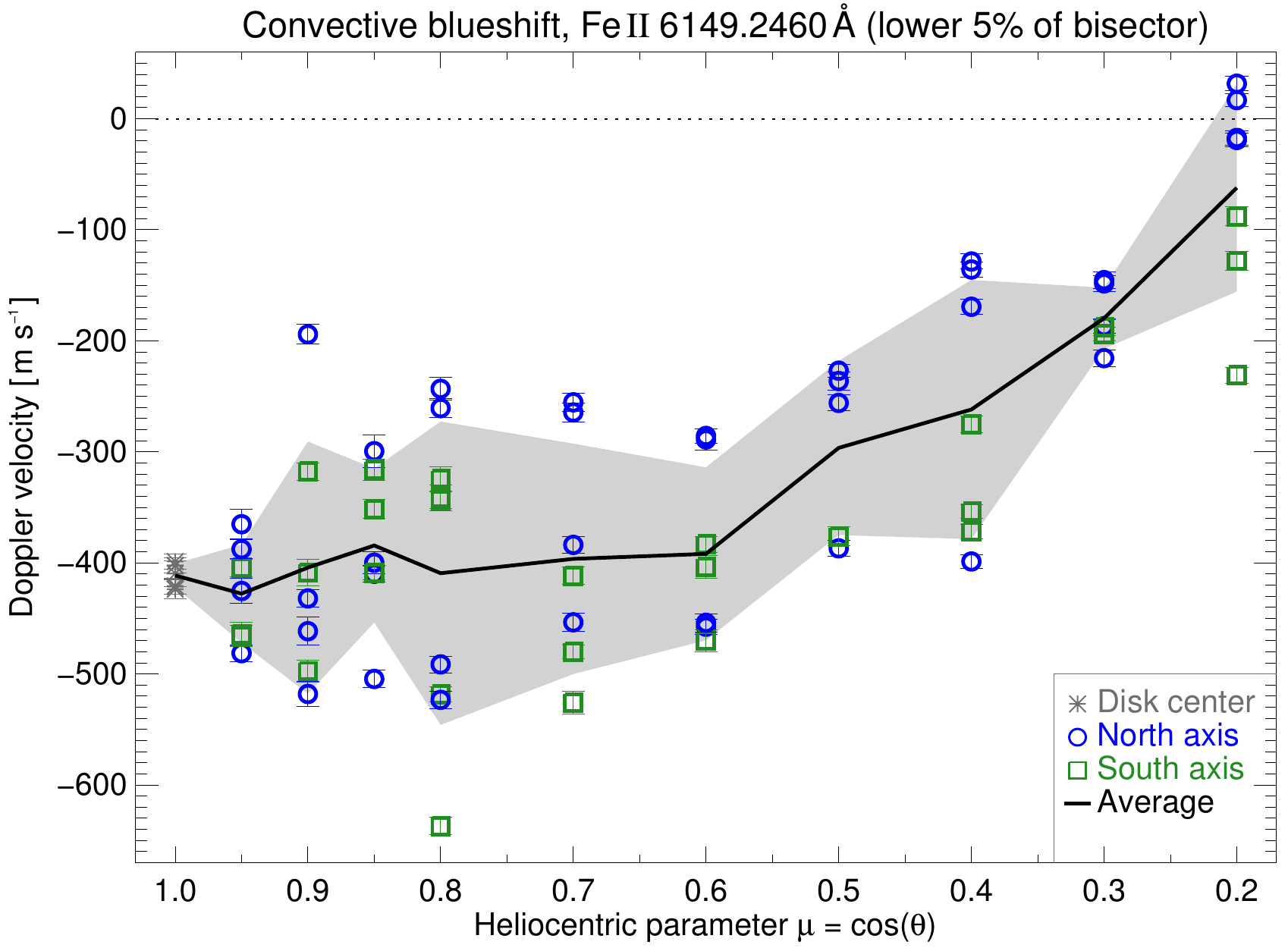}\\[0.2cm]
\textbf{c)}\\[-0.3cm]  \includegraphics[width=\columnwidth]{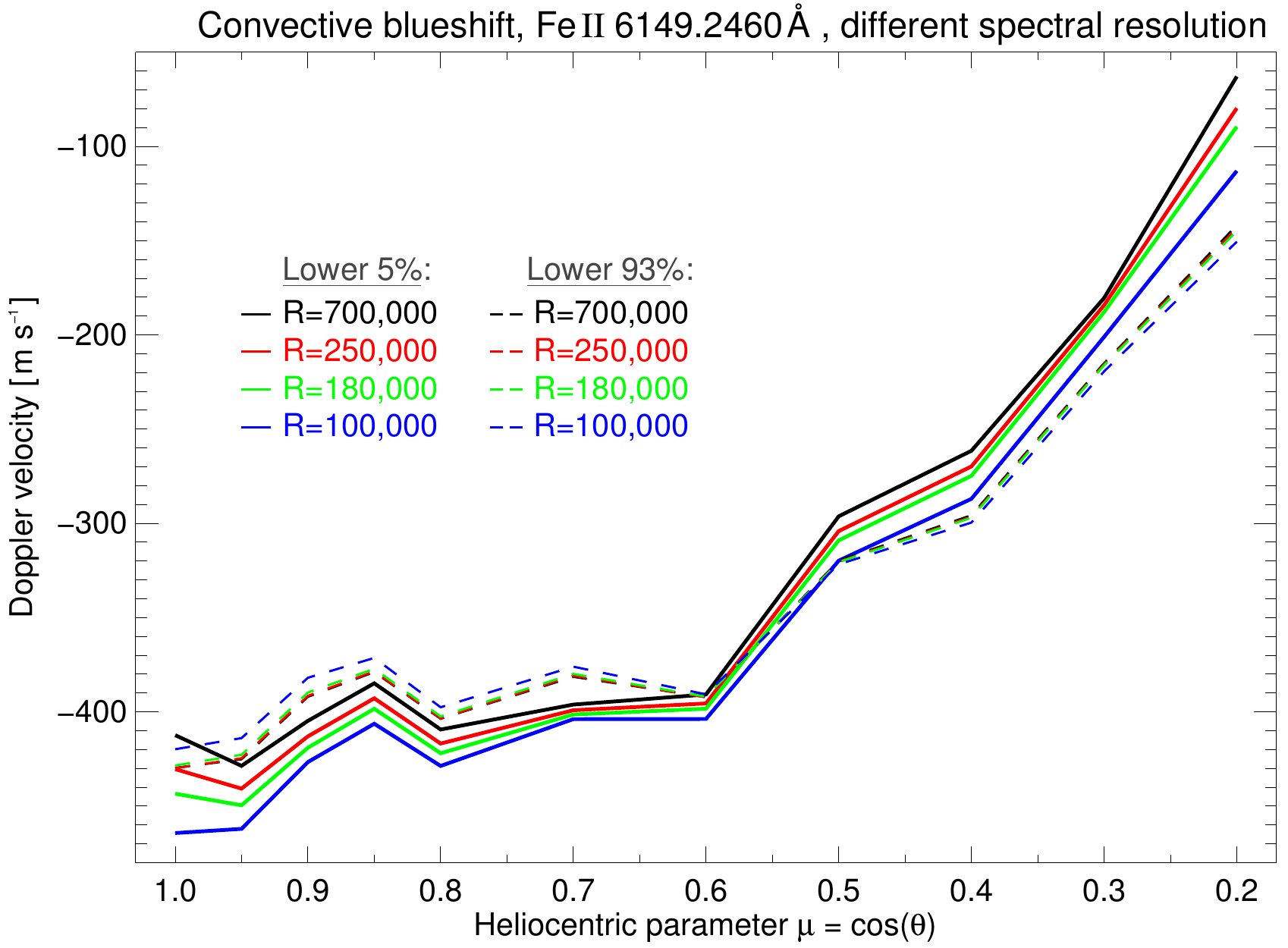}
\caption{Center-to-limb variation of the \ion{Fe}{II}\,6149.2\,\AA\ line. Panel a displays the average line bisectors from the disk center ($\mu=1.0$) toward the solar limb ($\mu=0.2$). Panel b shows the average convective blueshift of the line core (lower 5\,\% of the bisectors) for each observation. Colors and symbols indicate the axes. The average curve and its standard deviation are displayed as black solid line and gray shaded area. Panel c compares the observed convective blueshift for different spectral resolutions and line sections. Dashed lines are close to each other or even overlay.}
\label{fig_sec3_analysis_Fe6149}
\end{figure}

The center-to-limb variation of the convective blueshift of \ion{Fe}{II}\,6149.2\,\AA\ is displayed in Fig.\,\ref{fig_sec3_analysis_Fe6149}. The bisector (panel a) features a C-shape with a blueshift of up to $\mathrm{-500\,m\,s^{-1}}$ at disk center. After \ion{C}{I}\,5380.3\,\AA, this is the strongest blueshift measured. Toward the limb, the blueshift decreases and the bisector transforms to a more linear shape. The blueshift of the line core decrease only slowly from around $\mathrm{-420\,m\,s^{-1}}$ close to disk center to $\mathrm{-391\,m\,s^{-1}}$ at $\mu=0.6$. Toward $\mu=0.2$, we obtain a fast decrease in blueshift to $\mathrm{-63\,m\,s^{-1}}$. As shown in panel c, lowering the spectral resolution leads to {only} slightly different blueshifts. {In comparison}, we maintain similar velocities and center-to-limb variations {for the entire line average}. Detailed blueshifts are listed in Table\,\ref{table_sec3_resolution_comparison}. Only directly at the solar limb ($\mu\le0.2$) we expect deviations by more than $\mathrm{80\,m\,s^{-1}}$. We conclude that spectroscopy with \ion{Fe}{II}\,6149.2\,\AA\ yields accurate results, independent of the spectral resolution or velocity determination.

\subsection{Lines around 6173\,\AA}\label{sec_results_6173}
\begin{figure}[htbp]
\begin{center}
\includegraphics[width=\columnwidth]{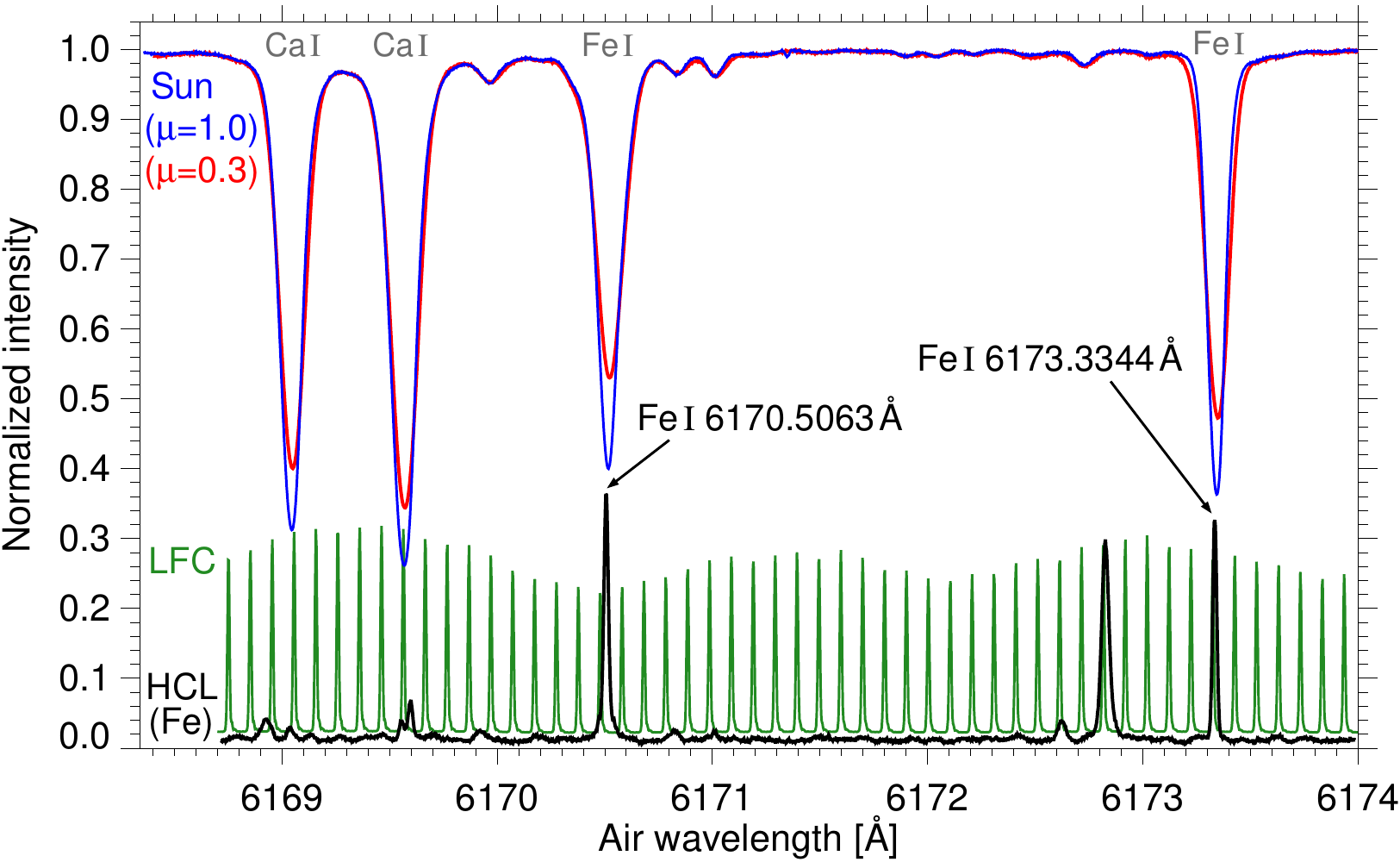}
\caption{Spectral region around 6173\,\AA, with the quiet Sun absorption spectra at the disk center ($\mu=1.0$, blue curve) and close to the solar limb ($\mu=0.3$, red curve). The {atomic species} are stated in gray. The emission spectra of the laser frequency comb (LFC, green curve) and the iron hollow cathode lamp (HCL, black curve) are displayed.}
\label{fig_sec3_spectra_6173}
\end{center}
\end{figure}

The 6173\,\AA\ region is shown in Fig.\,\ref{fig_sec3_spectra_6173}. In total, we performed 62 observation sequences from $\mu=1.0$ to $\mu=0.3$. We analyzed the Doppler shift of the spectral lines \ion{Ca}{I}\,6169.0\,\AA, \ion{Ca}{I}\,6169.6\,\AA, \ion{Fe}{I}\,6170.5\,\AA, and \ion{Fe}{I}\,6173.3\,\AA. The results for the latter have been discussed in Paper II \citep{2018arXiv181108685S}.

The region was selected because of \ion{Fe}{I}\,6173.3\,\AA. Its line core forms in the mid photosphere at around 270\,km above the solar surface at $\tau_{5000\,\AA}=1$ \citep{1991sopo.work.....N}. The unblended Zeeman-sensitive ($g=2.5$) line is one of the most widely used lines for spectro-polarimetric observations in the visible range. Moreover, \ion{Fe}{I}\,6173.3\,\AA\ is well suited for measurement of Doppler velocities \citep{2005A&A...439..687C}. This is why HMI applies the line to infer full-disk Dopplergrams and Magnetograms of the Sun. In Paper II, we have drawn a comparison of the measured convective blueshift between LARS and HMI. 

We measured the reference wavelengths of both \ion{Fe}{I} line with the iron hollow cathode lamp of LARS, and yield air wavelength of 6170.5063\,\AA\ and 6173.3344\,\AA\ with an uncertainty below 0.1\,m\AA. For both \ion{Ca}{I} lines, the uncertainty of $\mathrm{100}$\,m\AA\ for the observed air wavelength 6169.06\,\AA\ and 6169.56\,\AA\ provided by the NIST ASD was too large to draw meaningful conclusions on the resultant Doppler velocities. By aligning the bisectors with similar but well-characterized lines, we were able to refine the reference wavelengths to 6169.035\,\AA\ and 6169.557\,\AA\ with an estimated uncertainty of 1\,m\AA. The results of the convective blueshift analysis are displayed in Fig.\,\ref{fig_sec3_analysis_Fe6173} for \ion{Fe}{I}\,6173.3\,\AA, and in Figs.\,\ref{fig_A9} and \ref{fig_A10} for the other lines. 

\subsubsection{\ion{Fe}{I}\,6173.3\,\AA}

\begin{figure}[htbp]
\textbf{a)}\\[-0.3cm]  \includegraphics[width=\columnwidth]{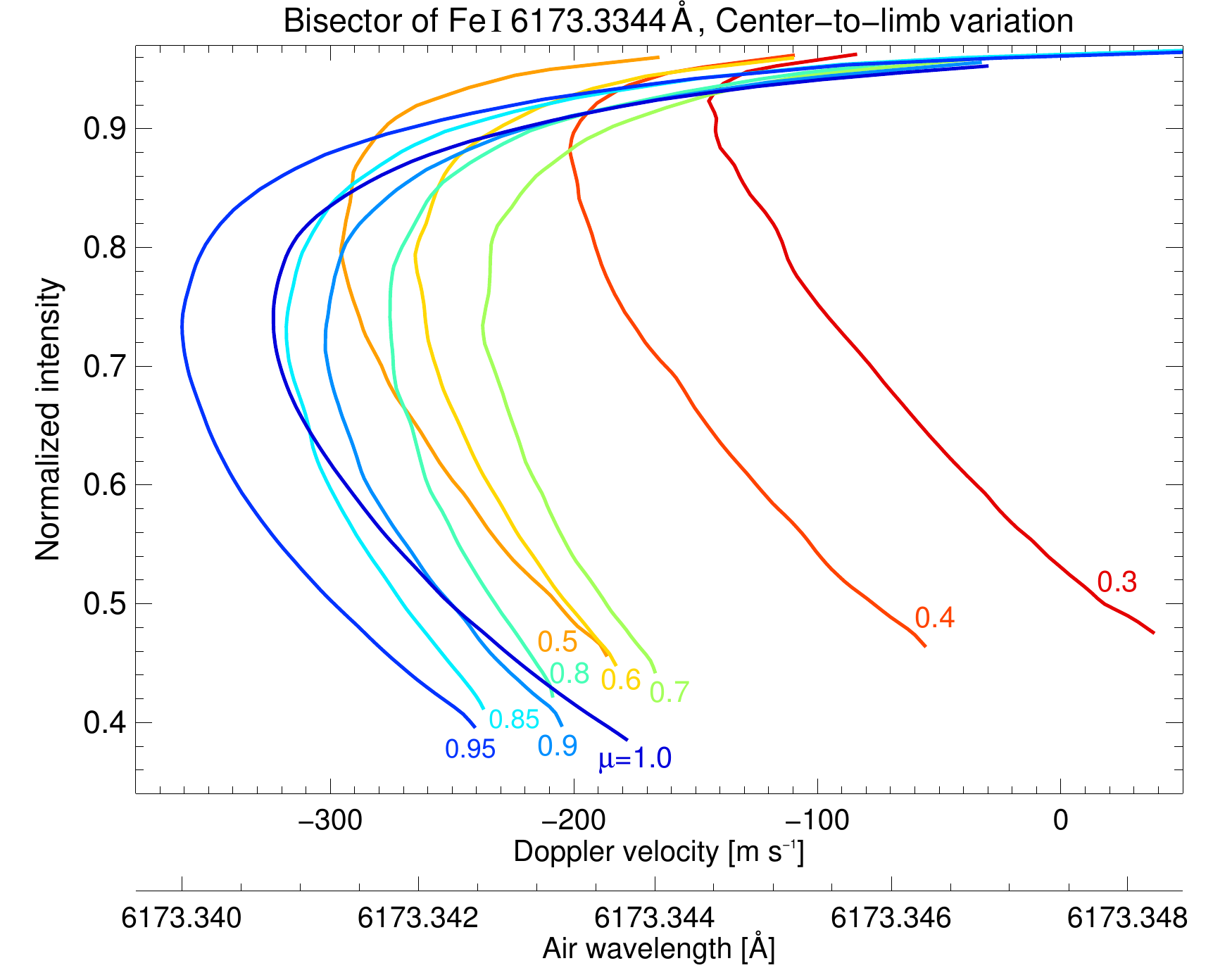}\\[0.2cm]
\textbf{b)}\\[-0.3cm]  \includegraphics[width=\columnwidth]{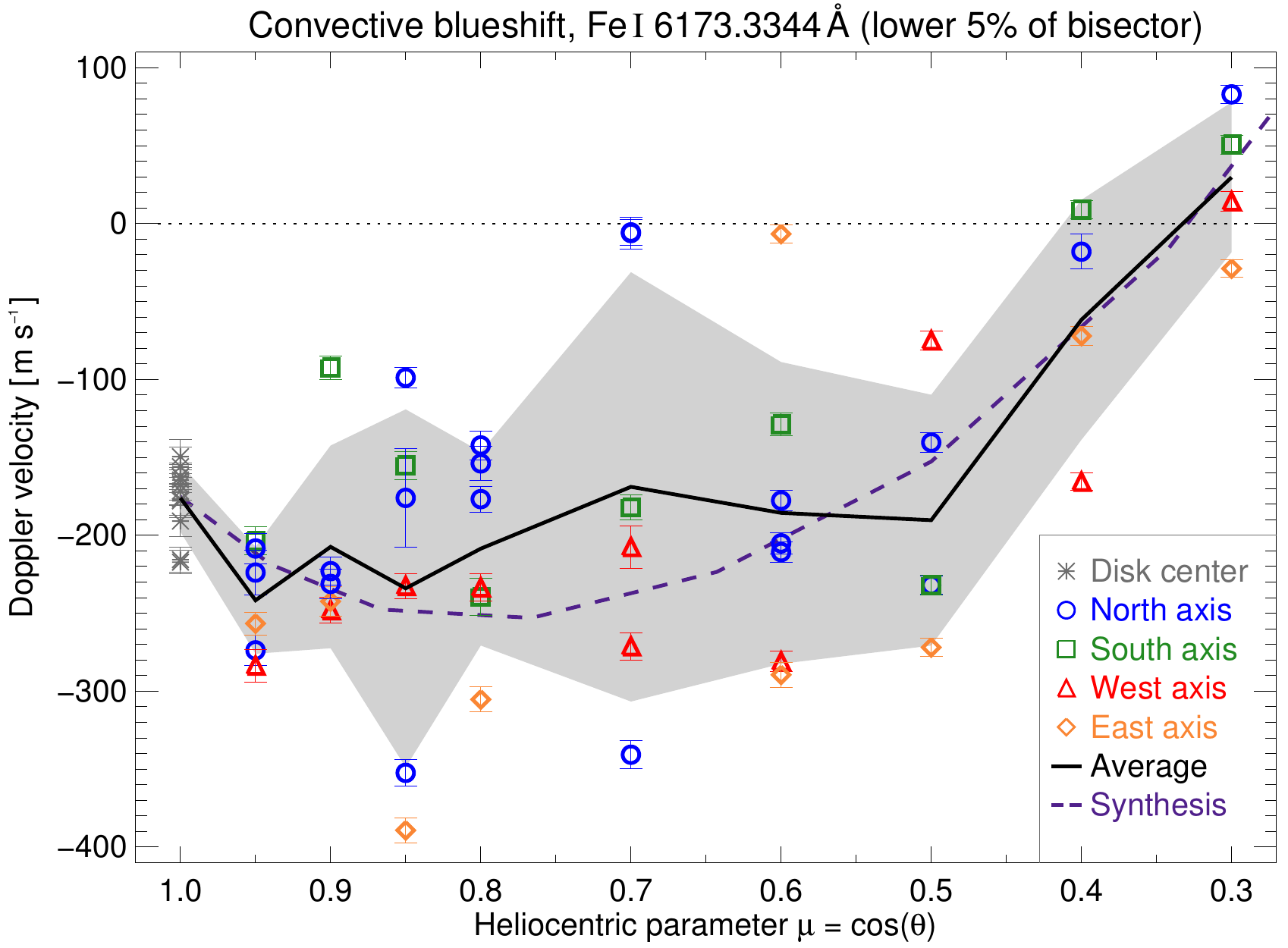}\\[0.2cm]
\textbf{c)}\\[-0.3cm]  \includegraphics[width=\columnwidth]{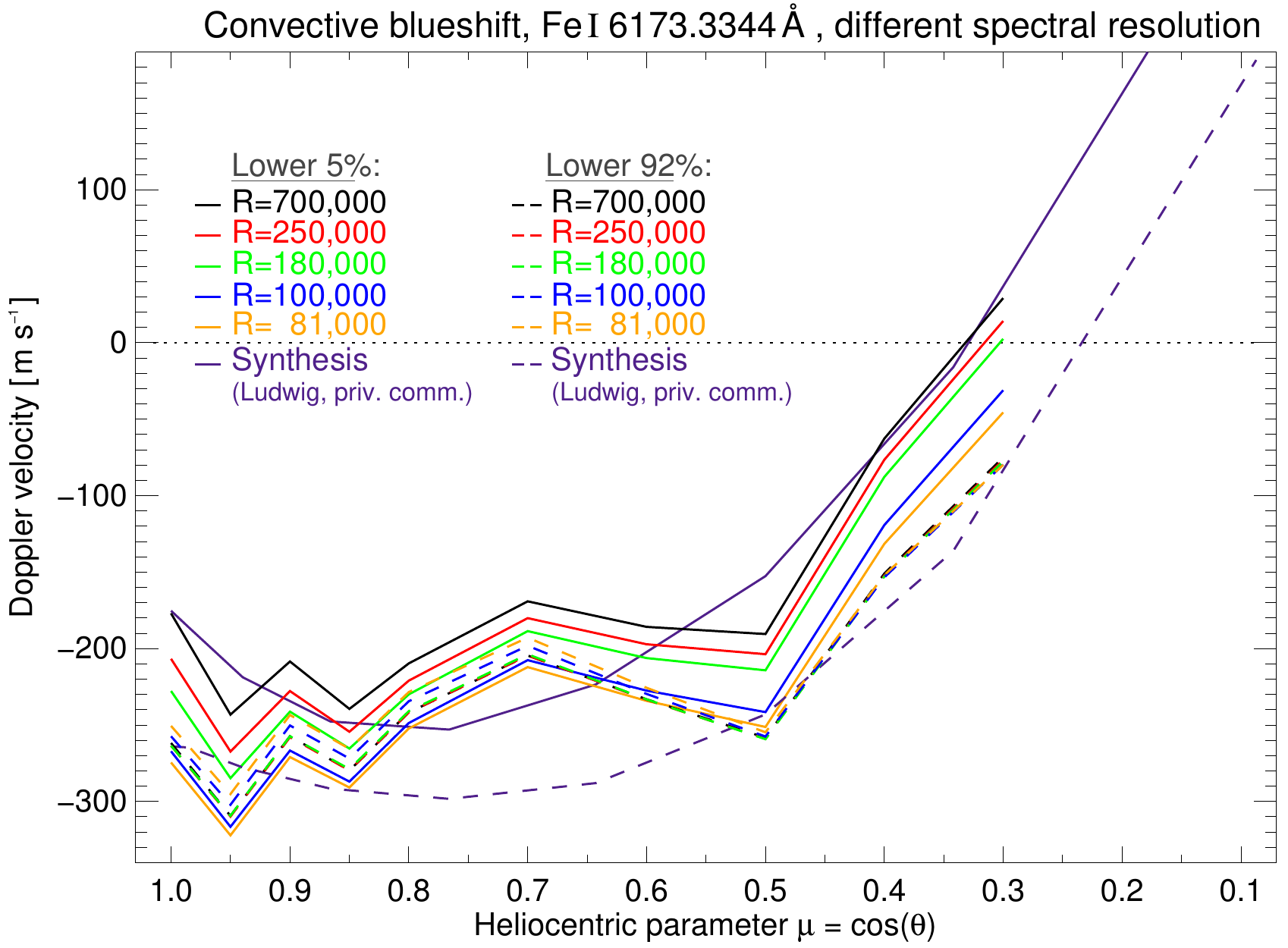}
\caption{Center-to-limb variation of the \ion{Fe}{I}\,6173.3\,\AA\ line. Panel a displays the average line bisectors from the disk center ($\mu=1.0$) toward the solar limb ($\mu=0.3$). Panel b shows the average convective blueshift of the line core (lower 5\,\% of the bisectors) for each observation. Colors and symbols indicate the axes. The average curve and its standard deviation are displayed as black solid line and gray shaded area. Panel c compares the observed convective blueshift for different spectral resolutions and line sections. Dashed lines are close to each other or even overlay. The observations are compared with the synthesis of H.-G. Ludwig (private communication).}
\label{fig_sec3_analysis_Fe6173}
\end{figure}

The line shift of \ion{Fe}{I}\,6173.3\,\AA\ and its center-to-limb variation are shown in Fig.\,\ref{fig_sec3_analysis_Fe6173}. The line asymmetry is described by the C-shaped bisector (panel a). From $\mu=1.0$ to $\mu=0.3$, the C-shape becomes less pronounced while the maximum blueshift decreases and shifts to higher intensities. The Doppler shift of the line core (panel b) increases slightly from disk center to $\mu=0.85$. From $\mu=0.5$ toward the limb, the blueshift decreases rapidly and turns into a redshift at $\mu=0.3$. The detailed Doppler shifts obtained for different spectral resolutions and averaged line segment (panel c) are listed in Table\,\ref{table_sec3_resolution_comparison}. A detailed interpretation of the results is given in Paper II \citep{2018arXiv181108685S}.

\subsection{Lines around 6302\,\AA}\label{sec_results_6302}
\begin{figure}[htbp]
\begin{center}
\includegraphics[width=\columnwidth]{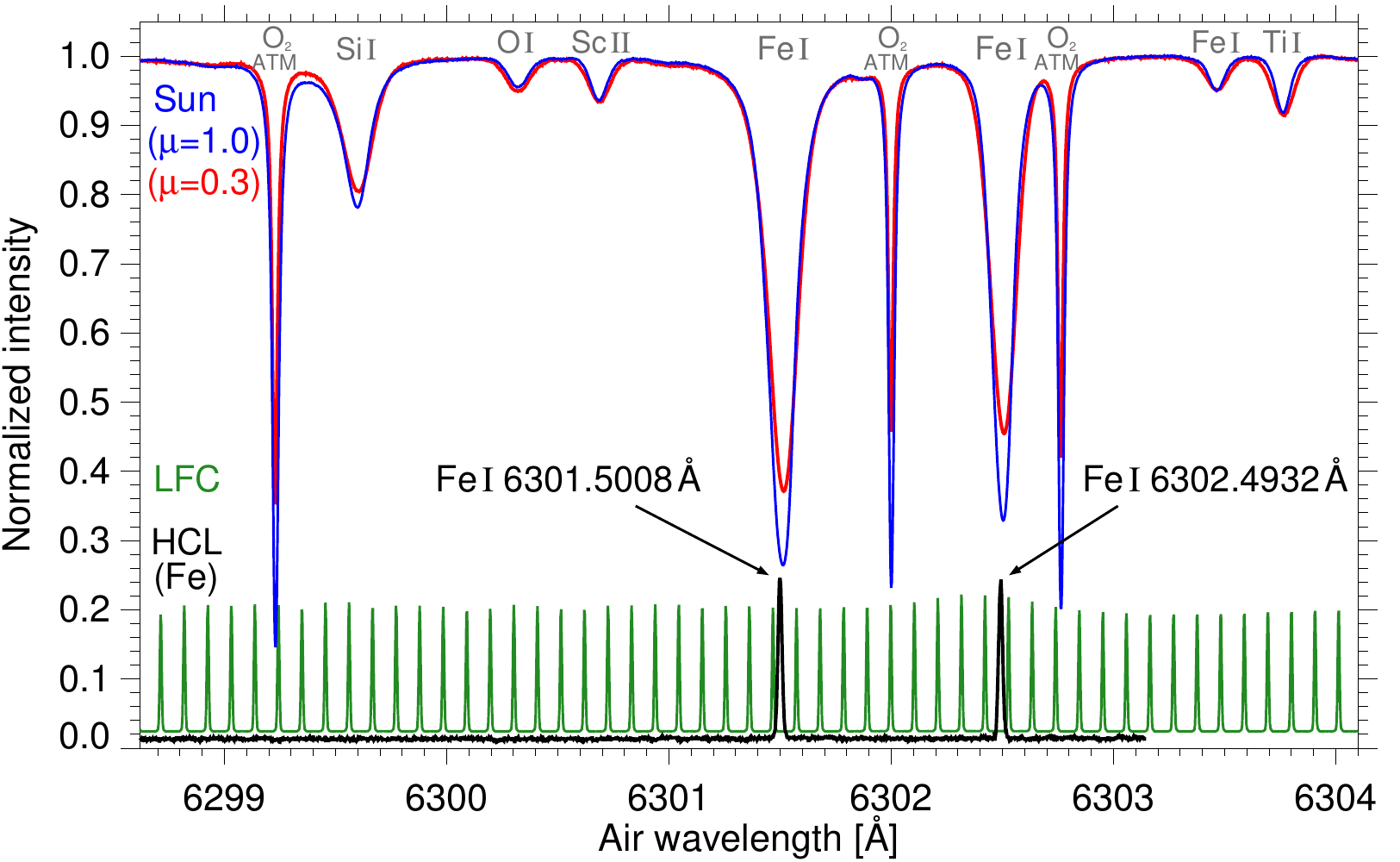}
\caption{Spectral region around 6302\,\AA, with the quiet Sun absorption spectra at the disk center ($\mu=1.0$, blue curve) and close to the solar limb ($\mu=0.3$, red curve). The {atomic species} are stated in gray. The emission spectra of the laser frequency comb (LFC, green curve) and the iron hollow cathode lamp (HCL, black curve) are displayed.}
\label{fig_sec3_spectra_6302}
\end{center}
\end{figure}

The last spectral region of this work is the 6302\,\AA\ region shown in Fig.\,\ref{fig_sec3_spectra_6302}. We have performed 99 observation sequences from $\mu=1.0$ to $\mu=0.3$. In Paper I \citep{2018A&A...611A...4L} of this series, we have presented a detailed analysis of the solar spectral lines and their convective blueshift. In this work, we include the results obtained for \ion{Fe}{I}\,6301.5\,\AA\ and \ion{Fe}{I}\,6302.5\,\AA.

\subsubsection{\ion{Fe}{I}\,6301.5\,\AA\ and \ion{Fe}{I}\,6302.5\,\AA}
The adjacent \ion{Fe}{I} lines at 6301.5\,\AA\ and 6302.5\,\AA\ are some of the most-studied lines in the visible part of the solar spectrum. Due to the proximity of telluric oxygen lines at 6302.0\,\AA\ and 6302.8\,\AA, both lines have been frequently used for the analysis of relative Doppler shifts in the solar photosphere. The formation height of the line core in the mid photosphere is around 340\,km (6301.5\,\AA) and 260\,km (6302.5\,\AA) above the solar surface at $\tau_{5000\,\AA}=1$ \citep{1991sopo.work.....N,1998A&A...332.1069K}. The \ion{Fe}{I}\,6302.5\,\AA\ line has a strong Zeeman-sensitivity ($g_\mathrm{eff}=2.50$) which is why it became the most popular line for spectro-polarimetric observations in the visible part of the solar spectrum, for example with the Hinode Spectro-Polarimeter \citep{2013SoPh..283..579L}. Beyond, VTF and ViSP will use the line for first-light high-resolution observations with DKIST.

\begin{figure}[htbp]
\textbf{a)}\\[-0.3cm] \includegraphics[width=\columnwidth]{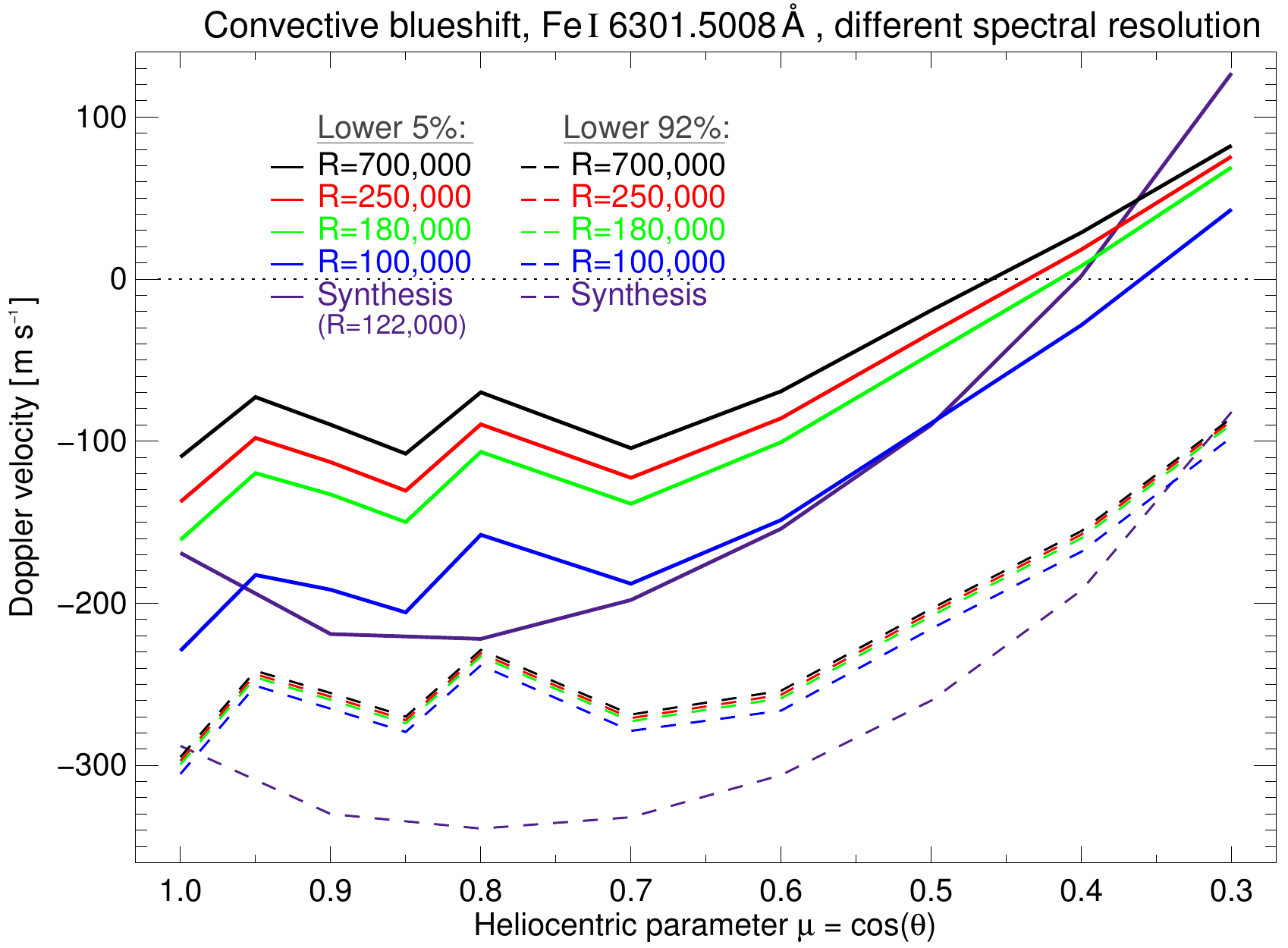}\\[0.2cm]
\textbf{b)}\\[-0.3cm] \includegraphics[width=\columnwidth]{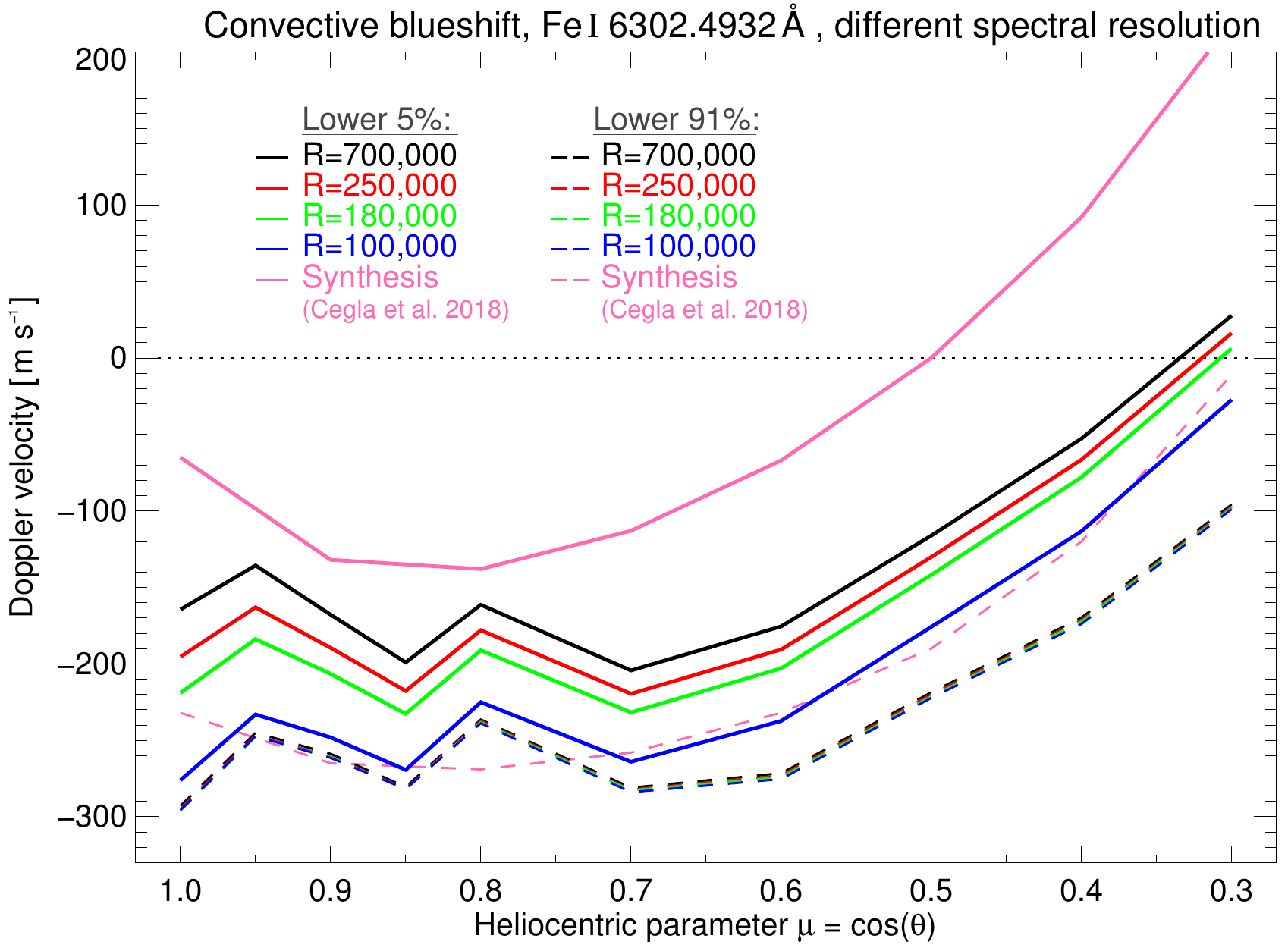}
\caption{Center-to-limb variation of the convective blueshift of \ion{Fe}{I}\,6301.5\,\AA\ (panel a) and \ion{Fe}{I}\,6302.5\,\AA\ (panel b) for different spectral resolutions and line sections. Dashed lines are close to each other or even overlay. The observations are compared with the synthesis for \ion{Fe}{I}\,6301.5\,\AA\ \citep{2011A&A...528A.113D} and \ion{Fe}{I}\,6302.5\,\AA\ \citep{2018ApJ...866...55C}.}
\label{fig_sec3_clv_resolution_Fe6302}
\end{figure}

The observed air wavelengths of 6301.5008\,\AA\ and 6302.4932\,\AA\ from the NIST ASD were taken as laboratory references with an uncertainty of 1.2\,m\AA. Various measurements with the iron hollow cathode lamp of LARS are in agreement within the given uncertainty. 

The bisectors of \ion{Fe}{I}\,6301.5\,\AA\ and \ion{Fe}{I}\,6302.5\,\AA\ and their center-to-limb variation were initially displayed in Fig.\,8 of Paper I. The distribution and center-to-limb variation of the line core shifts were presented in Fig.\,10 of Paper I. For reasons of clarity and completeness, the adapted figures are displayed in Fig.\,\ref{fig_A11} and \ref{fig_A12} of this work. For a comprehensive discussion of the results, we refer to Paper I.

The effect of the spectral resolution on the obtained Doppler velocities is highlighted in Fig.\,\ref{fig_sec3_clv_resolution_Fe6302}. By lowering the spectral resolution from 700\,000 to 100\,000, the convective blueshift of the line core increases by more than $\mathrm{100\,m\,s^{-1}}$ at disk center for both spectral lines. Toward the solar limb, the difference diminishes monotonically to around $\mathrm{50\,m\,s^{-1}}$. Detailed Doppler shifts are listed in Table\,\ref{table_sec3_resolution_comparison}. The average of the entire bisector yields the strongest blueshift. Moreover, both \ion{Fe}{I} lines exhibit the same center-to-limb variation of the convective blueshift. The blueshift at disk center amount to around $\mathrm{-295\,m\,s^{-1}}$. Toward $\mu=0.6$, the blueshift decreases slightly to around $\mathrm{-260\,m\,s^{-1}}$. When approaching the solar limb, we obtain a rapid decrease in blueshift to around $\mathrm{-90\,m\,s^{-1}}$ at $\mu=0.3$. A change of the spectral resolution hardly affects the average Doppler shift of both line. In case of \ion{Fe}{I}\,6301.5\,\AA, differences are below $\mathrm{15\,m\,s^{-1}}$. In case of \ion{Fe}{I}\,6302.5\,\AA, differences are negligible. 

In Fig.\,\ref{fig_sec3_clv_resolution_Fe6302}, we compare the observed line shifts with theoretical syntheses for \ion{Fe}{I}\,6301.5\,\AA\ \citep{2011A&A...528A.113D} and \ion{Fe}{I}\,6302.5\,\AA\ \citep{2018ApJ...866...55C}. The synthesized line shifts agree with the observations within a margin of around $\mathrm{\pm100\,m\,s^{-1}}$. In case of \ion{Fe}{I}\,6301.5\,\AA, the synthesis is consistent with the observed shifts at disk center and around $\mu=0.4$. However, it fails to reproduce the slope of the center-to-limb variation. In case of \ion{Fe}{I}\,6302.5\,\AA, the synthesis of \citet{2018ApJ...866...55C} manages to largely reproduce the slope of the observed center-to-limb variation. In anticipation of Section \ref{sec_discussion_depth}, the apparent systematic offset of the line core shift to the blue and the misplacement of its maximum blueshift of the slope at $\mu=0.8$ suggest a slight overestimation of the formation height of \ion{Fe}{I}\,6302.5\,\AA\ for the synthesis.


\section{Discussion}\label{sec_discussion}
The observation of absolute Doppler shifts of various spectral lines with the same unprecedented accuracy allows for their direct comparison. In this section, we assemble the bisectors, the inferred line core velocities, and their center-to-limb variations. On this basis, we are able to conclude on the systematic behavior of the convective Doppler shift in the solar atmosphere.

\subsection{Comparison of spectral lines}\label{sec_discussion_lines}

\begin{figure}[htbp]
\textbf{a)}\\[-0.25cm]  \includegraphics[trim=0cm 0cm 0.5cm 0cm,clip,width=\columnwidth]{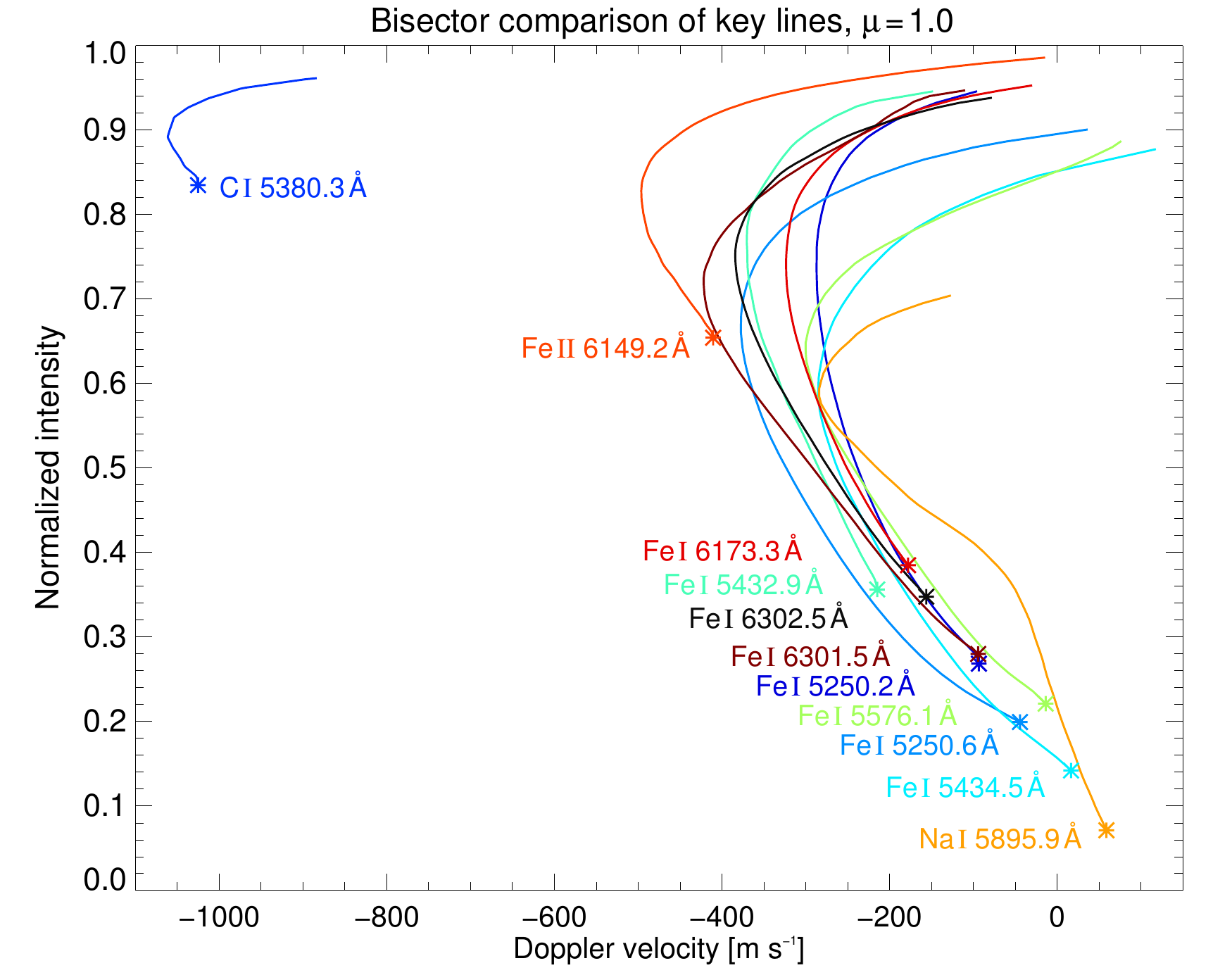}\\[0.1cm]
\textbf{b)}\\[-0.25cm]  \includegraphics[width=\columnwidth]{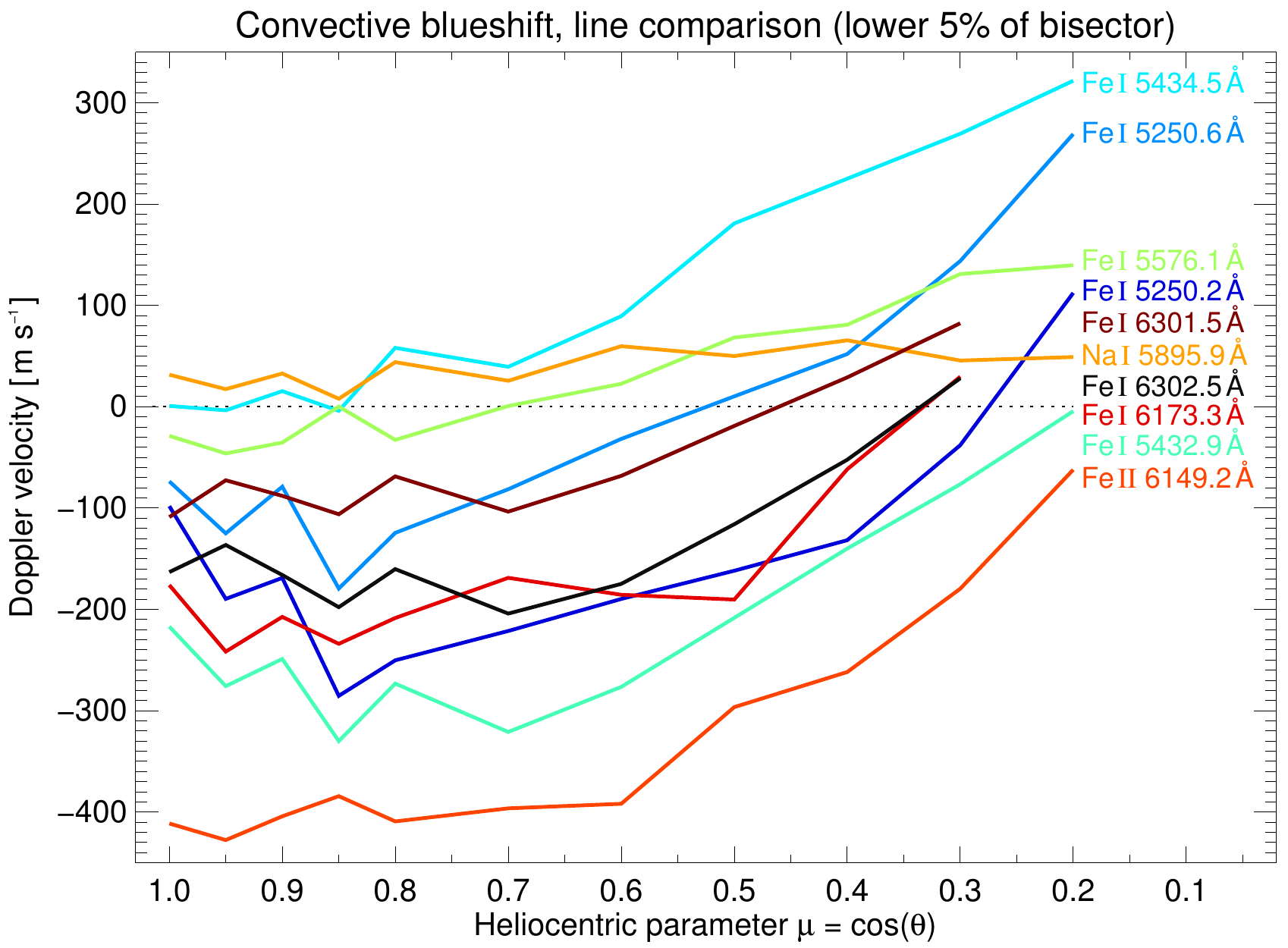}
\caption{Comparison of the convective blueshift of important solar lines. Panel a displays the bisectors at the solar disk center ($\mu=1.0$), in absolute velocities. Panel b shows the center-to-limb variation of the average convective blueshift of the line cores (lower 5\,\% of the bisectors).}
\label{fig_sec4_bisectors_keylines}
\end{figure}

First, we draw a comparison for the selection of important spectral lines as listed in Table\,\ref{table1}. The comparative view of their bisectors at disk center is provided in Fig.\,\ref{fig_sec4_bisectors_keylines} (panel a). The bisectors were normalized to the respective continuum intensity and are shown on the same absolute Doppler scale. The Doppler shifts range from maximum blueshifts of $\mathrm{-1050\,m\,s^{-1}}$ to slight redshifts of $\mathrm{+50\,m\,s^{-1}}$. The trend shows that weaker lines exhibit stronger blueshifts. But noticeably, the light \ion{C}{I}\,5380.3\,\AA\ line deviates significantly from the linear trend by an additional blueshift. In Section \ref{sec_results_5381}, we have reported on the particular characteristics of the line and its formation in the deep photosphere. We exclude the \ion{C}{I}\,5380.3\,\AA\ line for the further discussion of systematic Doppler shifts. The second deviation in the comparative view lies in the shape of the \ion{Na}{I}\,5895.9\,\AA\ bisector. Due to its transition into the chromosphere, it exhibits a C-shape proceeding into a saturation toward the line core. In contrast, all iron lines feature a distinct simple C-shape. The line minima, marked by asterisks, reveal a conspicuous relation of the Doppler shift with the line depth, and hence the formation height. Whereas the \ion{Fe}{II}\,6149.2\,\AA\ line core is formed in the lower photosphere yields a blueshift of around $\mathrm{-400\,m\,s^{-1}}$, the \ion{Fe}{I}\,5434.5\,\AA\ line core, which is formed around 400\,km higher up in the photosphere (around 550\,km), ends in zero Doppler shift. In between, the iron lines assembled according to their line depth and Doppler shift of the line core. \ion{Fe}{I}\,5250.6\,\AA\ and \ion{Fe}{I}\,5576.1\,\AA\ are the next strongest lines with a formation height of around 360\,km and a slight blueshift of around $\mathrm{-50\,m\,s^{-1}}$. The \ion{Fe}{I}\,5250.2\,\AA\ and \ion{Fe}{I}\,6301.5\,\AA\ lines form at around 330\,km in the middle photosphere with a blueshift of around $\mathrm{-100\,m\,s^{-1}}$. Another step deeper in the photosphere, the \ion{Fe}{I}\,6302.5\,\AA, \ion{Fe}{I}\,6173.3\,\AA\, and \ion{Fe}{I}\,5432.9\,\AA\ lines form at a height of around 260\,km and yield a blueshift of around $\mathrm{-200\,m\,s^{-1}}$.

The comparison of the line core shifts and their center-to-limb variations are shown in Fig.\,\ref{fig_sec4_bisectors_keylines} (panel b). For this selection of lines, the ionized \ion{Fe}{II}\,6149.2\,\AA\ line yields the overall strongest blueshift. On the contrary, the \ion{Fe}{I}\,5434.5\,\AA\ and \ion{Na}{I}\,5895.9\,\AA\ line cores are redshifted throughout the solar disk. Toward the solar limb, all line cores exhibit a decreasing blueshift (or increasing redshift). However, the slopes of the center-to-limb variations differ significantly. Between disk center and heliocentric positions around $\mu=0.6$, the blueshift remains almost constant or decreases slightly in some cases (\ion{Fe}{II}\,6149.2\,\AA, \ion{Fe}{I}\,6301.5\,\AA, \ion{Fe}{I}\,5576.1\,\AA, \ion{Fe}{I}\,5434.5\,\AA, \ion{Na}{I}\,5895.9\,\AA). For these lines the formation height of the core is either below 150\,km or above 360\,km. For the spectral lines with the core formed in between this photospheric layer (\ion{Fe}{I}\,5432.9\,\AA, \ion{Fe}{I}\,6173.3\,\AA, \ion{Fe}{I}\,5250.2\,\AA, \ion{Fe}{I}\,6302.5\,\AA, \ion{Fe}{I}\,5250.6\,\AA), the blueshift increases by up to $\mathrm{100\,m\,s^{-1}}$ toward heliocentric positions around $\mu=0.8$. Evidently, the effect of horizontal flows in granulation leading to this initial center-to-limb increase in blueshift is best measured in the mid photosphere. The \ion{Fe}{I}\,5432.9\,\AA\ line core formed at a height of around 250\,km yields the most pronounced reversal point of the center-to-limb variation. 

\subsection{Systematic convective blueshift}\label{sec_discussion_systematics}

\begin{figure}[htbp]
\textbf{a)}\\[-0.25cm]  \includegraphics[width=\columnwidth]{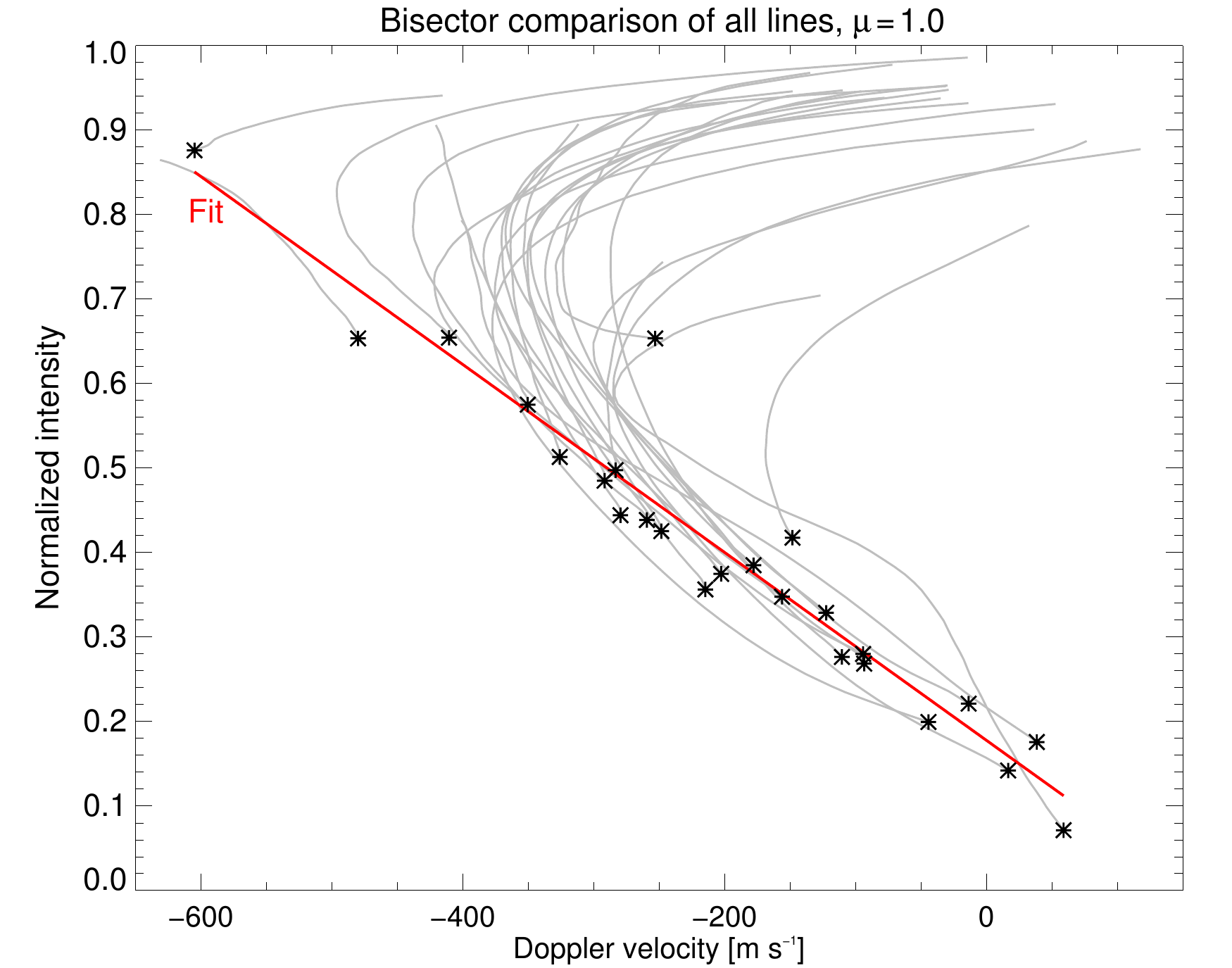}\\[0.1cm]
\textbf{b)}\\[-0.25cm]  \includegraphics[width=\columnwidth]{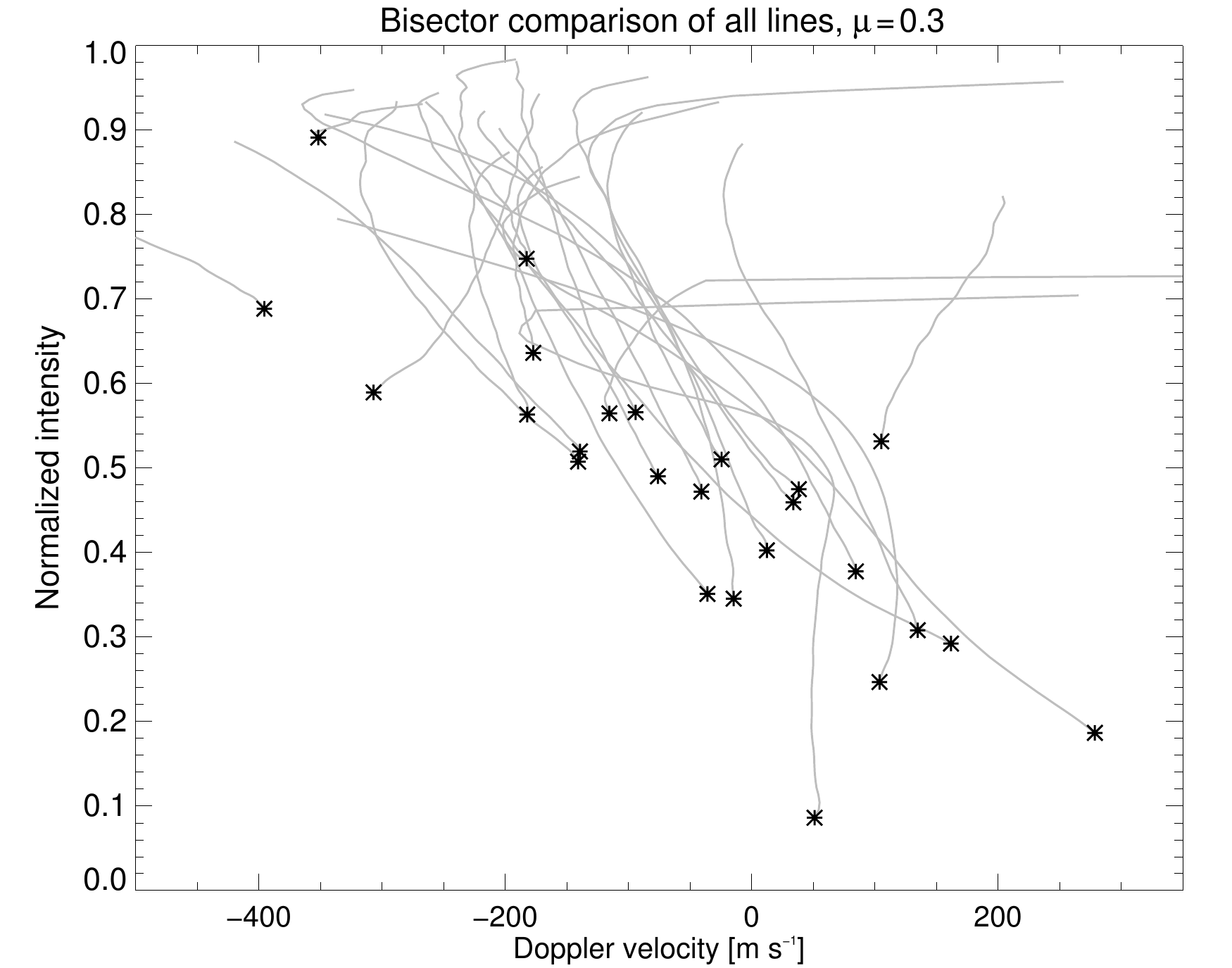}
\caption{Bisector comparison of all analyzed spectral lines at the disk center (panel a) and at $\mu=0.3$ (panel b) close to the solar limb. }
\label{fig_sec4_bisectors_alllines}
\end{figure}

From the above selection of 11 spectral lines in Fig.\,\ref{fig_sec4_bisectors_keylines} (panel a), we have inferred a noticeable relation between the line depth and the Doppler shift of the line core. Deeper lines feature stronger convective blueshifts. To review this dependence, we now want to compare the bisectors of all 26 analyzed spectral lines. In Fig.\,\ref{fig_sec4_bisectors_alllines}, we display their bisectors at disk center ($\mu=1.0$, panel a) and close to the solar limb ($\mu=0.3$, panel b). We exclude the \ion{C}{I}\,5380.3\,\AA\ for reasons of clarity. The comparative view at disk center reveals that most bisectors follow the same trend. The maximum blueshift of the C-shape amounts to around $\mathrm{-350\,m\,s^{-1}}$ and is typically reached at a normalized intensity between 0.7 and 0.8. From 0.6 to below 0.2, the bisectors follow a monotonic decrease from blueshifts around $\mathrm{-320\,m\,s^{-1}}$ toward slight redshifts. We mark the line minima by asterisks to highlight the significant relation between the Doppler shift $\mathrm{v_{los}}$ of the line core and its intensity $\mathrm{I_{min}}$ (or line depth). The linear fit 
\begin{equation}
\mathrm{v_{los}=-831.7\,m\,s^{-1}\cdot I_{min}+132.7\,m\,s^{-1}}\label{eq2}
\end{equation}
appears to be a believable model. The respective 1$\sigma$-uncertainties of $\mathrm{49.5\,m\,s^{-1}}$ for the slope and $\mathrm{21.9\,m\,s^{-1}}$ for the intercept are small. 
We interpret this trend as the vertically decelerating convective upflow in the solar atmosphere. From the deepest to the highest layers of the photosphere, the systematic convective blueshift decreases by approximately $\mathrm{700\,m\,s^{-1}}$. 

Toward the solar limb, the accordances of the bisector profiles weakens. For $\mu=0.3$ close to the solar limb, Fig.\,\ref{fig_sec4_bisectors_alllines} (panel b) shows a larger scatter of bisector shapes and Doppler shifts. The relation of decreasing blueshifts (or increasing redshifts) for deeper lines still exists, but a linear fit through the minimum positions of all lines would have a much larger uncertainty than at disk center. We attribute the larger scatter of Doppler shifts to the widened region of line formation in the photosphere.

\subsection{Center-to-limb variation with line depth}\label{sec_discussion_depth}

\begin{figure}[htbp]
\includegraphics[width=\columnwidth]{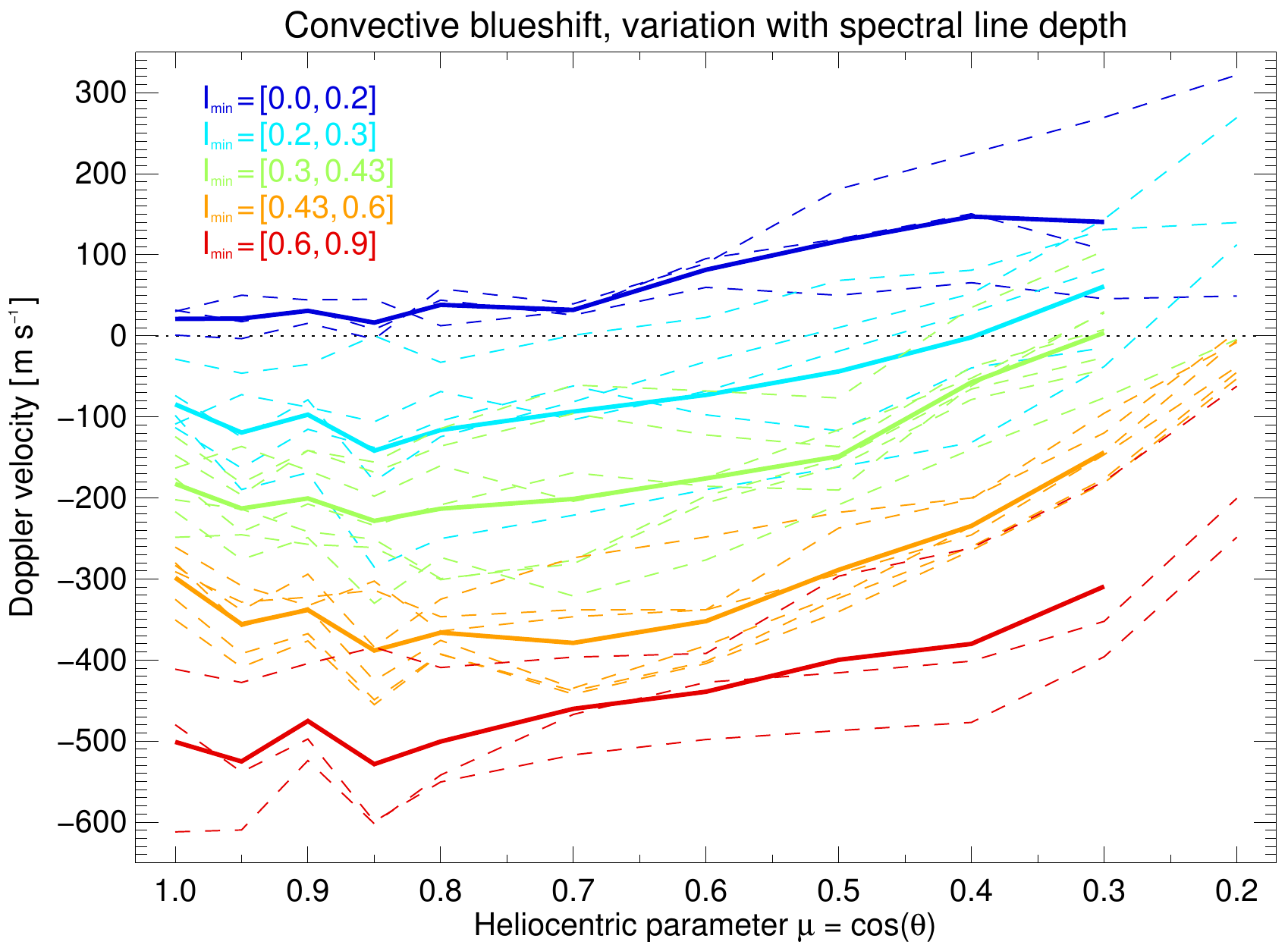}
\caption{Center-to-limb variation of the convective blueshift as a function of spectral line depths. The Doppler shifts of the line core of 25 analyzed spectral lines (dashed curves) are plotted in the respective color of the normalized intensity $\mathrm{I_{min}}$ range of the line core. The average curve of the respective range is highlighted as solid thick line.}
\label{fig_sec4_clv_linedepth}
\end{figure}

As demonstrated in Fig.\,\ref{fig_sec4_bisectors_keylines} (panel b), the center-to-limb variation of the convective blueshift and its slope differs from line to line. To capture any systematic behavior of the center-to-limb variation, we extend the comparative view to all 25 analyzed spectral lines (excluding \ion{C}{I}\,5380.3\,\AA). As illustrated in Fig.\,\ref{fig_sec4_clv_linedepth}, the Doppler shifts of the line cores at disk center range from strong blueshift of $\mathrm{-610\,m\,s^{-1}}$ to slight redshifts of $\mathrm{+30\,m\,s^{-1}}$. Toward the solar limb around $\mu=0.2$, Doppler velocities have shifted to a range from $\mathrm{-240\,m\,s^{-1}}$ to $\mathrm{+320\,m\,s^{-1}}$. A classification of the convective blueshift according to the spectral line depth (respectively the intensity $\mathrm{I_{min}}$ of the line minimum normalized to the continuum) yields the systematic trend of its center-to-limb variation. We found five classes of line depth for which the comprised center-to-limb variations are in good agreement. The average center-to-limb variations of the five classes with line minima at intensities within [0,\,0.2], [0.2,\,0.3], [0.3,\,0.43], [0.43,\,0.6], and [0.6,\,0.9], are highlighted in Fig.\,\ref{fig_sec4_clv_linedepth}. The weakest lines with intensity minima between 0.6 and 0.9 which are attributed to the lower photosphere feature the strongest blueshifts. The blueshift of around $\mathrm{-500\,m\,s^{-1}}$ is stable from disk center to heliocentric positions around $\mu=0.8$. Toward $\mu=0.3$ close to the limb, the blueshift decreases monotonically to around $\mathrm{-300\,m\,s^{-1}}$. As opposed to this, the strongest lines with intensity minima below 0.2 are exclusively redshifted. Ascribed to the high photosphere to lower chromosphere, these lines show an almost constant redshift from disk center to heliocentric positions around $\mu=0.7$. When approaching the solar limb, the redshift slightly increases first and saturates at around $\mathrm{+140\,m\,s^{-1}}$ close to the limb. Neither very strong lines nor very weak lines display a reversal point in the slope of their center-to-limb variation. This initial increase in blueshift by around $\mathrm{50-80\,m\,s^{-1}}$ when departing from disk center appears to be only present for spectral lines with line minima at intensities between 0.2 and 0.6. The effect becomes strongest for line minima at intensities between 0.43 and 0.6, attributed to a formation in the mid photosphere. Moreover, the reversal point of the center-to-limb variation shifts from $\mu=0.85$ for line minima intensities between 0.2 and 0.43, toward $\mu=0.7$ for intensities between 0.43 and 0.6. We conclude that the maximum line-of-sight effect of horizontal flows in granular motion shifts by almost one fifth of the disk radius toward the solar limb when sampling slightly lower layers ($\mathrm{300\,km\,\rightarrow\,200\,km}$) in the mid photosphere.

\section{Conclusions}\label{sec_conclusions}
In this observational study, we have performed the most accurate analysis of the convective blueshift in the quiet Sun at highest spectral resolution. The LARS instrument at the German Vacuum Tower Telescope provided an accuracy at the $\mathrm{m\,s^{-1}}$, given by the wavelength calibration using a laser frequency comb. This guaranteed the unrestricted repeatability of the measurements, in total 203 hours. This repeatability also ensures the unlimited comparability of measurements taken at very different times. We selected eight spectral regions in the visible range including 11 important spectral lines in the field of solar physics. The formation heights of these lines covered several hundred km in the solar atmosphere, from the lowermost photosphere up to the lower chromosphere. To obtain the center-to-limb variation of the asymmetry and Doppler shift of each line, we scanned the quiet Sun along four radial axes, each axis at up to 11 strategically chosen heliocentric positions. We have provided the systematic transition of the line bisectors, which typically transform from a C-shape at disk center into a \textbackslash-shape at the solar limb. Further, we restricted the analysis to the Doppler shifts of the line core, and the average shift of the entire line. The results are listed in Table \ref{table_sec3_resolution_comparison} and are intended to serve as reference values for future observations. To allow for a comparison with observations from other instruments, we have transformed the results to the respective lower spectral resolution. We thus facilitate the indirect calibration of absolute Dopplergrams with an accuracy of a few ten $\mathrm{m\,s^{-1}}$. Given that the Dopplergram contains a portion of quiet Sun, one could set the reference shift for its temporally and spatially averaged line profile, at the given spectral resolution and heliocentric position. An accurate Doppler calibration will enable the unambiguous identification of small-scale dynamics in active and quiet Sun regions, as planned with high-resolution observations with the new 4\,m-class Daniel K. Inouye Solar Telescope \citep{2012ASPC..463..377R,2015IAUGA..2257167R,2016AN....337.1064T} and its first-light instruments VTF \citep{2012SPIE.8446E..77K}, ViSP \citep{2012SPIE.8446E..6XD} and DL-NIRSP \citep{2014SPIE.9147E..07E}, or with IMAX \citep{2011SoPh..268...57M} aboard the third Sunrise \citep{2018cosp...42E.215B} flight.

We have performed a comprehensive and meaningful analysis of the systematic convective blueshift by including in total 26 spectral lines, most of them \ion{Fe}{I} lines. In comparison, the average Doppler shifts of the entire line features a stronger blueshift than the line core itself. In conclusion, the uppermost layer of the formation region exhibits a weaker blueshift than the average atmosphere below. The direct comparison of the line bisectors at disk center has revealed a linear decrease of the convective blueshift of the line core with increasing line depth. This systematic behavior represents the deceleration of the average upflow with increasing height in the photosphere. Starting from a strong blueshift of potentially more than $\mathrm{-700\,m\,s^{-1}}$ at the surface, it ends in a slight redshift at the transition into the chromosphere. From disk center toward the solar limb, the convective shifts describe a systematic variation of decreasing blueshifts, or increasing redshifts. Model syntheses by \citet{2011A&A...528A.113D} and \citet{2018ApJ...866...55C} yield deviations by less than $\mathrm{100\,m\,s^{-1}}$, but need refinement to better reproduce the observed slope of the center-to-limb variation. We found the center-to-limb variation to be highly dependent on the line depth and the linked atmospheric height. Only in the mid photosphere, the center-to-limb variation displays an initial increase in blueshift when departing the solar disk center. In line with \citet{1985SoPh...99...31B}, the reversal point of the center-to-limb curve at heliocentric positions around $\mu=0.8$ marks the maximum contribution of horizontal granular motions to the observed Doppler shift. Moreover, we were able to measure a dependence of the reversal point on the photospheric layer. We attribute this heliocentric shift to the three-dimensional geometry of granular motions and its implied optical line-of-sight effect. 

\section{Outlook: Sun as a Star}\label{sec_outlook}
Beyond, the detection of Earth-like exoplanets requires stellar radial velocity measurements with a precision of better than $\mathrm{1\,m\,s^{-1}}$. Only accurately-calibrated high-resolution spectrographs, like ESPRESSO \citep[Echelle SPectrograph for Rocky Exoplanets and Stable Spectroscopic Observations,][]{2014AN....335....8P} at ESO's Very Large Telescope (VLT) at the Paranal Observatory, HARPS \citep[High Accuracy Radial velocity Planet Searcher,][]{2003Msngr.114...20M} at ESO's 3.6-metre telescope at the La Silla Observatory, or HARPS-North \citep{2012SPIE.8446E..1VC,2012Msngr.149....2L} at the Italian Telescopio Nazionale Galileo (TNG) at the Roque de los Muchachos Observatory, can provide the necessary precision and repeatability to disentangle the weak oscillatory signature of potentially habitable planets in stellar spectra.
Taking our solar system as a reference, observations of the Sun as a Star \citep{2015ApJ...814L..21D} would have to be performed at the $\mathrm{cm\,s^{-1}}$ precision over several decades to identify the planetary contribution to the radial velocity signal of the Sun. For example, the signal of Jupiter has an amplitude of $\mathrm{12.4\,m\,s^{-1}}$ over 11.9 years, while Saturn causes an oscillation of $\mathrm{2.8\,m\,s^{-1}}$ over 29.5 years. To recover the Earth or Venus, observations have to be capable to resolve the radial velocity signal of $\mathrm{9\,cm\,s^{-1}}$.
To this end, the detection of Earth-like exoplanets is essentially based on the profound determination and distinction of all radial velocity perturbations induced by the stellar activity itself \citep{2011A&A...525A.140D,2015PhDT.......193H}. However, an optimal characterization and correction is extremely challenging for an unresolved Star. As we learn from the Sun as our unique test case, such noise components include acoustic waves with amplitudes of few hundred $\mathrm{m\,s^{-1}}$ on the time scale of minutes, as well as non-periodic surface motions such as supergranular flows with variations on the time scale of hours. Active regions with dark spots and bright faculae can suppress the local convection by a few hundred $\mathrm{m\,s^{-1}}$ \citep{2018A&A...617A..19L}. Their temporal evolution and their rotation across the integrated disk lead to Doppler variations on the time scale of days to weeks. Thereby, the rotational imbalance due to active regions causes an overall variation at the order of a few ten $\mathrm{cm\,s^{-1}}$. In comparison, the suppression of convective blueshift, mostly by faculae, induces a radial-velocity variation of a few $\mathrm{m\,s^{-1}}$ \citep{2010A&A...512A..39M,2016MNRAS.457.3637H}. On the time scale of several months to years, magnetic cycles can result in periodic variations of the convective blueshift by a few $\mathrm{m\,s^{-1}}$ \citep{2018A&A...610A..52B}. These activity-induced variations can easily be misinterpreted as the signal of an orbiting planet. Thus, the crucial step toward the unambiguous detection of Earth-like exoplanets lies in the optimal modeling of the convective blueshift of the stellar atmosphere. Most accurate measurements of the solar convective blueshift and its center-to-limb variation, as presented in the work at hand, have to serve as the basis for an optimized model of the solar convective blueshift. The final translation to the spatially unresolved case of the solar-like Star will advance the required modeling of its convective blueshift and its contribution to the radial velocity signal. 

As in the solar case, we do not expect average Doppler shifts of stellar spectral lines to depend on the spectral resolution of the instrument. But it is crucial to understand that Doppler shifts vary from line to line according to their formation in the lower stellar atmosphere \citep{2018arXiv180901548D}. Therefore, the radial velocity analysis of the stellar spectrum has to consider line-dependent Doppler shifts, presumably relative to the given line depth.

\begin{acknowledgements} We thank our colleagues at the Kiepenheuer Institute for Solar Physics, at Menlo Systems GmbH, and at the Max Planck Institute of Quantum Optics for their work on the LARS instrument. We especially acknowledge Dr. Hans-Peter Doerr for his work on the LARS prototype and his support on the operation of the instrument. We thank our observing assistants for their support during the observation campaigns. The operation of the Vacuum Tower Telescope at the Observatorio del Teide on Tenerife was performed by the Kiepenheuer Institute for Solar Physics Freiburg, which is a public law foundation of the State of Baden-W\"urttemberg. This work was funded by the Deutsche Forschungsgemeinschaft (DFG, Ref.-No. Schm-1168/10). We thank Catherine Fischer the internal proofreading, and the referee for the constructive comments on the manuscript. The National Center for Atmospheric Research is sponsored by the National Science Foundation.
\end{acknowledgements}

\bibliographystyle{aa} 
\bibliography{LARS} 

\begin{appendix}
\section{Additional table and figures}

\begin{table*}[ht]
\caption{Convective shifts of spectral lines in the Quiet Sun. The Doppler velocities (in $\mathrm{ m\,s^{-1}}$) from the disk center ($\mu=1.0$) to the solar limb ($\mu=0.2$) are listed for different spectral resolutions $\mathrm{R}$. Average velocities are given for the full line profile (from the line minimum to an upper threshold close to the continuum intensity level) and only the line core (lower 5\% of the bisector).}
\label{table_sec3_resolution_comparison}
\centering
\tabcolsep=0.19cm
\begin{tabular}{c c c r r r r r r r r r r r}
\hline\hline
Spectral line&Line&$\mathrm{R}$&\multicolumn{11}{c}{Heliocentric position $\mu = \cos\theta$}\\ 
($\lambda_0$ in \AA)&section&&1.0&0.95&0.9&0.85&0.8&0.7&0.6&0.5&0.4&0.3&0.2\\ 
\hline\\[-0.30cm]
\ion{Fe}{I} 5250.2084&Full&700\,000&$-226$ &$-286$ &$-241$ &$-346$ &$-300$ &$-271$ &$-253$
&$-237$ &$-226$ &$-157$ &$-7$ \\ &Core&700\,000 &$-99$ &$-191$ &$-170$
&$-286$ &$-251$ &$-222$ &$-190$ &$-162$ &$-132$ &$-38$ &$111$ \\ &&250\,000
&$-141$ &$-222$ &$-193$ &$-303$ &$-264$ &$-233$ &$-201$ &$-174$ &$-145$ &$-52$
&$96$ \\ &&180\,000 &$-169$ &$-243$ &$-210$ &$-315$ &$-274$ &$-242$ &$-210$
&$-184$ &$-157$ &$-64$ &$83$ \\ &&100\,000 &$-221$ &$-283$ &$-241$ &$-342$
&$-297$ &$-263$ &$-235$ &$-211$ &$-188$ &$-101$ &$46$\\[0.05cm]
\ion{Fe}{I} 5250.6453&Full&700\,000 &$-274$ &$-318$ &$-257$ &$-349$ &$-287$ &$-236$ &$-202$
&$-170$ &$-147$ &$-69$ &$89$ \\ &Core&700\,000 &$-74$ &$-126$ &$-80$
&$-180$ &$-126$ &$-82$ &$-32$ &$10$ &$52$ &$144$ &$268$ \\ &&250\,000 &$-113$
&$-165$ &$-116$ &$-212$ &$-157$ &$-109$ &$-59$ &$-15$ &$27$ &$120$ &$250$ \\
&&180\,000 &$-145$ &$-197$ &$-146$ &$-239$ &$-182$ &$-131$ &$-81$ &$-37$ &$5$
&$99$ &$234$ \\ &&100\,000 &$-228$ &$-276$ &$-219$ &$-307$ &$-247$ &$-191$
&$-144$ &$-101$ &$-61$ &$34$ &$180$\\[0.05cm]
\ion{C}{I} 5380.3308&Full&700\,000 &$-1028$ &$-988$ &$-974$ &$-940$ &$-959$ &$-887$ &$-802$
&$-764$ &$-708$ &$-679$ &\\ &Core&700\,000 &$-1015$ &$-980$ &$-977$ &$-935$
&$-950$ &$-890$ &$-796$ &$-783$ &$-738$ &$-688$ &\\ &&250\,000 &$-1022$ &$-986$
&$-981$ &$-940$ &$-957$ &$-898$ &$-808$ &$-785$ &$-742$ &$-714$ &\\ &&180\,000
&$-1027$ &$-990$ &$-983$ &$-944$ &$-962$ &$-901$ &$-814$ &$-788$ &$-743$ &$-718$
&\\ &&100\,000 &$-1040$ &$-1000$ &$-989$ &$-953$ &$-973$ &$-906$ &$-821$ &$-790$
&$-743$ &$-719$ &\\[0.05cm]
\ion{Fe}{I} 5434.5232&Full&700\,000 &$-179$ &$-204$ &$-175$ &$-242$ &$-178$ &$-211$ &$-172$ &$-103$
&$-40$ &$22$ &$115$ \\ &Core&700\,000 &$1$ &$-2$ &$16$ &$-4$ &$55$ &$39$ &$90$
&$181$ &$225$ &$269$ &$321$ \\ &&250\,000 &$-15$ &$-21$ &$-1$ &$-24$ &$36$ &$20$
&$72$ &$165$ &$211$ &$258$ &$312$ \\ &&180\,000 &$-32$ &$-39$ &$-19$ &$-44$
&$17$ &$1$ &$55$ &$149$ &$197$ &$247$ &$302$ \\ &&100\,000 &$-93$ &$-106$ &$-81$
&$-114$ &$-50$ &$-67$ &$-6$ &$87$ &$145$ &$203$ &$265$\\[0.05cm]
\ion{Fe}{I} 5432.9470&Full&700\,000 &$-312$ &$-357$ &$-313$ &$-397$ &$-337$ &$-381$ &$-341$ &$-289$
&$-222$ &$-171$ &$-74$ \\ &Core&700\,000 &$-217$ &$-277$ &$-249$ &$-331$ &$-274$
&$-323$ &$-276$ &$-208$ &$-140$ &$-77$ &$-5$ \\ &&250\,000 &$-237$ &$-292$
&$-261$ &$-342$ &$-284$ &$-332$ &$-284$ &$-218$ &$-149$ &$-84$ &$-7$ \\
&&180\,000 &$-253$ &$-305$ &$-271$ &$-351$ &$-292$ &$-339$ &$-290$ &$-226$
&$-156$ &$-91$ &$-10$ \\ &&100\,000 &$-292$ &$-338$ &$-298$ &$-378$ &$-316$
&$-360$ &$-312$ &$-249$ &$-180$ &$-115$ &$-24$\\[0.05cm]
\ion{Fe}{I} 5576.0881&Full&700\,000 &$-158$ &$-166$ &$-160$ &$-108$ &$-150$ &$-110$ &$-91$ &$-54$
&$-17$ &$37$ &$98$ \\ &Core&700\,000 &$-31$ &$-47$ &$-36$ &$0$ &$-33$ &$1$ &$22$
&$68$ &$81$ &$131$ &$140$ \\ &&250\,000 &$-59$ &$-75$ &$-63$ &$-23$ &$-56$
&$-18$ &$7$ &$55$ &$72$ &$125$ &$139$ \\ &&180\,000 &$-84$ &$-98$ &$-85$ &$-42$
&$-75$ &$-34$ &$-7$ &$43$ &$63$ &$119$ &$138$ \\ &&100\,000 &$-150$ &$-159$
&$-145$ &$-95$ &$-128$ &$-80$ &$-47$ &$4$ &$34$ &$98$ &$132$\\[0.05cm]
\ion{Na}{I} 5895.92424&Full&700\,000 &$-99$ &$-110$ &$-81$ &$-101$ &$-84$ &$-78$ &$-33$ &$-42$ &$11$
&$19$ &$37$ \\ &Core&700\,000 &$32$ &$18$ &$35$ &$9$ &$45$ &$26$ &$61$ &$51$
&$65$ &$45$ &$52$ \\ &&250\,000 &$30$ &$15$ &$31$ &$7$ &$43$ &$25$ &$59$ &$48$
&$65$ &$45$ &$47$ \\ &&180\,000 &$27$ &$12$ &$29$ &$5$ &$41$ &$24$ &$58$ &$47$
&$64$ &$44$ &$45$ \\ &&100\,000 &$19$ &$3$ &$23$ &$-2$ &$34$ &$20$ &$56$ &$43$
&$62$ &$43$ &$41$\\[0.05cm]
\ion{Fe}{II} 6149.2460&Full&700\,000 &$-430$ &$-425$ &$-392$ &$-379$ &$-404$ &$-381$ &$-392$ &$-320$
&$-296$ &$-215$ &$-142$ \\ &Core&700\,000 &$-412$ &$-429$ &$-405$ &$-385$
&$-409$ &$-396$ &$-391$ &$-296$ &$-261$ &$-180$ &$-63$ \\ &&250\,000 &$-430$
&$-441$ &$-413$ &$-393$ &$-417$ &$-399$ &$-396$ &$-304$ &$-270$ &$-184$ &$-80$
\\ &&180\,000 &$-443$ &$-450$ &$-419$ &$-398$ &$-422$ &$-401$ &$-398$ &$-309$
&$-275$ &$-188$ &$-90$ \\ &&100\,000 &$-464$ &$-462$ &$-427$ &$-406$ &$-429$
&$-404$ &$-404$ &$-320$ &$-287$ &$-201$ &$-113$\\[0.05cm]
\ion{Fe}{I} 6173.3344&Full&700\,000 &$-262$ &$-309$ &$-257$ &$-280$ &$-242$ &$-205$ &$-233$ &$-258$
&$-151$ &$-75$ &\\ &Core&700\,000 &$-177$ &$-243$ &$-208$ &$-239$ &$-210$
&$-169$ &$-186$ &$-191$ &$-63$ &$29$ &\\ &&250\,000 &$-207$ &$-267$ &$-228$
&$-254$ &$-221$ &$-180$ &$-197$ &$-204$ &$-77$ &$14$ &\\ &&180\,000 &$-228$
&$-285$ &$-241$ &$-265$ &$-230$ &$-189$ &$-206$ &$-214$ &$-88$ &$2$ &\\
&&100\,000 &$-267$ &$-317$ &$-267$ &$-287$ &$-249$ &$-208$ &$-227$ &$-242$
&$-119$ &$-31$ &\\ &&81\,000 &$-275$ &$-322$ &$-271$ &$-291$ &$-252$ &$-212$
&$-234$ &$-251$ &$-132$ &$-46$ &\\[0.05cm]
\ion{Fe}{I} 6301.5008&Full&700\,000 &$-295$ &$-242$ &$-255$ &$-270$ &$-229$ &$-269$ &$-254$ &$-203$
&$-155$ &$-86$ &\\ &Core&700\,000 &$-110$ &$-73$ &$-90$ &$-108$ &$-70$ &$-104$
&$-69$ &$-19$ &$29$ &$82$ &\\ &&250\,000 &$-137$ &$-98$ &$-113$ &$-131$ &$-90$
&$-123$ &$-86$ &$-33$ &$18$ &$76$ &\\ &&180\,000 &$-161$ &$-120$ &$-133$ &$-150$
&$-107$ &$-139$ &$-101$ &$-46$ &$8$ &$69$ &\\ &&100\,000 &$-229$ &$-183$ &$-192$
&$-206$ &$-158$ &$-188$ &$-149$ &$-89$ &$-28$ &$43$ &\\[0.05cm]
\ion{Fe}{I} 6302.4932&Full&700\,000 &$-293$ &$-245$ &$-259$ &$-280$ &$-236$ &$-281$ &$-272$ &$-219$
&$-170$ &$-96$ &\\ &Core&700\,000 &$-164$ &$-136$ &$-168$ &$-199$ &$-161$
&$-204$ &$-176$ &$-116$ &$-53$ &$28$ &\\ &&250\,000 &$-195$ &$-163$ &$-189$
&$-218$ &$-178$ &$-220$ &$-191$ &$-130$ &$-66$ &$16$ &\\ &&180\,000 &$-219$
&$-184$ &$-206$ &$-233$ &$-191$ &$-232$ &$-203$ &$-142$ &$-78$ &$6$ &\\
&&100\,000 &$-276$ &$-233$ &$-248$ &$-269$ &$-225$ &$-264$ &$-237$ &$-176$
&$-113$ &$-27$ &\\[0.05cm]
\hline
\end{tabular}
\end{table*}

\clearpage
\subsection{Lines around 5250\,\AA}

\begin{figure}[htbp]
\vspace{-0.2cm}
\textbf{a)}\\[-0.3cm]  \includegraphics[width=\columnwidth]{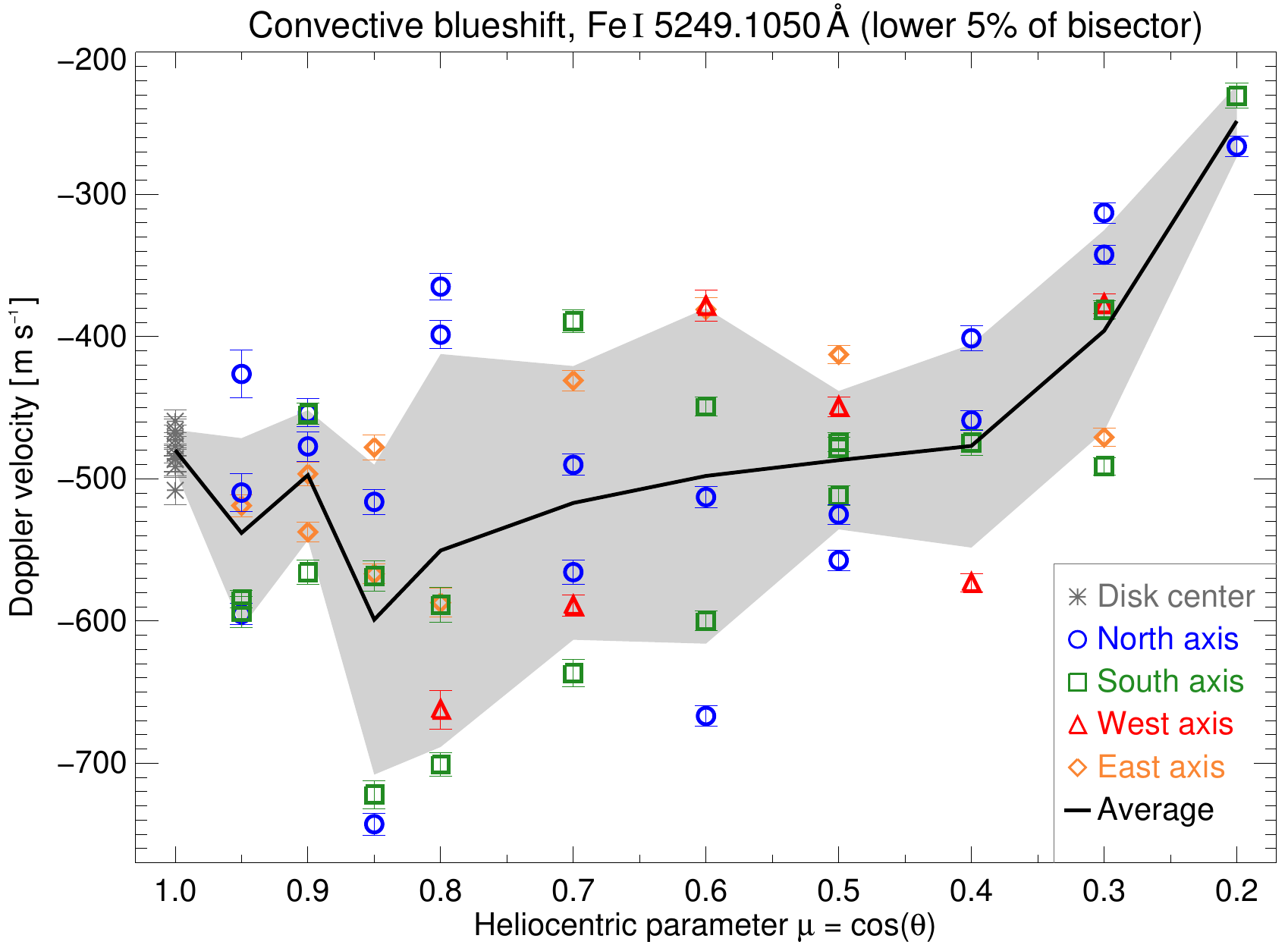}\\[0.1cm]
\textbf{b)}\\[-0.3cm]  \includegraphics[width=\columnwidth]{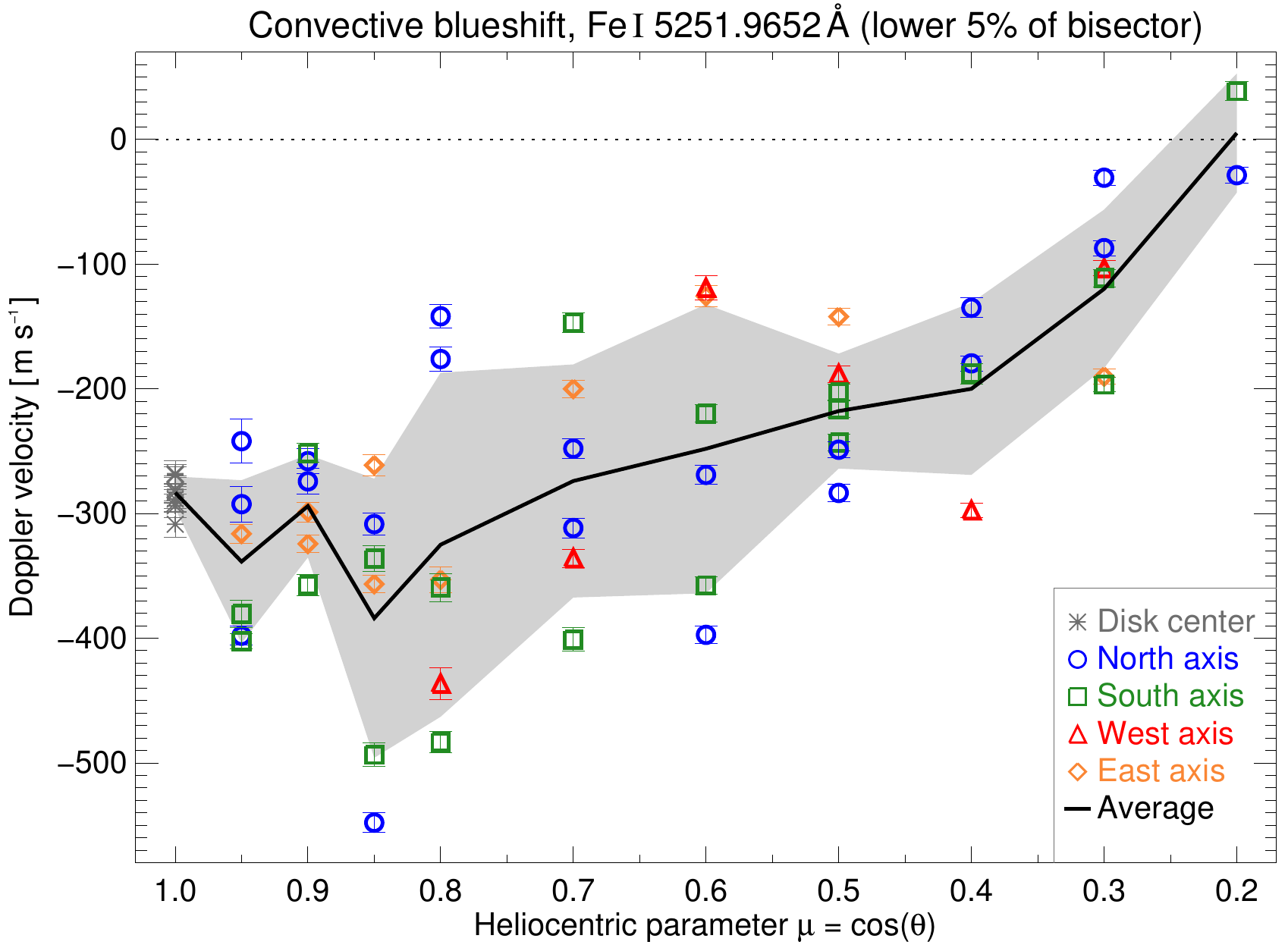}\\[0.1cm]
\textbf{c)}\\[-0.3cm]  \includegraphics[width=\columnwidth]{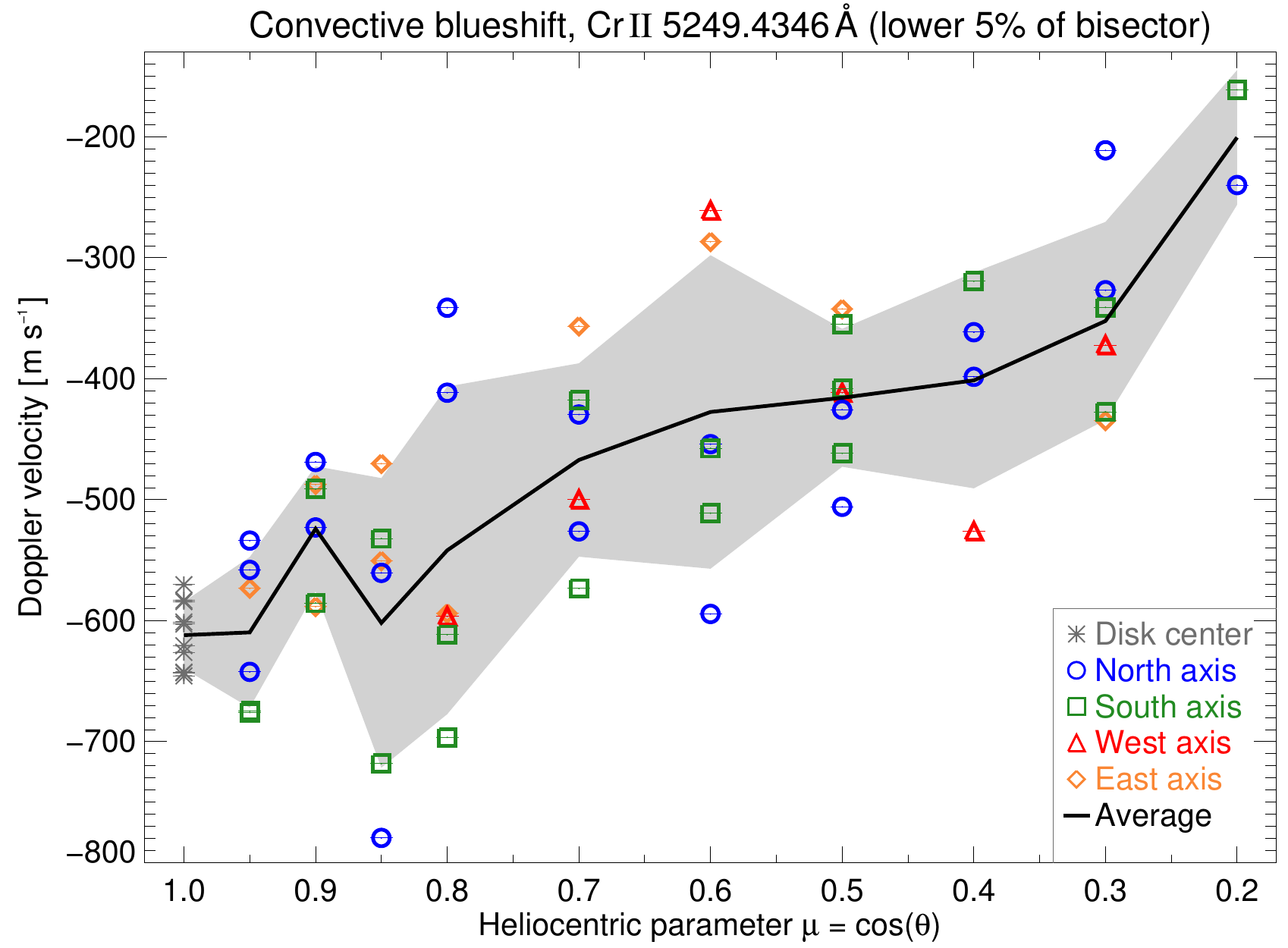}
\caption{Center-to-limb variation of the convective blueshift of the \ion{Fe}{I}\,5249.1\,\AA\ (panel a), \ion{Fe}{I}\,5251.9\,\AA\ (panel b), and \ion{Cr}{II}\,5249.4\,\AA\ (panel c) line. Each data point represents the mean Doppler velocity of the lower 5\,\% of the bisector of the temporally averaged observation sequence. Error bars indicate the mean error. Radial axes are indicated by colors and symbols. The black solid line and gray shaded area display the average center-to-limb variation and its standard deviation.}
\label{fig_A1}
\end{figure}

\begin{figure}[htbp]
\vspace{0.87cm}
\textbf{a)}\\[-0.3cm]  \includegraphics[width=0.937\columnwidth]{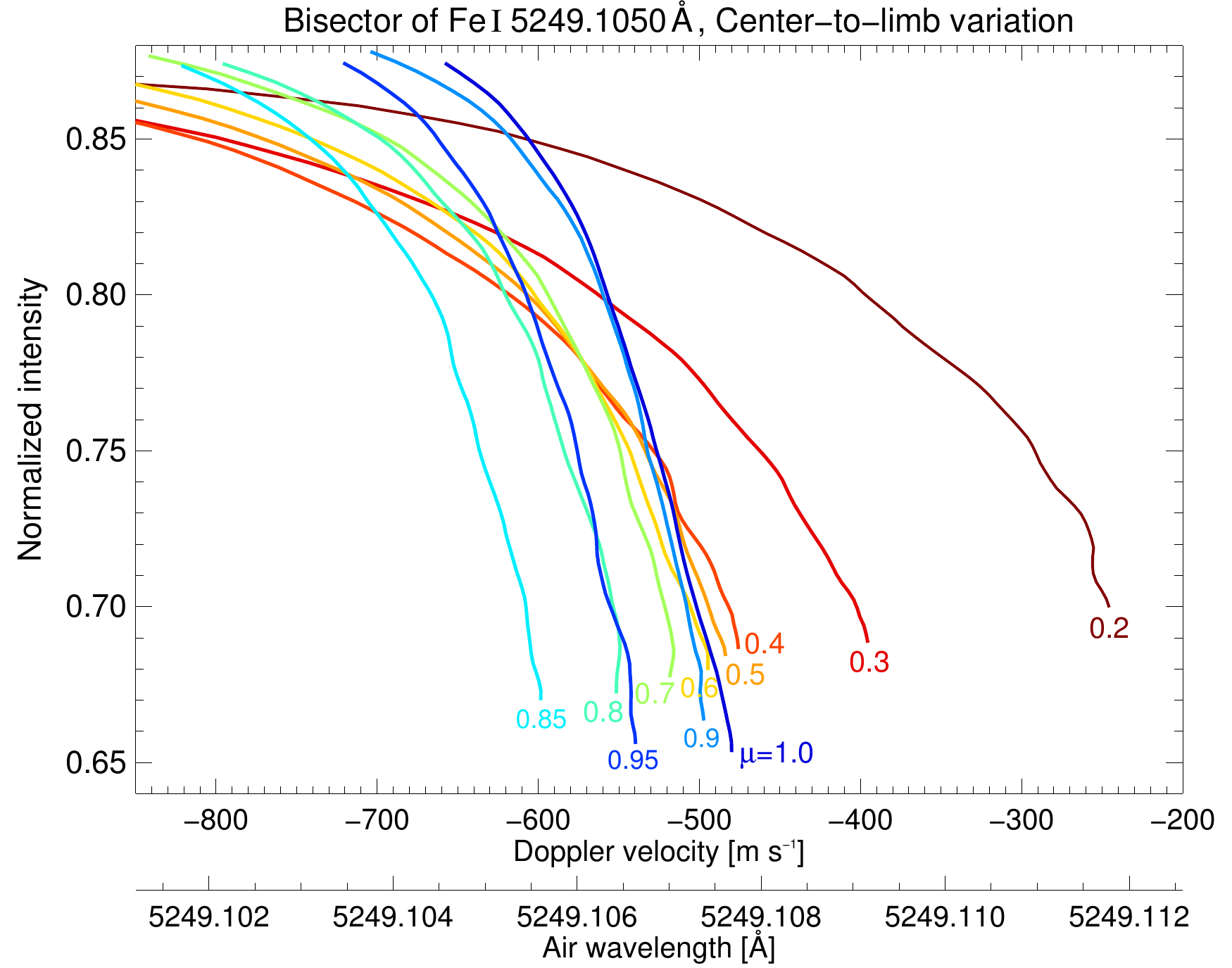}\\[0.1cm]
\textbf{b)}\\[-0.3cm]  \includegraphics[width=0.937\columnwidth]{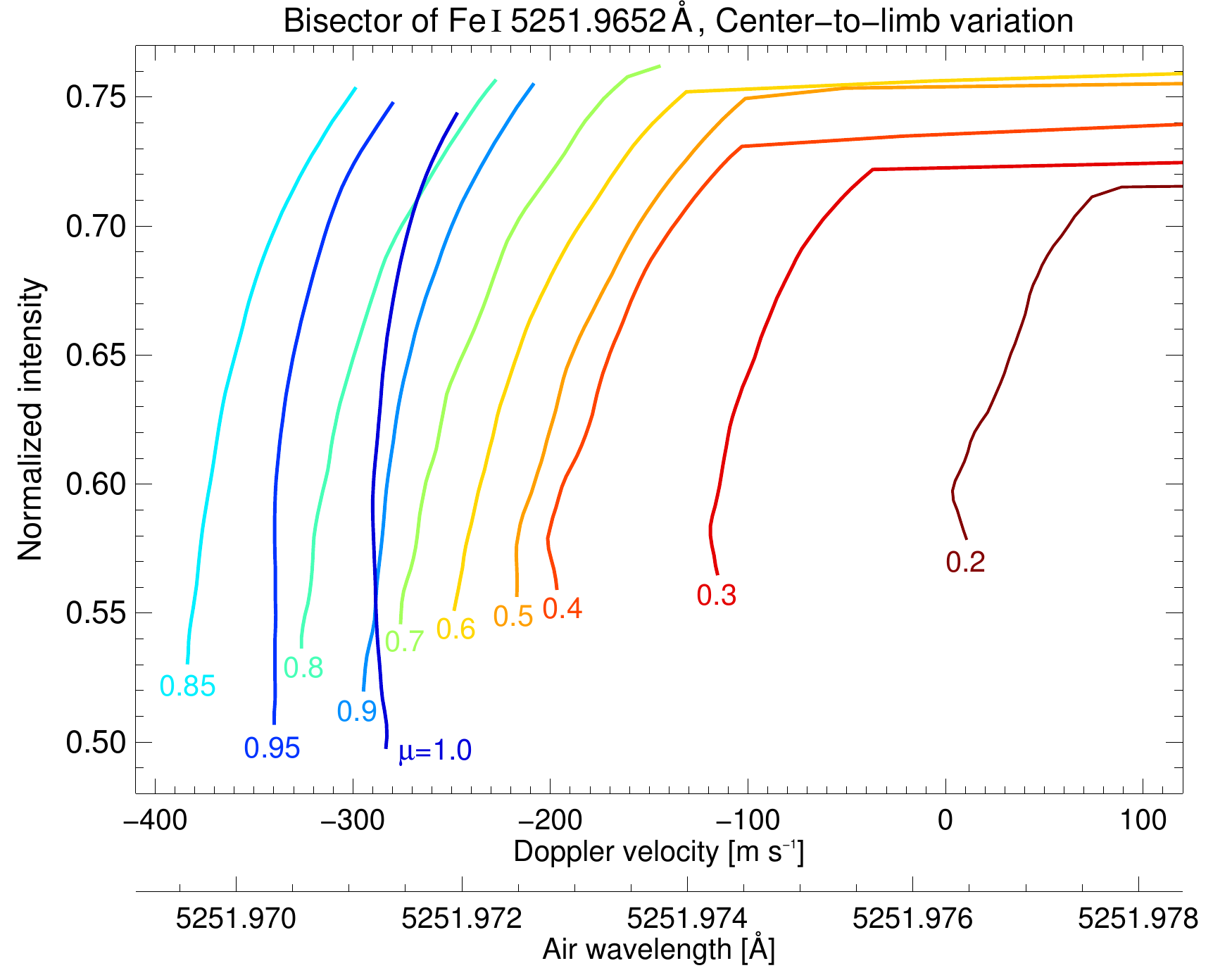}\\[0.1cm]
\textbf{c)}\\[-0.3cm]  \includegraphics[width=0.937\columnwidth]{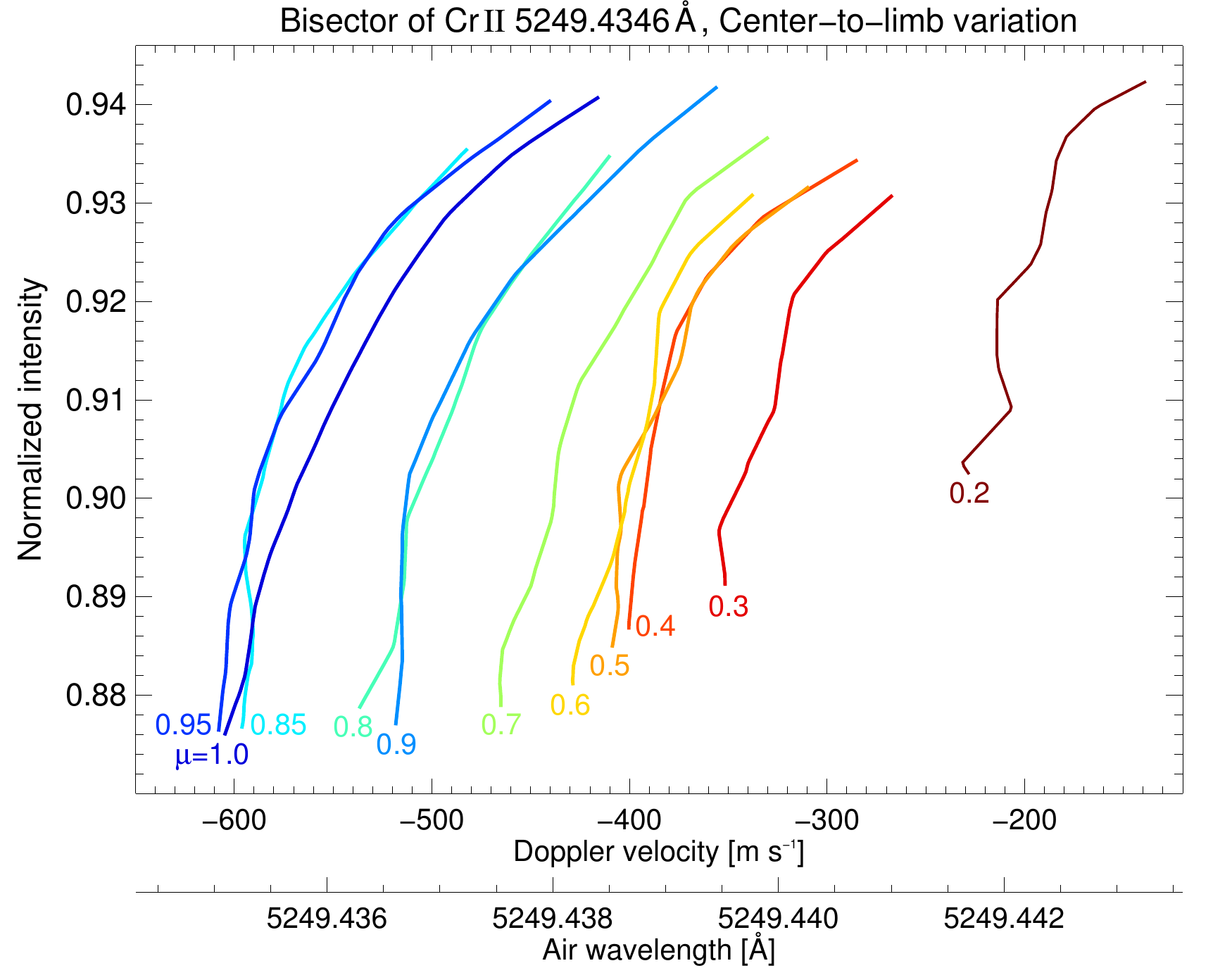}
\caption{Center-to-limb variation of the \ion{Fe}{I}\,5249.1\,\AA\ (panel a), \ion{Fe}{I}\,5251.9\,\AA\ (panel b), and \ion{Cr}{II}\,5249.4\,\AA\ (panel c) line bisector, from the solar disk center ($\mu=1.0$, blue curve) toward the limb ($\mu=0.2$, dark red curve). The normalized intensity is displayed against the absolute air wavelength and Doppler velocity. Each curve represents the average bisector for all measurements at the respective heliocentric position. \ion{Fe}{I}\,5249.1\,\AA\ exhibits a reverse bisector shape caused the blend in the blue line wing. The analysis was thus limited to the line core.}
\label{fig_A2}
\end{figure}

\clearpage
\subsection{Lines around 5381\,\AA}

\begin{figure}[htbp]
\vspace{-0.2cm}
\textbf{a)}\\[-0.3cm]  \includegraphics[width=\columnwidth]{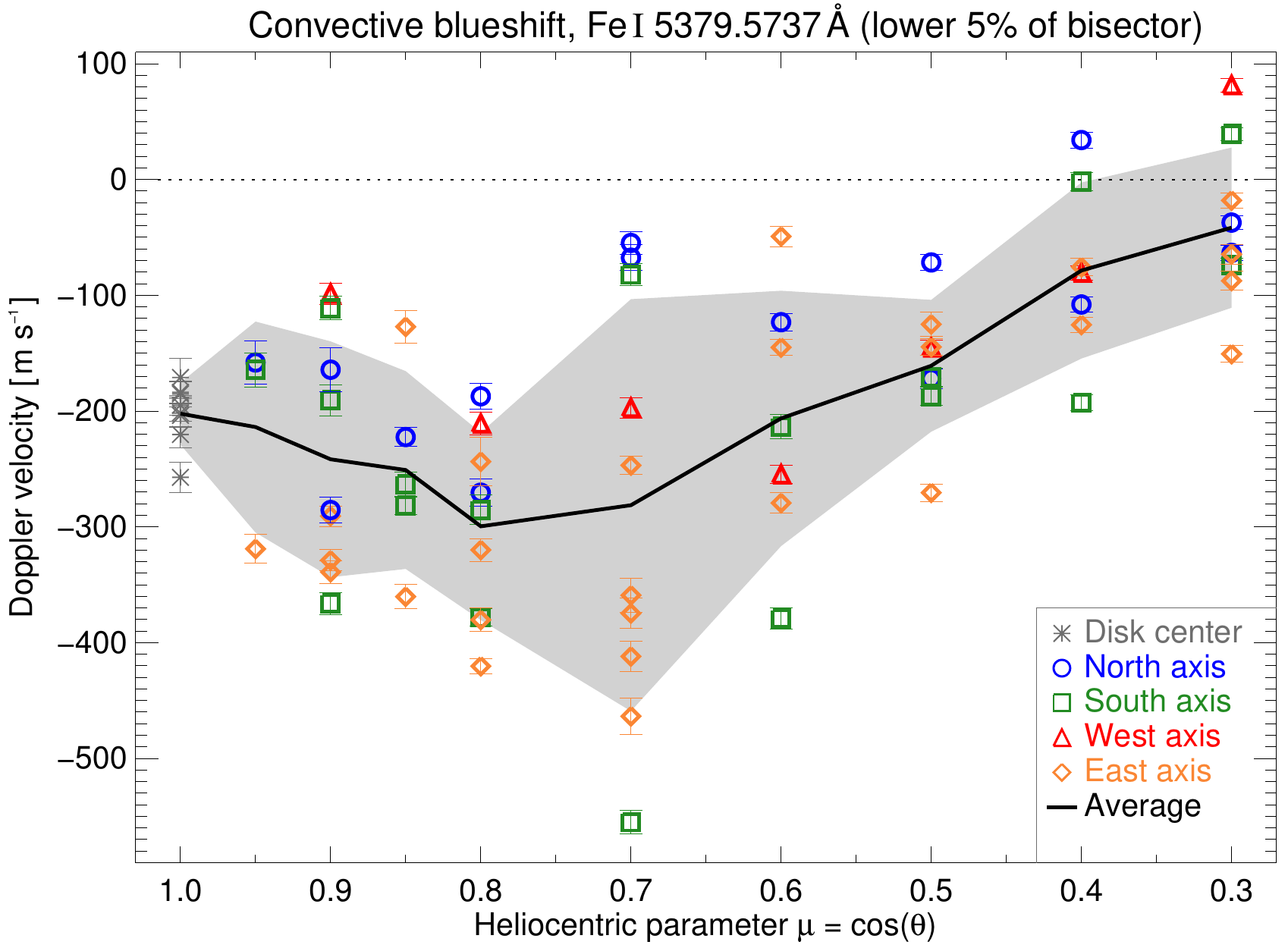}\\[0.1cm]
\textbf{b)}\\[-0.3cm]  \includegraphics[width=\columnwidth]{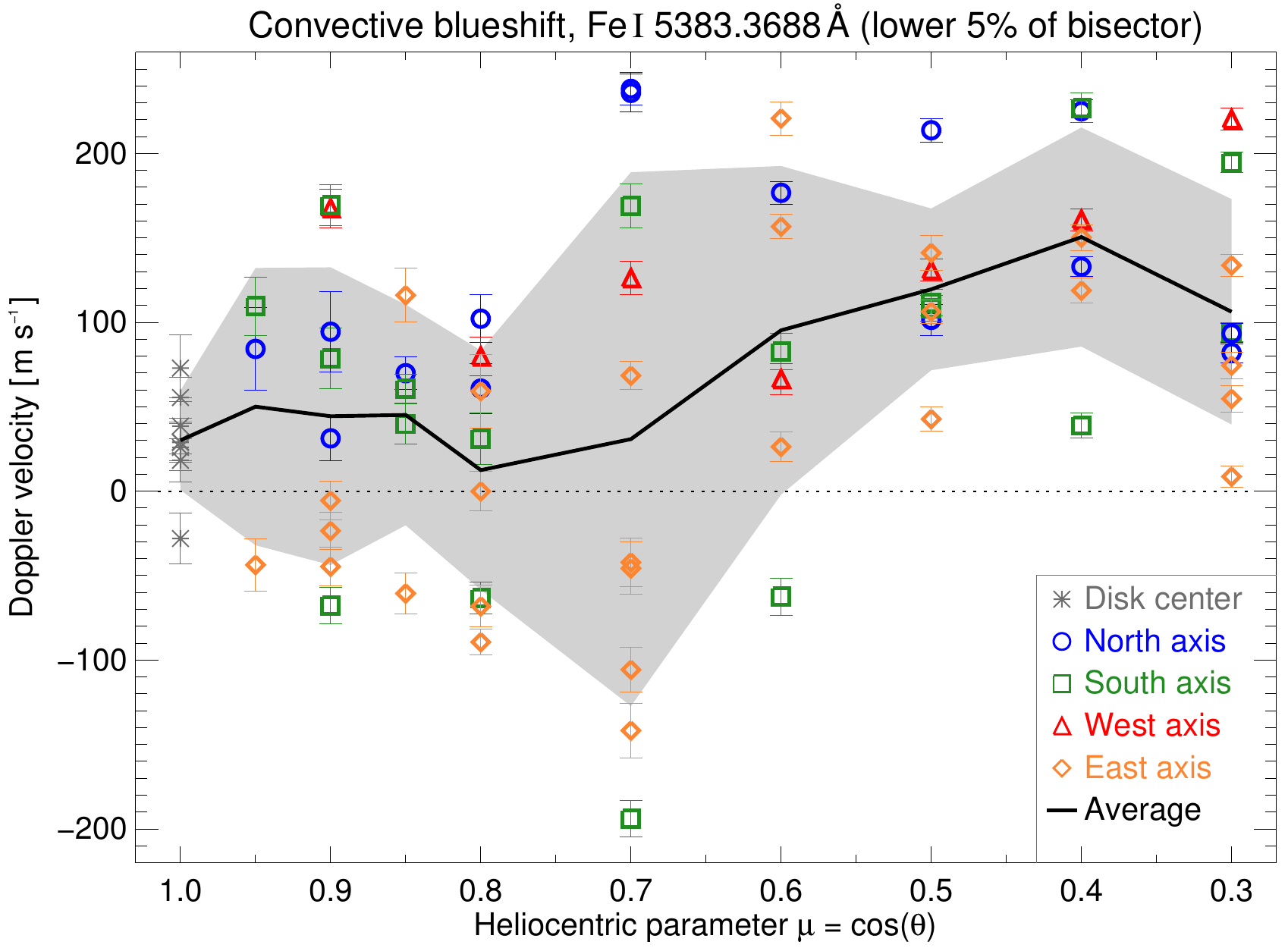}\\[0.1cm]
\textbf{c)}\\[-0.3cm]  \includegraphics[width=\columnwidth]{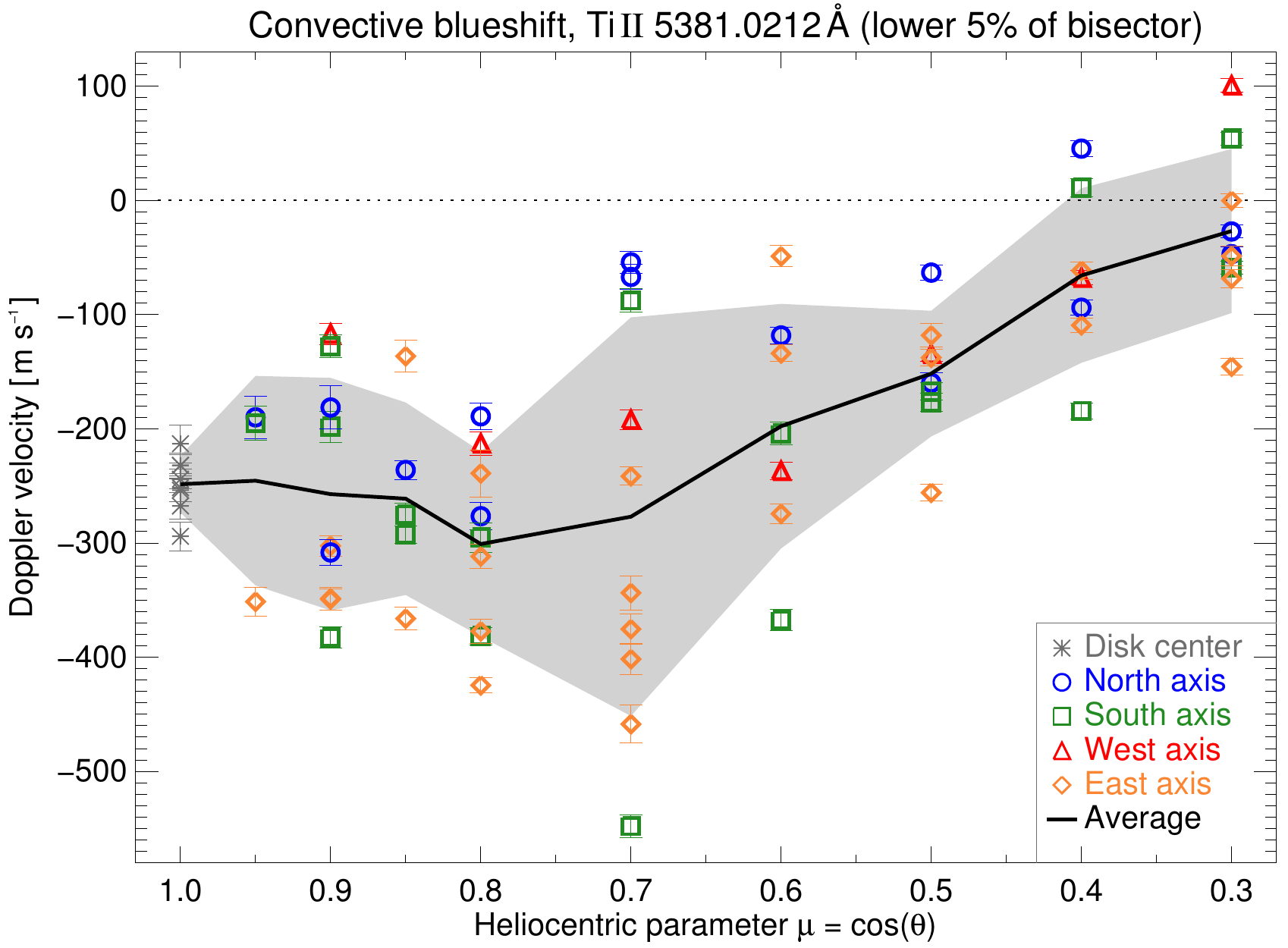}
\caption{Center-to-limb variation of the convective blueshift of the \ion{Fe}{I}\,5379.6\,\AA\ (panel a), \ion{Fe}{I}\,5383.4\,\AA\ (panel b), and \ion{Ti}{II}\,5381.0\,\AA\ (panel c) line. Each data point represents the mean Doppler velocity of the lower 5\,\% of the bisector of the temporally averaged observation sequence. Error bars indicate the mean error. Radial axes are indicated by colors and symbols. The black solid line and gray shaded area display the average center-to-limb variation and its standard deviation.}
\label{fig_A3}
\end{figure}

\begin{figure}[htbp]
\vspace{0.87cm}
\textbf{a)}\\[-0.3cm]  \includegraphics[width=0.937\columnwidth]{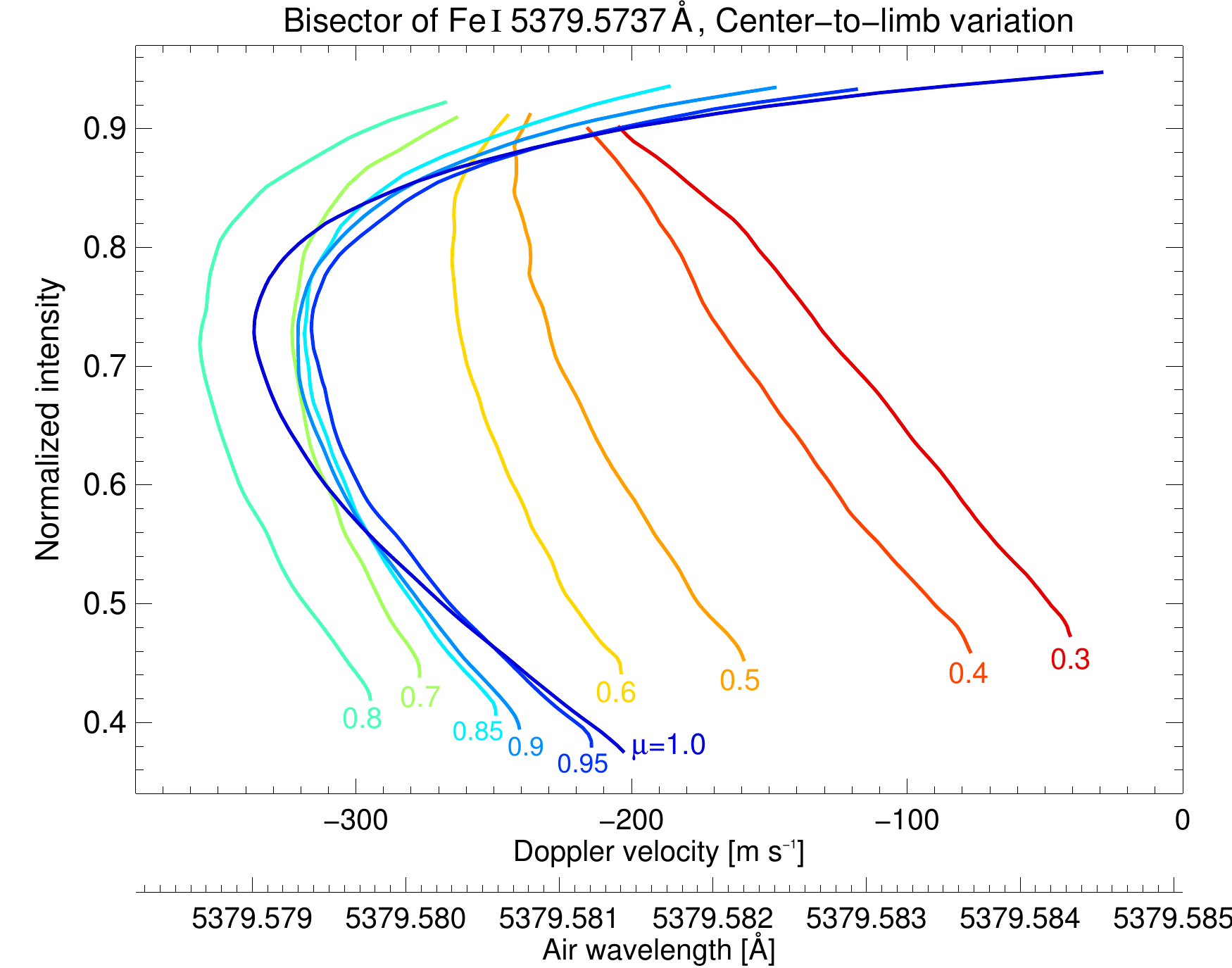}\\[0.1cm]
\textbf{b)}\\[-0.3cm]  \includegraphics[width=0.937\columnwidth]{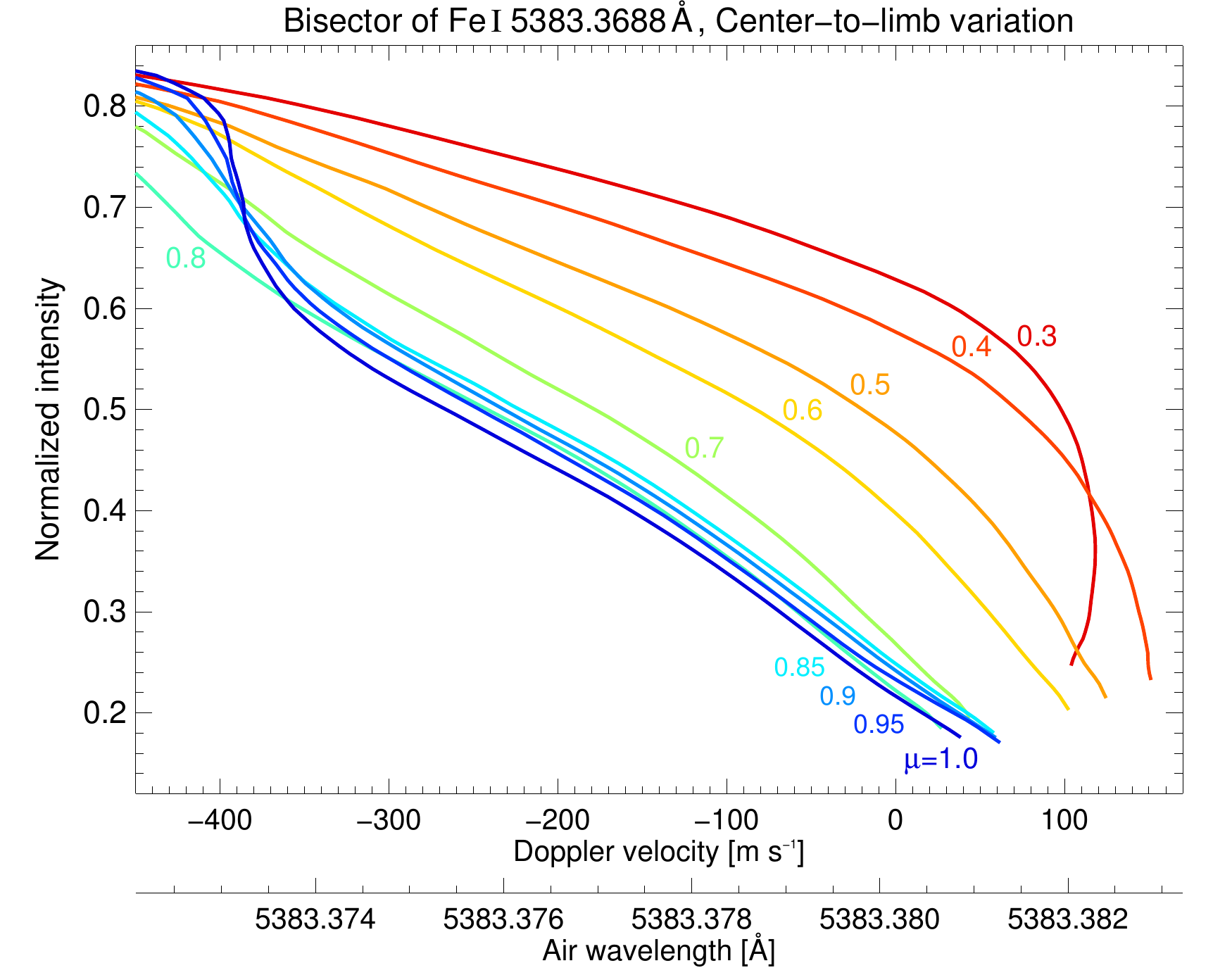}\\[0.1cm]
\textbf{c)}\\[-0.3cm]  \includegraphics[width=0.937\columnwidth]{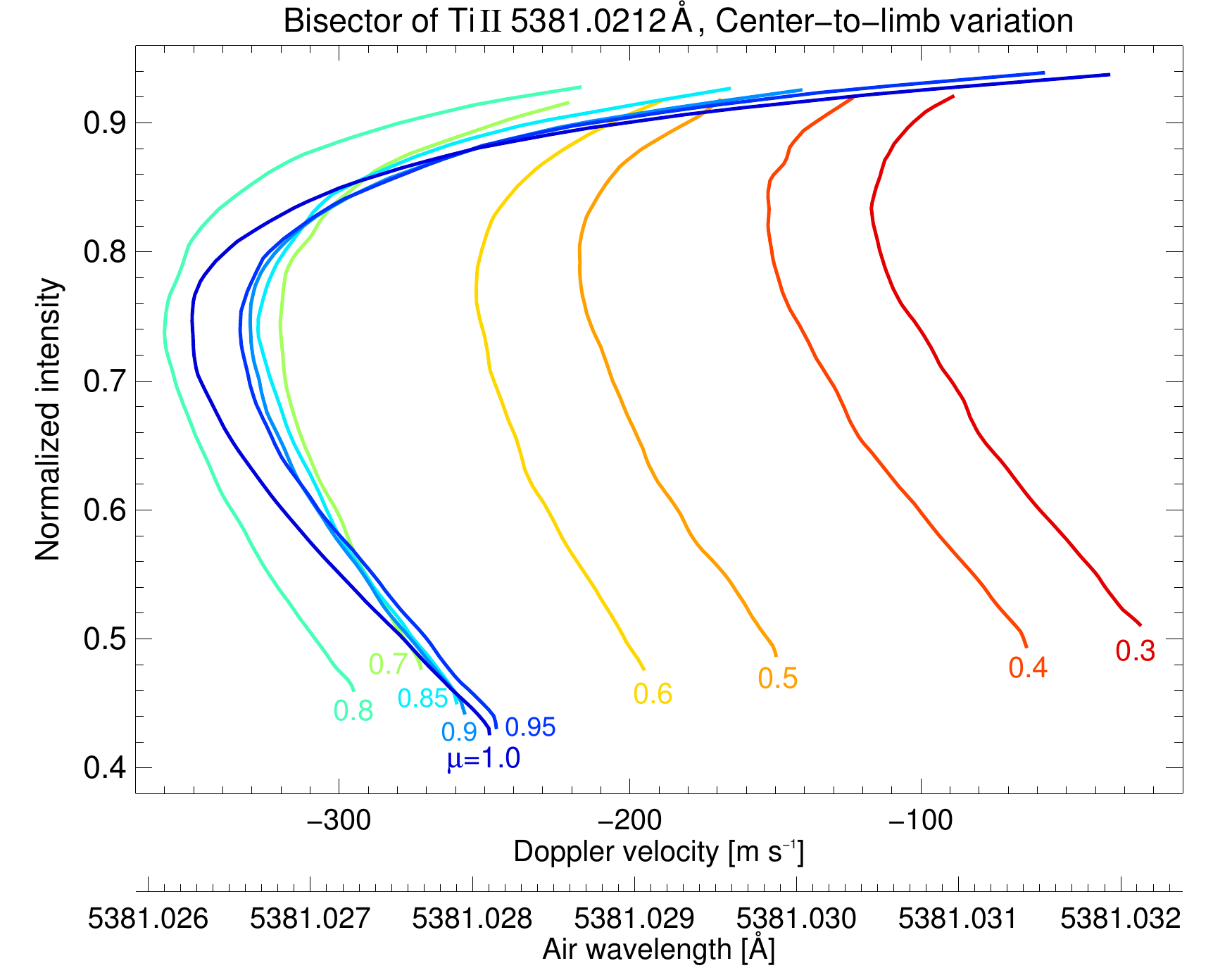}
\caption{Center-to-limb variation of the \ion{Fe}{I}\,5379.6\,\AA\ (panel a), \ion{Fe}{I}\,5383.4\,\AA\ (panel b), and \ion{Ti}{II}\,5381.0\,\AA\ (panel c) line bisector, from the solar disk center ($\mu=1.0$, blue curve) toward the limb ($\mu=0.3$, red curve). The normalized intensity is displayed against the absolute air wavelength and Doppler velocity. Each curve represents the average bisector for all measurements at the respective heliocentric position. The strong blueshift of the \ion{Fe}{I}\,5383.4\,\AA\ bisector toward the continuum is caused by blends in the blue line wing. The analysis was thus limited to lower half of the line.}
\label{fig_A4}
\end{figure}

\clearpage
\subsection{Lines around 5434\,\AA, 5576\,\AA, and 6149\,\AA}

\begin{figure}[htbp]
\vspace{-0.2cm}
\textbf{a)}\\[-0.3cm]  \includegraphics[width=\columnwidth]{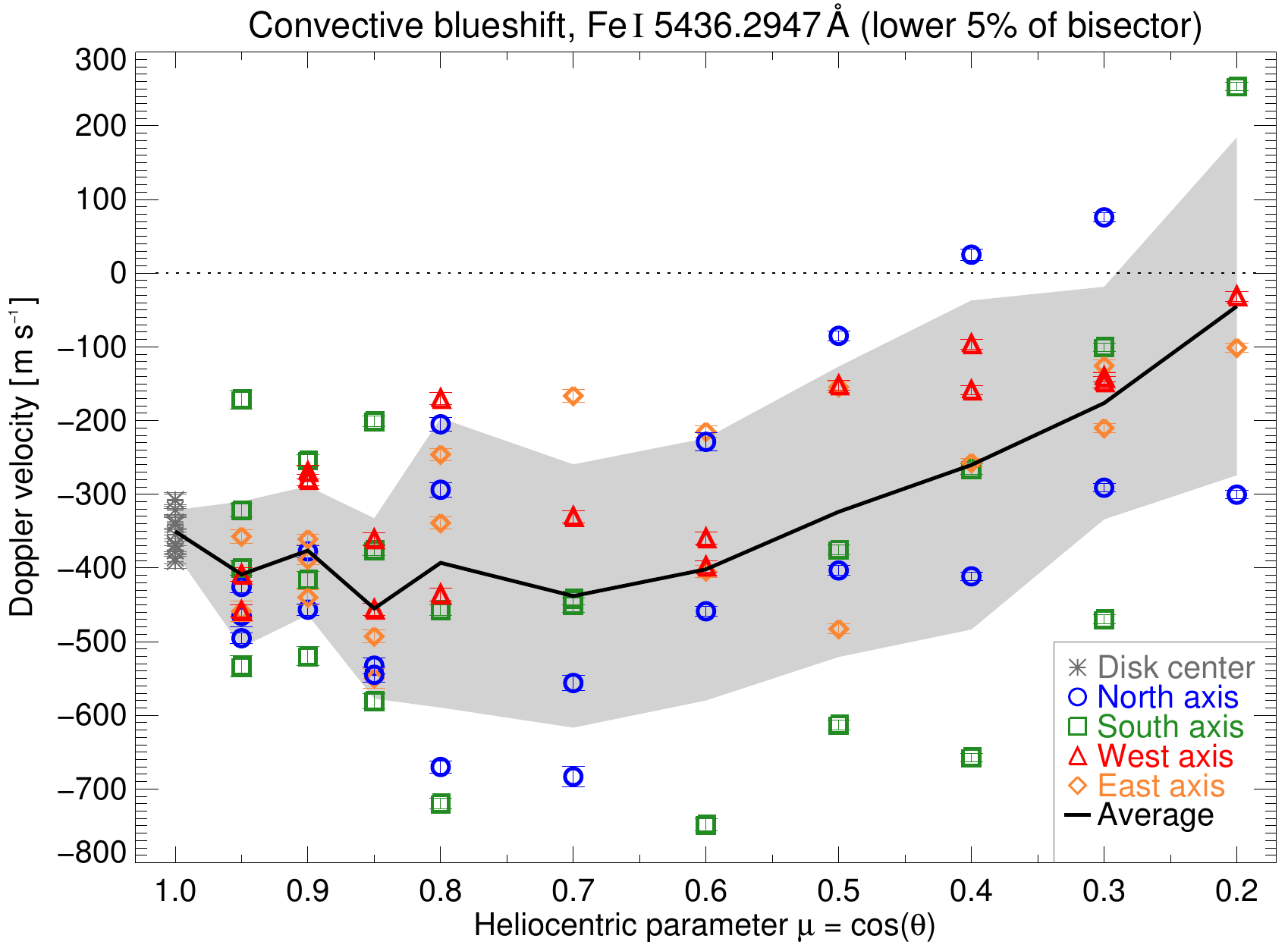}\\[0.1cm]
\textbf{b)}\\[-0.3cm]  \includegraphics[width=\columnwidth]{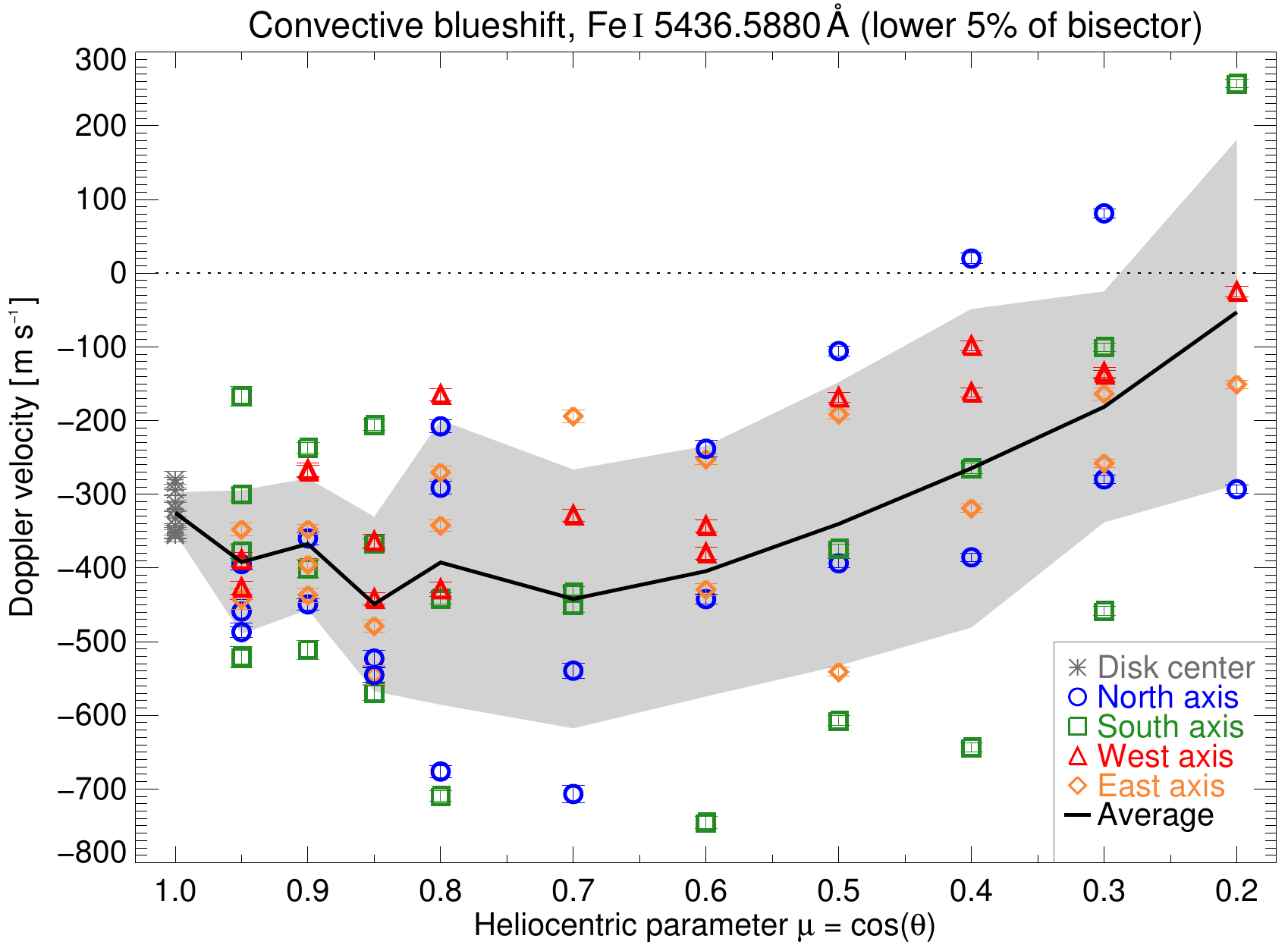}\\[0.1cm]
\textbf{c)}\\[-0.3cm]  \includegraphics[width=\columnwidth]{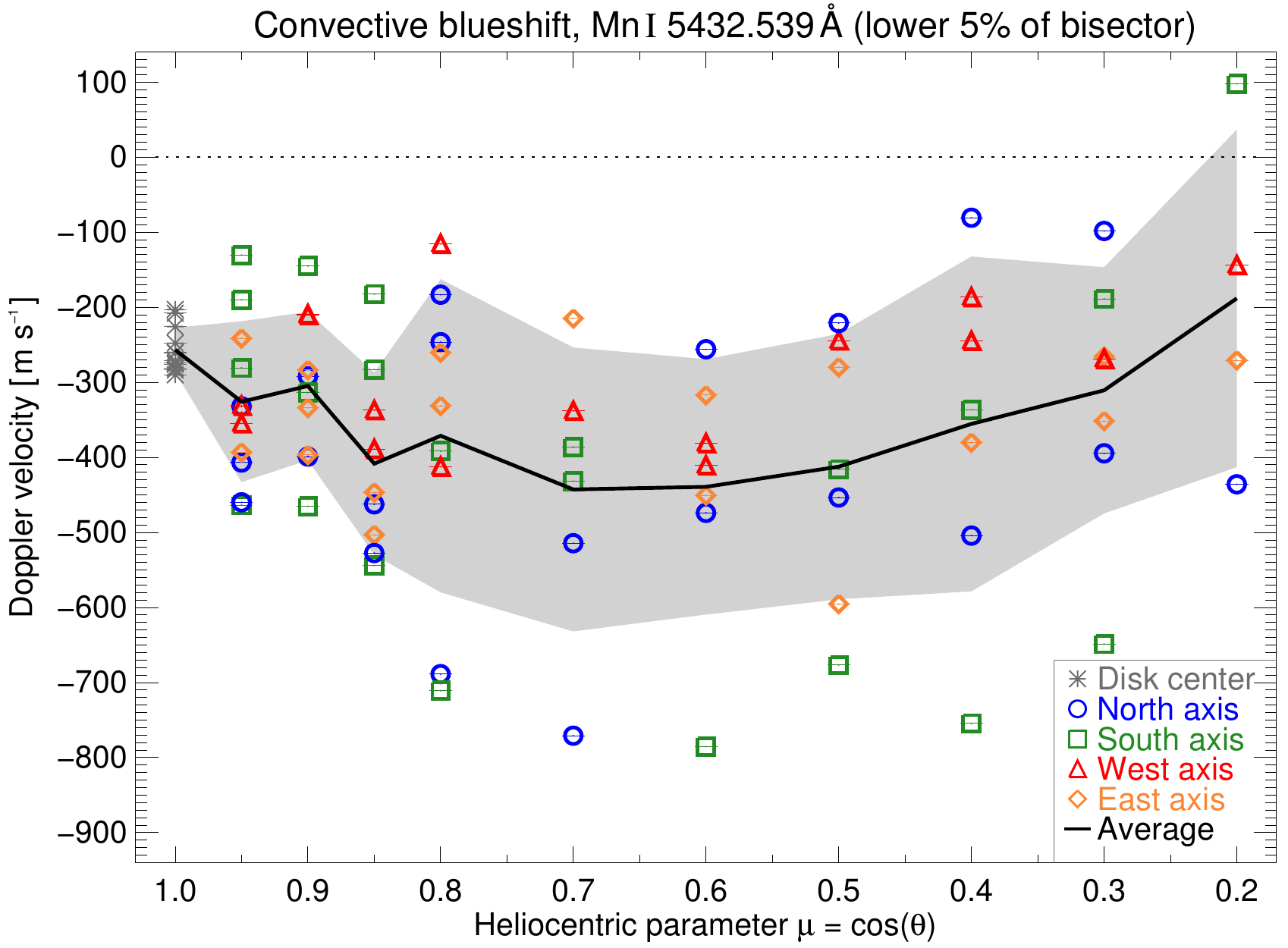}
\caption{Center-to-limb variation of the convective blueshift of the \ion{Fe}{I}\,5436.3\,\AA\ (panel a), \ion{Fe}{I}\,5436.6\,\AA\ (panel b), and \ion{Mn}{I}\,5432.6\,\AA\ (panel c) line. Each data point represents the mean Doppler velocity of the lower 5\,\% of the bisector of the temporally averaged observation sequence. Error bars indicate the mean error. Radial axes are indicated by colors and symbols. The black solid line and gray shaded area display the average center-to-limb variation and its standard deviation.}
\label{fig_A5}
\end{figure}

\begin{figure}[htbp]
\vspace{0.87cm}
\textbf{a)}\\[-0.3cm]  \includegraphics[width=0.937\columnwidth]{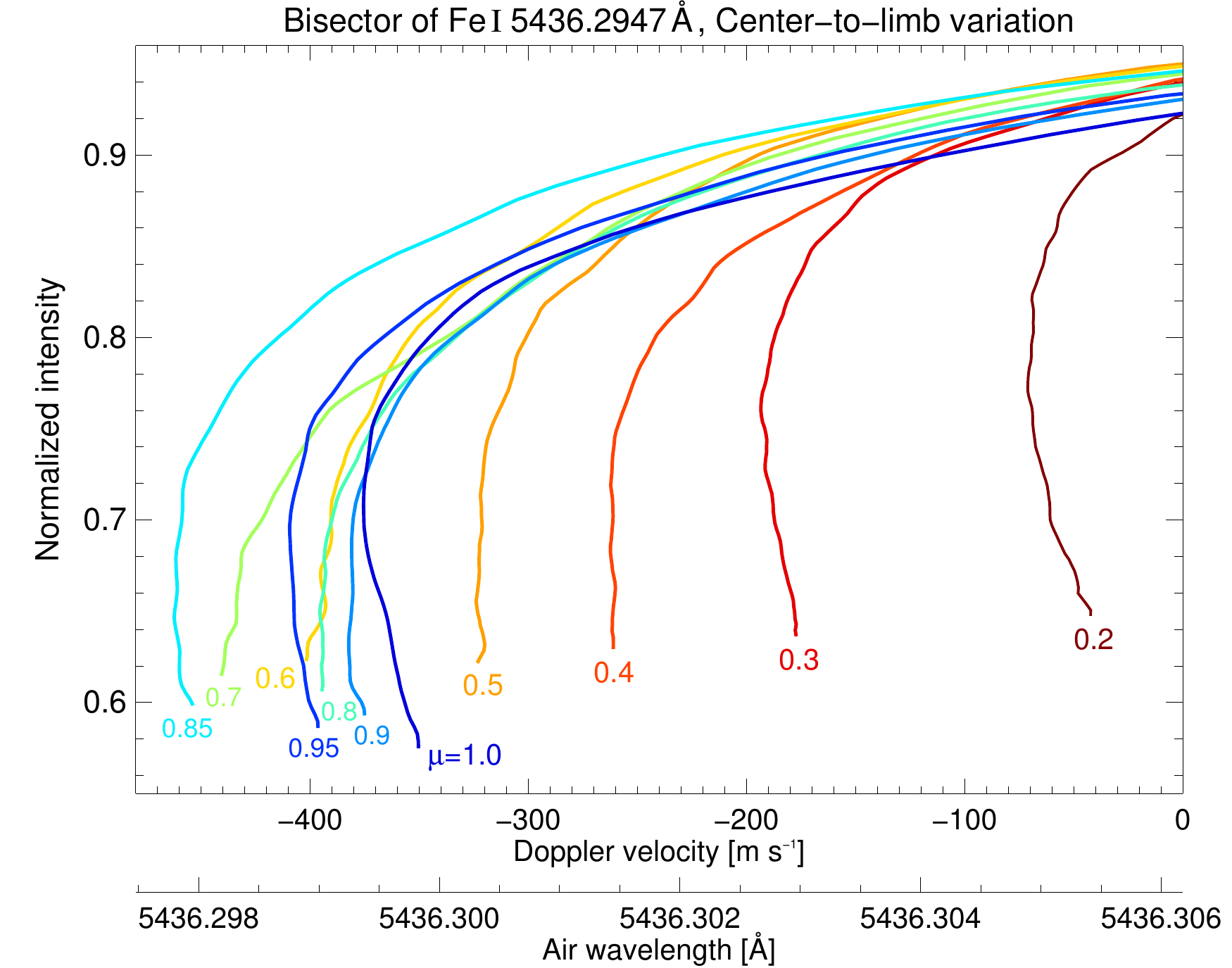}\\[0.1cm]
\textbf{b)}\\[-0.3cm]  \includegraphics[width=0.937\columnwidth]{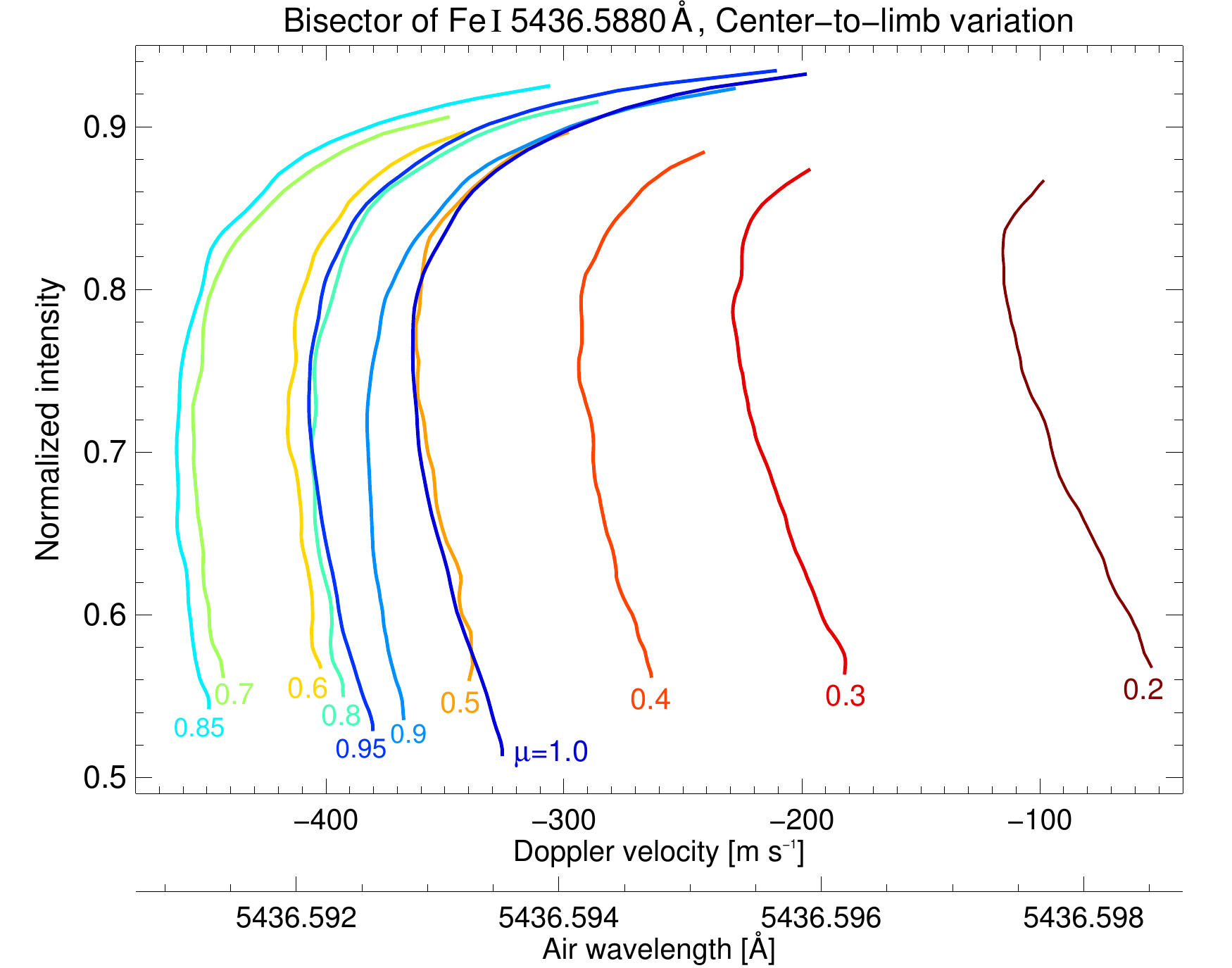}\\[0.1cm]
\textbf{c)}\\[-0.3cm]  \includegraphics[width=0.937\columnwidth]{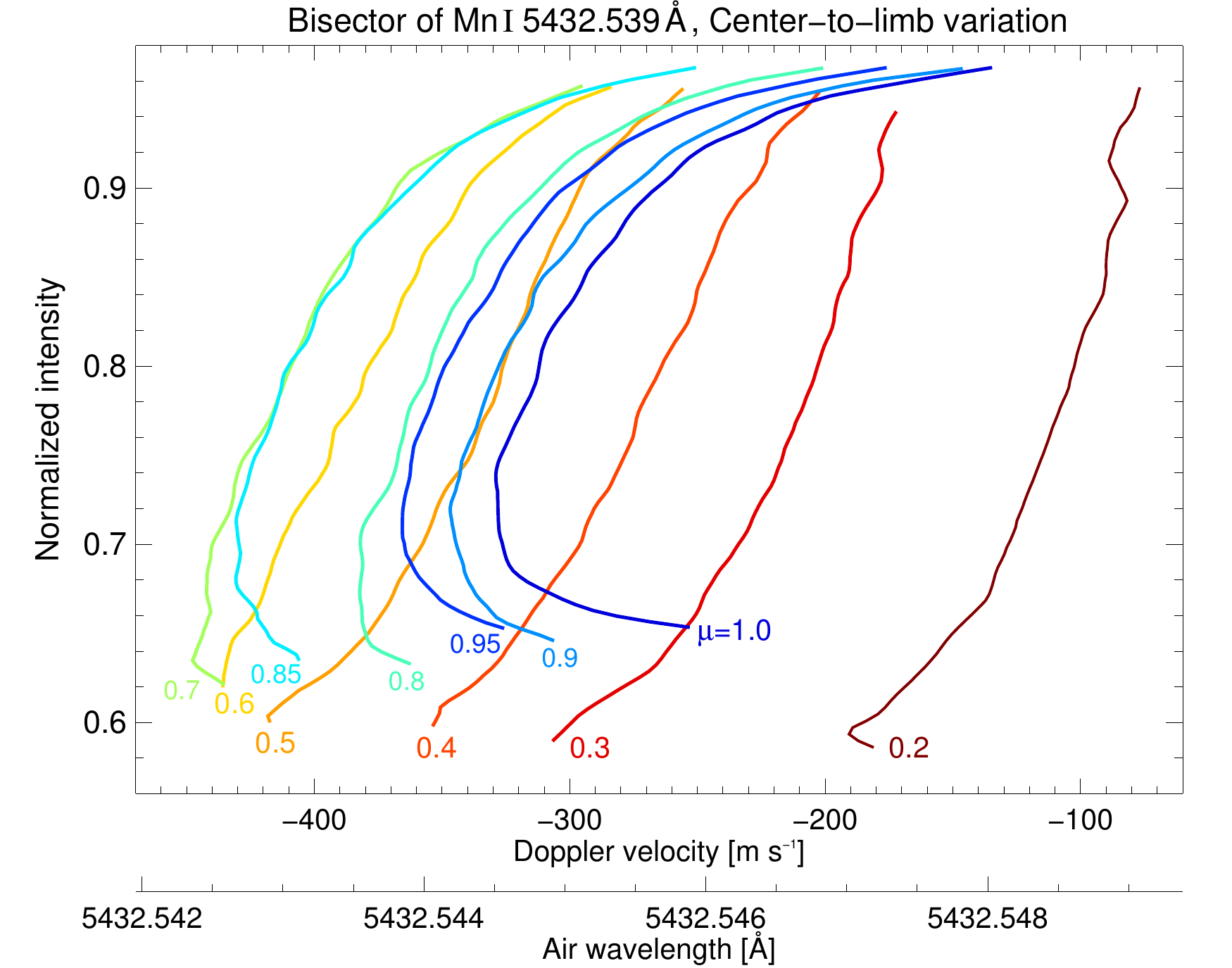}
\caption{Center-to-limb variation of the \ion{Fe}{I}\,5436.3\,\AA\ (panel a), \ion{Fe}{I}\,5436.6\,\AA\ (panel b), and \ion{Mn}{I}\,5432.6\,\AA\ (panel c) line bisector, from the solar disk center ($\mu=1.0$, blue curve) toward the limb ($\mu=0.2$, dark red curve). The normalized intensity is displayed against the absolute air wavelength and Doppler velocity. Each curve represents the average bisector for all measurements at the respective heliocentric position.}
\label{fig_A6}
\end{figure}

\clearpage

\begin{figure}[htbp]
\vspace{0.0cm}
\textbf{a)}\\[-0.3cm]  \includegraphics[width=\columnwidth]{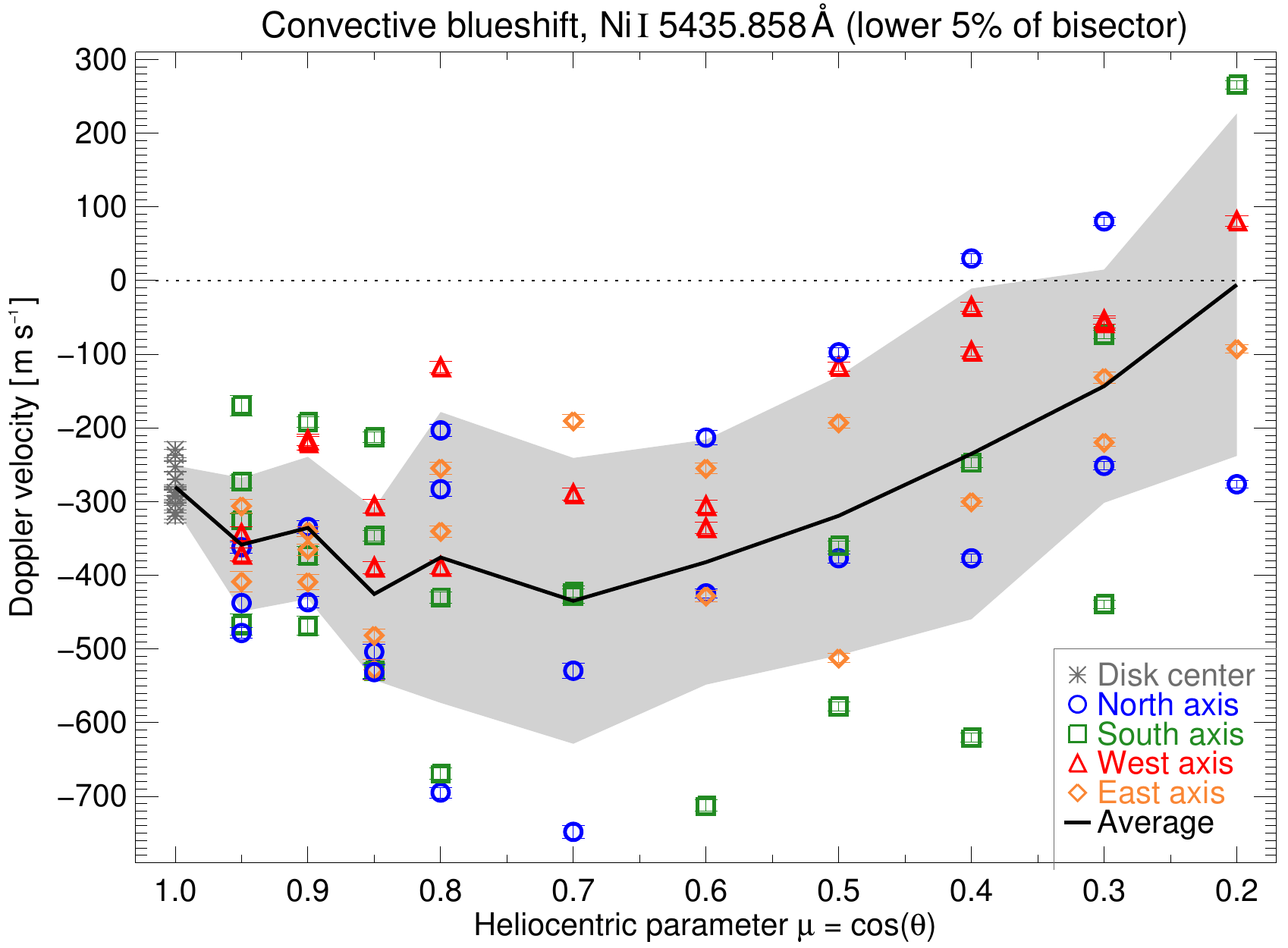}\\[0.1cm]
\textbf{b)}\\[-0.3cm]  \includegraphics[width=\columnwidth]{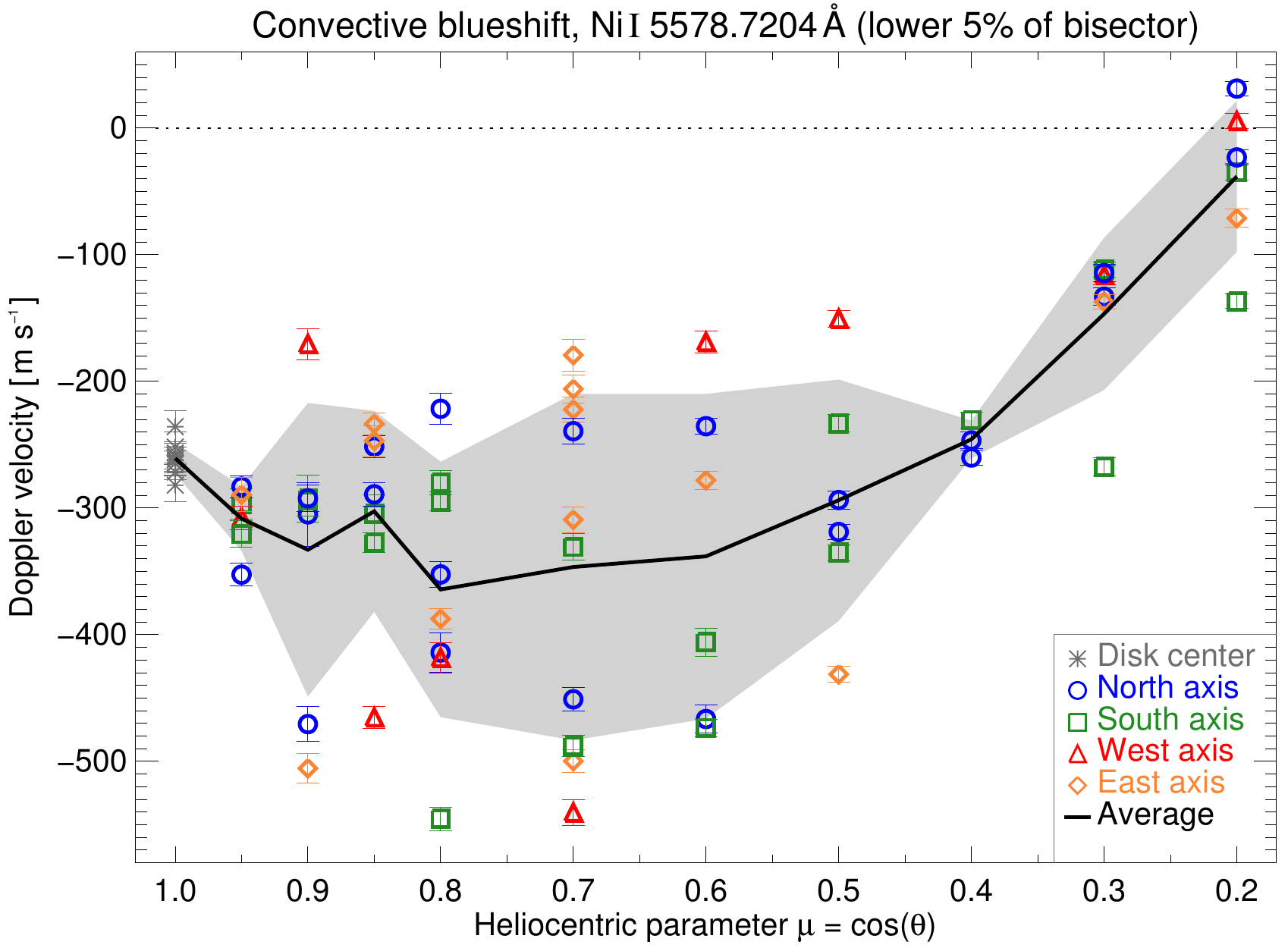}\\[0.1cm]
\textbf{c)}\\[-0.3cm]  \includegraphics[width=\columnwidth]{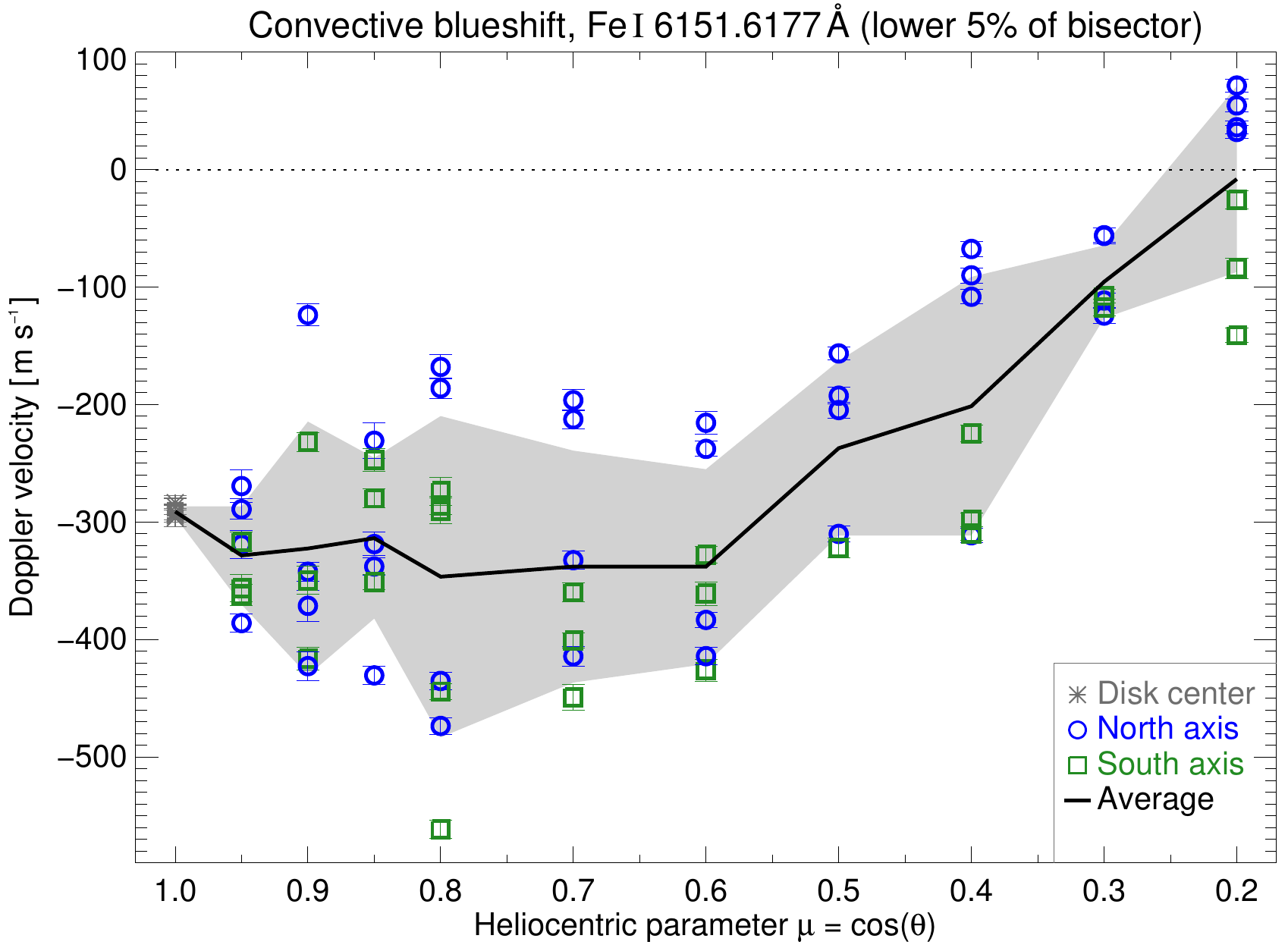}
\caption{Center-to-limb variation of the convective blueshift of the \ion{Ni}{I}\,5435.9\,\AA\ (panel a), \ion{Ni}{I}\,5578.7\,\AA\ (panel b), and \ion{Fe}{I}\,6151.6\,\AA\ (panel c) line. Each data point represents the mean Doppler velocity of the lower 5\,\% of the bisector of the temporally averaged observation sequence. Error bars indicate the mean error. Radial axes are indicated by colors and symbols. The black solid line and gray shaded area display the average center-to-limb variation and its standard deviation.}
\label{fig_A7}
\end{figure}

\begin{figure}[htbp]
\vspace{0.0cm}
\textbf{a)}\\[-0.3cm]  \includegraphics[width=0.937\columnwidth]{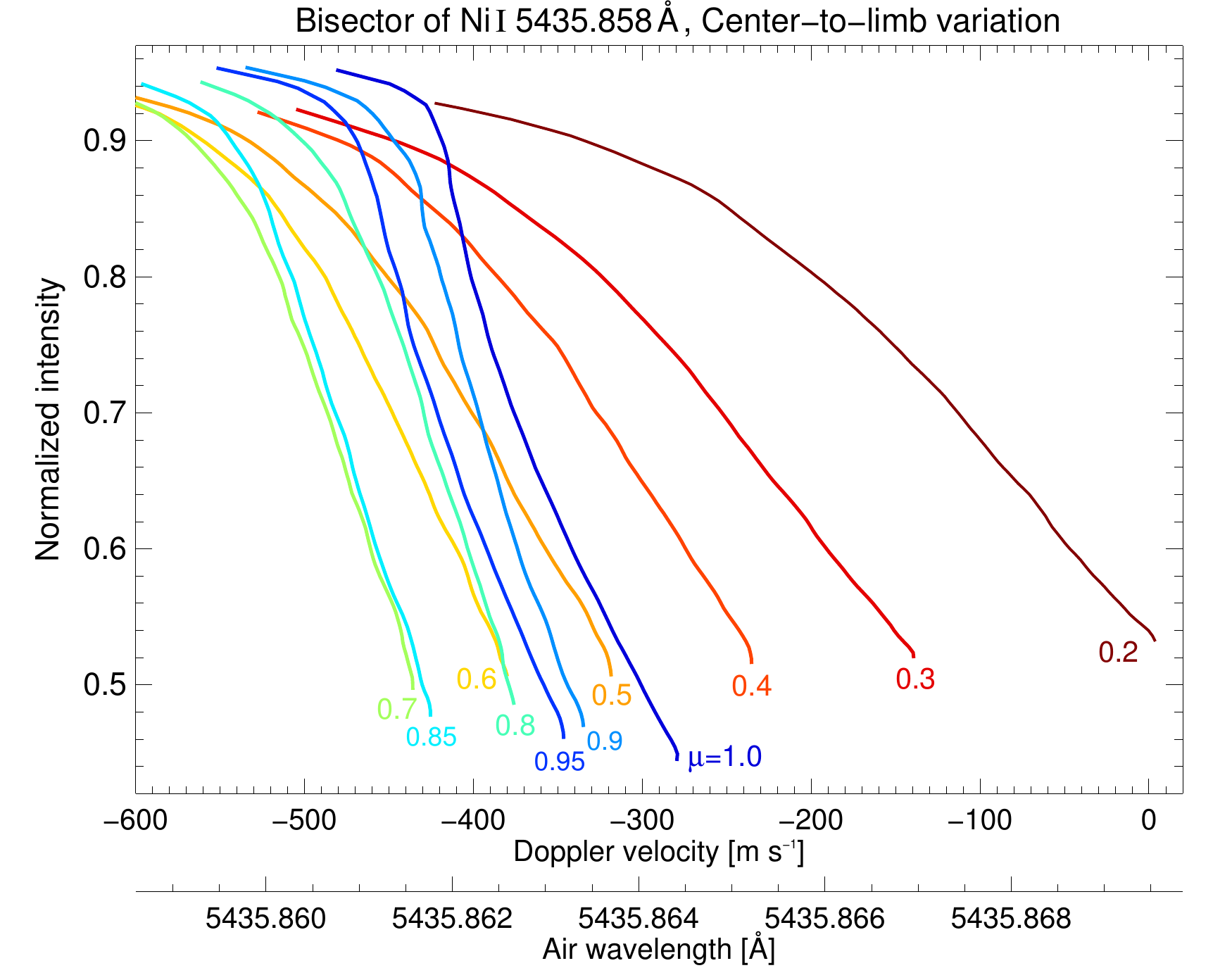}\\[0.1cm]
\textbf{b)}\\[-0.3cm]  \includegraphics[width=0.937\columnwidth]{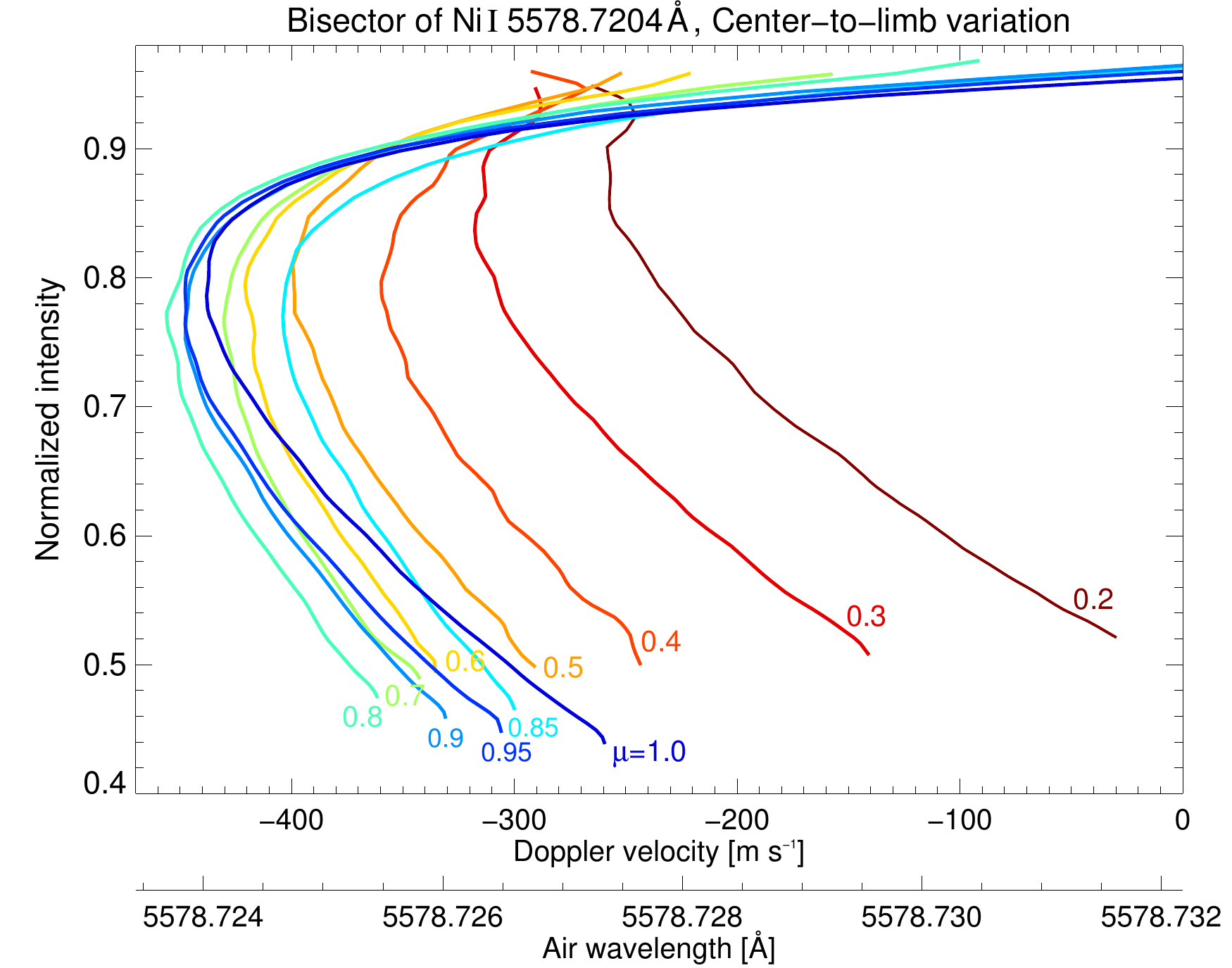}\\[0.1cm]
\textbf{c)}\\[-0.3cm]  \includegraphics[width=0.937\columnwidth]{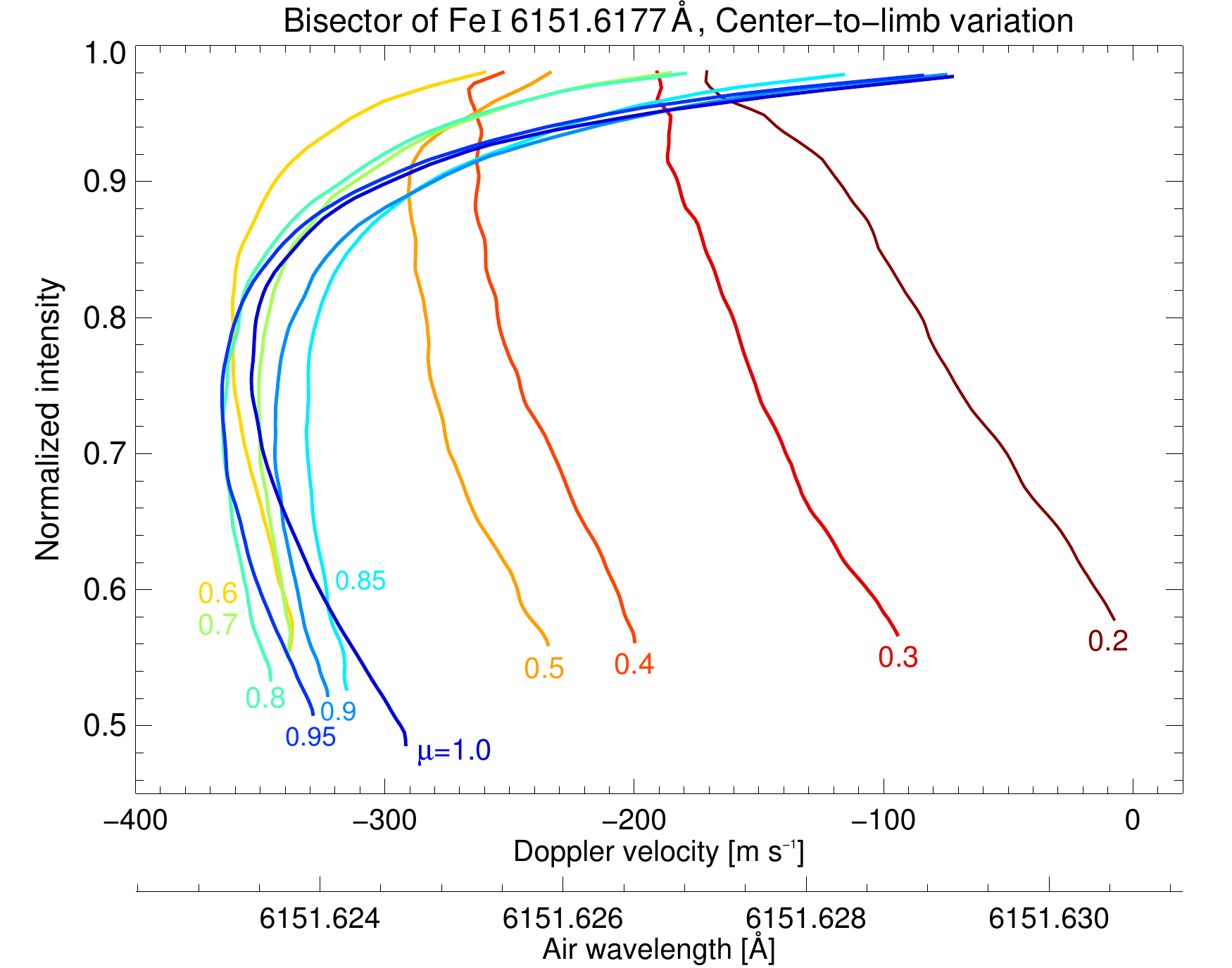}
\caption{Center-to-limb variation of the \ion{Ni}{I}\,5435.9\,\AA\ (panel a), \ion{Ni}{I}\,5578.7\,\AA\ (panel b), and \ion{Fe}{I}\,6151.6\,\AA\ (panel c) line bisector, from the solar disk center ($\mu=1.0$, blue curve) toward the limb ($\mu=0.2$, dark red curve). The normalized intensity is displayed against the absolute air wavelength and Doppler velocity. Each curve represents the average bisector for all measurements at the respective heliocentric position. The strong blueshift of the \ion{Ni}{I}\,5435.9\,\AA\ bisector toward the continuum is caused by blends in the blue line wing. The analysis was thus limited to the line core.}
\label{fig_A8}
\end{figure}

\clearpage
\subsection{Lines around 6173\,\AA}

\begin{figure}[htbp]
\vspace{-0.2cm}
\textbf{a)}\\[-0.3cm]  \includegraphics[width=\columnwidth]{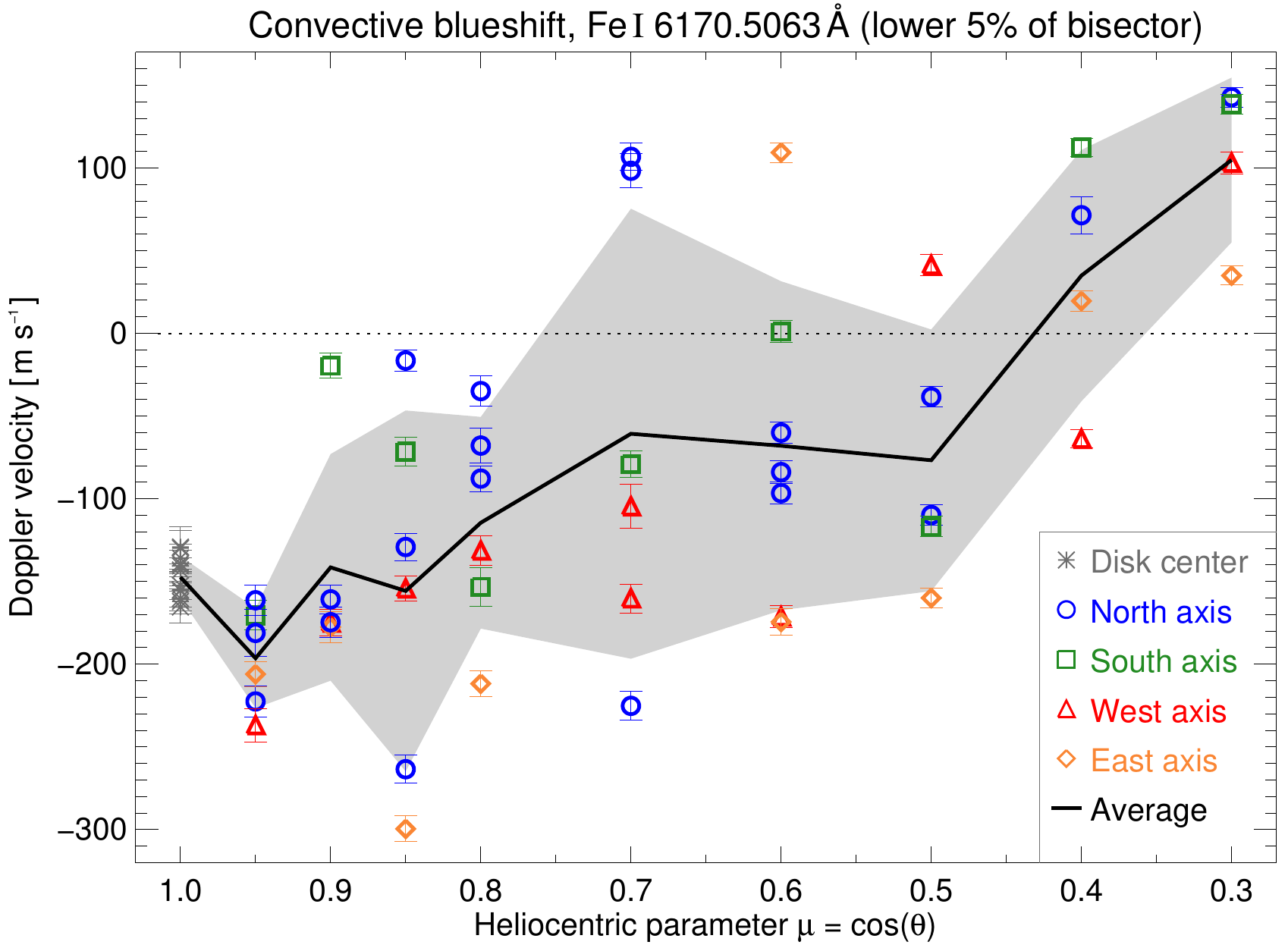}\\[0.1cm]
\textbf{b)}\\[-0.3cm]  \includegraphics[width=\columnwidth]{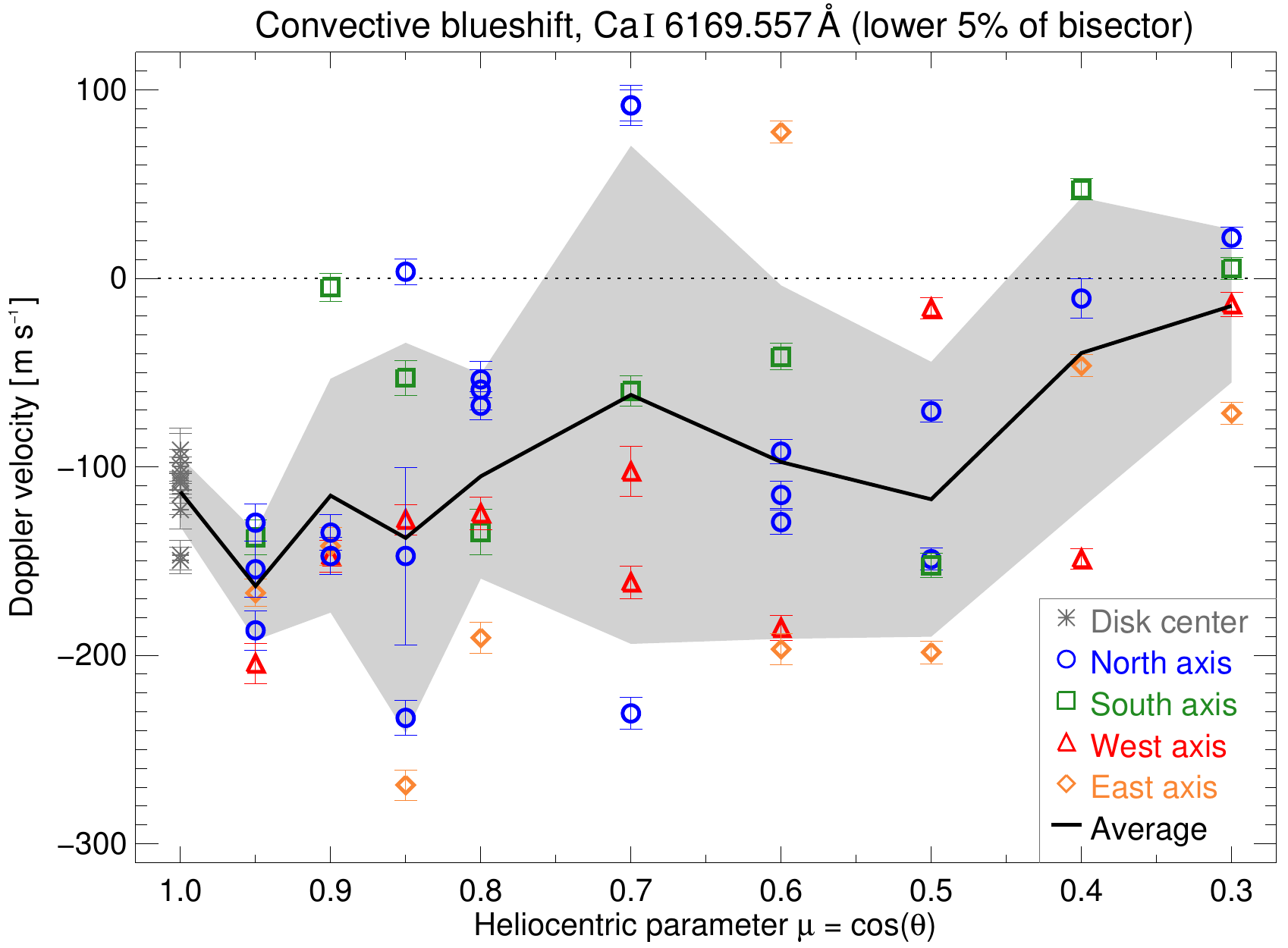}\\[0.1cm]
\textbf{c)}\\[-0.3cm]  \includegraphics[width=\columnwidth]{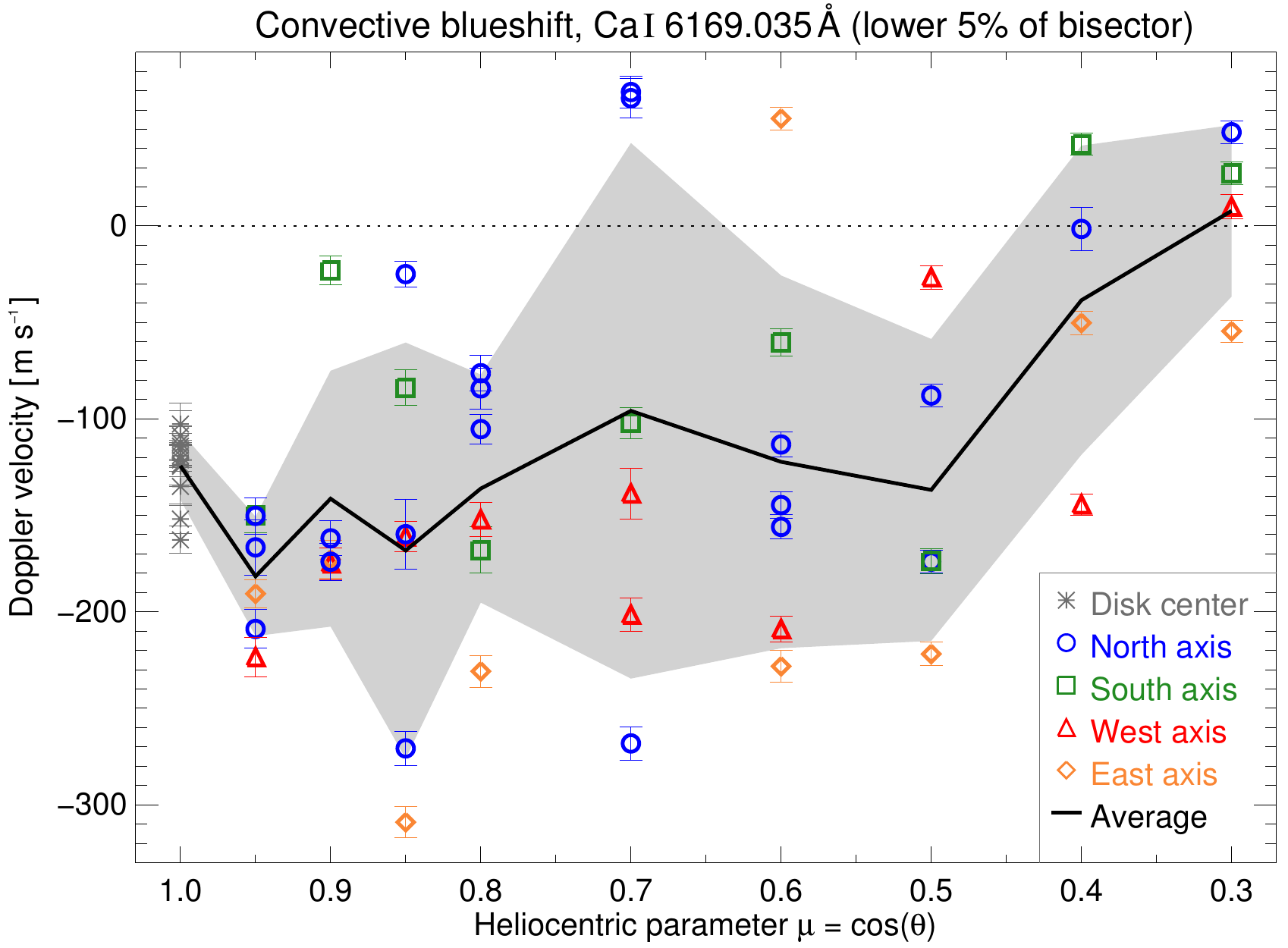}
\caption{Center-to-limb variation of the convective blueshift of the \ion{Fe}{I}\,6170.5\,\AA\ (panel a), \ion{Ca}{I}\,6169.6\,\AA\ (panel b), and \ion{Ca}{I}\,6169.0\,\AA\ (panel c) line. Each data point represents the mean Doppler velocity of the lower 5\,\% of the bisector of the temporally averaged observation sequence. Error bars indicate the mean error. Radial axes are indicated by colors and symbols. The black solid line and gray shaded area display the average center-to-limb variation and its standard deviation.}
\label{fig_A9}
\end{figure}

\begin{figure}[htbp]
\vspace{0.87cm}
\textbf{a)}\\[-0.3cm]  \includegraphics[width=0.937\columnwidth]{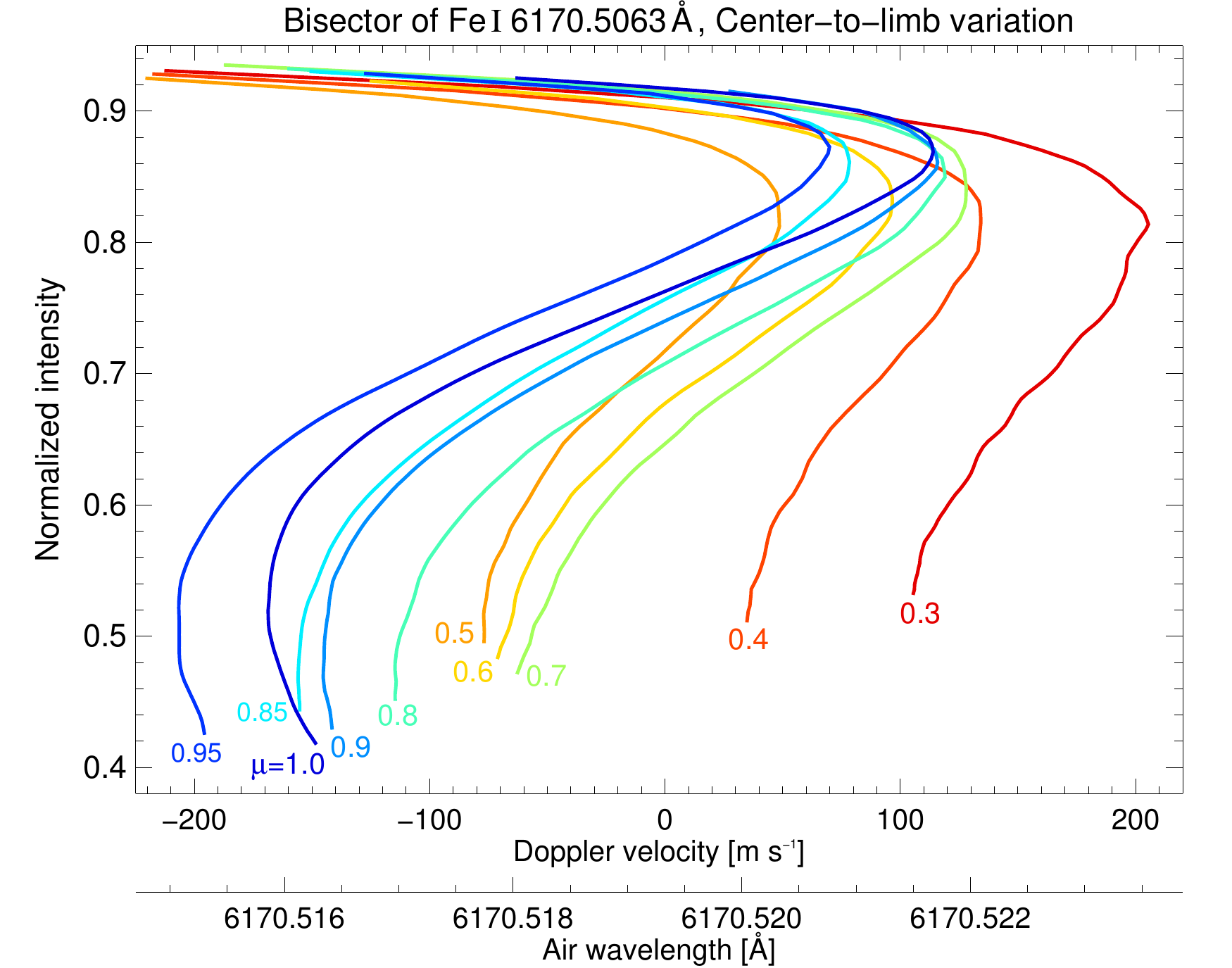}\\[0.1cm]
\textbf{b)}\\[-0.3cm]  \includegraphics[width=0.937\columnwidth]{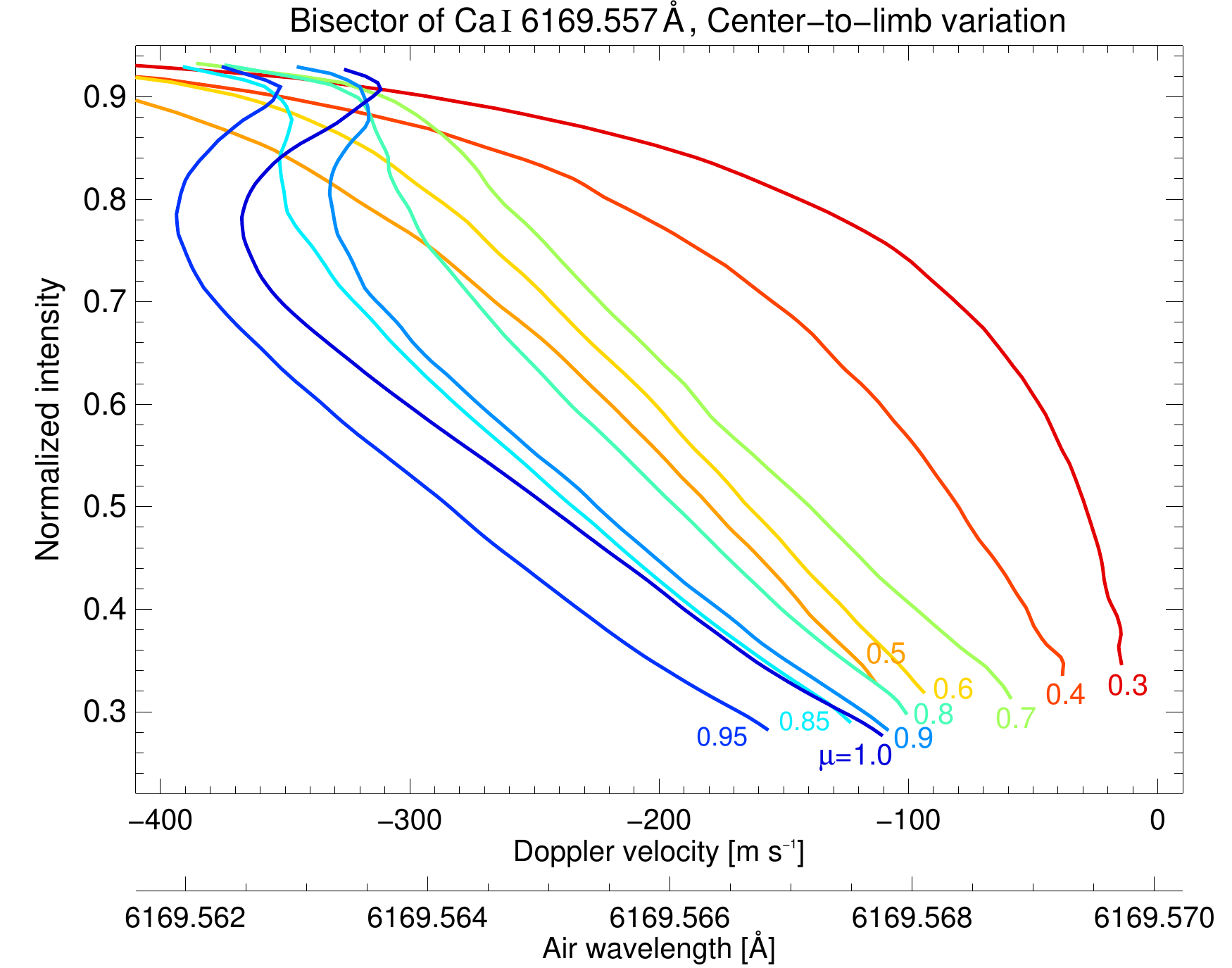}\\[0.1cm]
\textbf{c)}\\[-0.3cm]  \includegraphics[width=0.937\columnwidth]{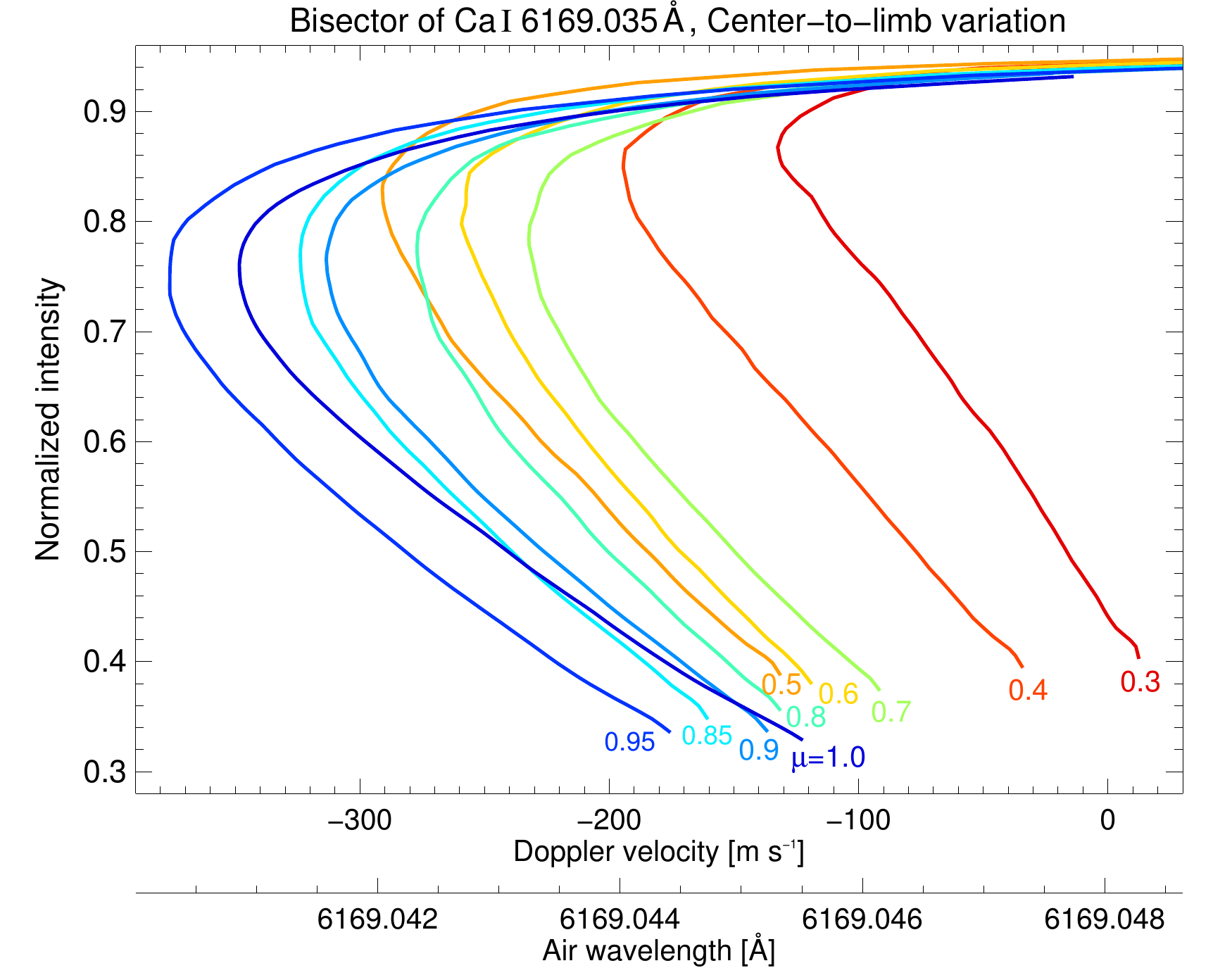}
\caption{Center-to-limb variation of the \ion{Fe}{I}\,6170.5\,\AA\ (panel a), \ion{Ca}{I}\,6169.6\,\AA\ (panel b), and \ion{Ca}{I}\,6169.0\,\AA\ (panel c) line bisector, from the solar disk center ($\mu=1.0$, blue curve) toward the limb ($\mu=0.3$, red curve). The normalized intensity is displayed against the absolute air wavelength and Doppler velocity. Each curve represents the average bisector for all measurements at the respective heliocentric position.\ion{Fe}{I}\,6170.5\,\AA\ exhibits a reverse bisector shape toward the continuum is caused by a blend in the blue line wing. The analysis was thus limited to the lower half of the line.}
\label{fig_A10}
\end{figure}

\clearpage
\subsection{Lines around 6302\,\AA}

\begin{figure}[htbp]
\vspace{-0.2cm}
\textbf{a)}\\[-0.3cm]  \includegraphics[width=\columnwidth]{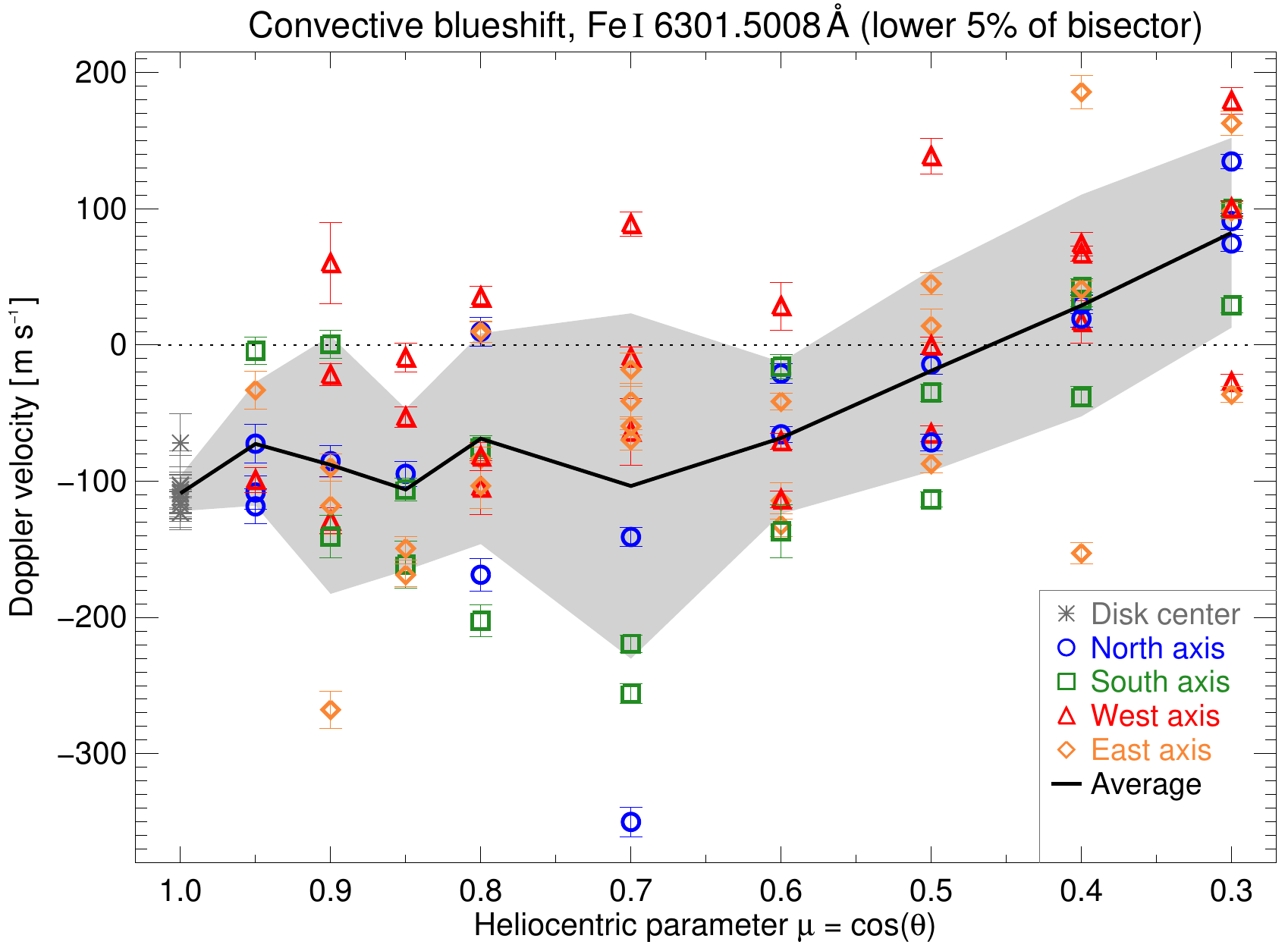}\\[0.1cm]
\textbf{b)}\\[-0.3cm]  \includegraphics[width=\columnwidth]{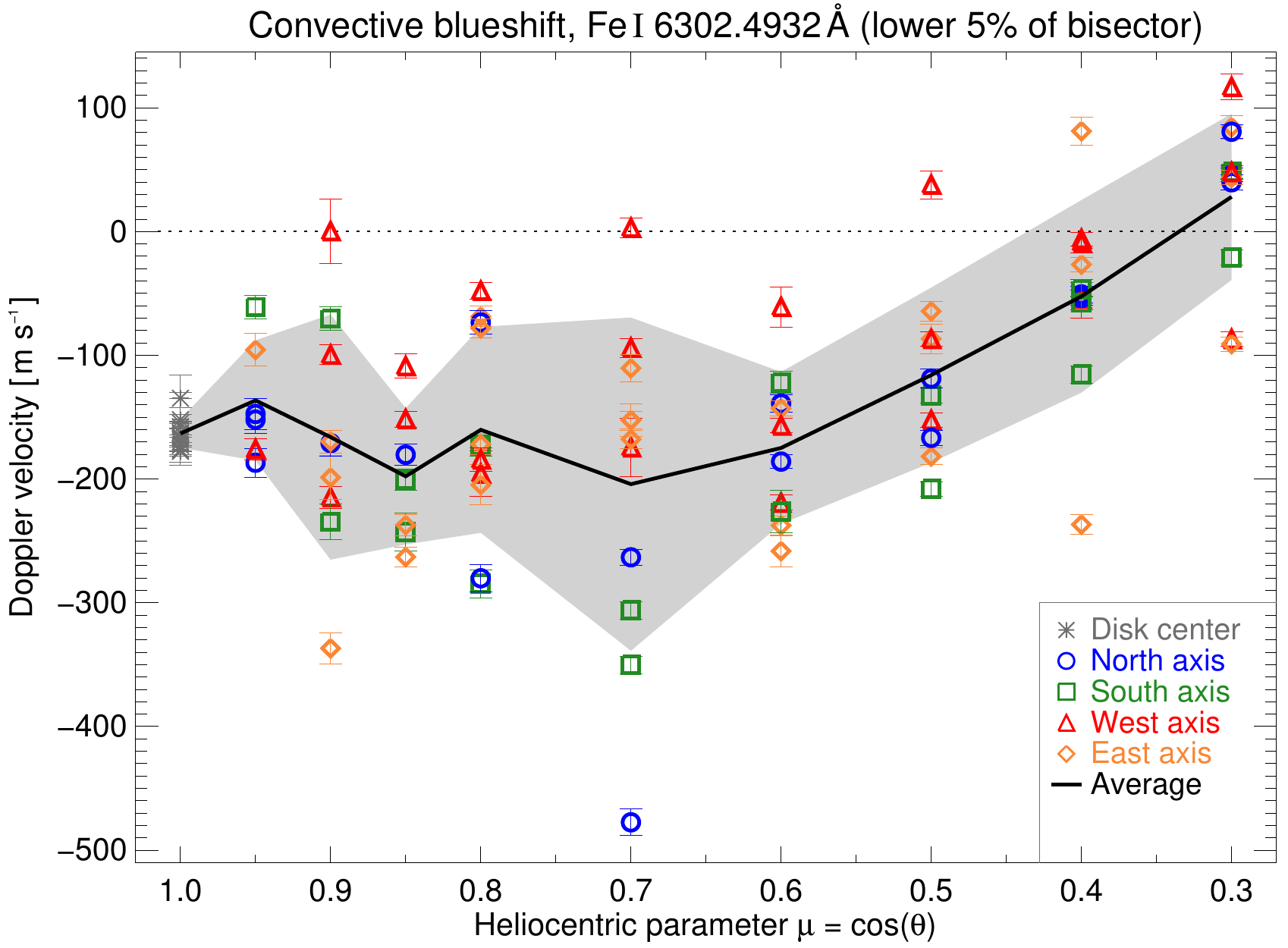}
\caption{Center-to-limb variation of the convective blueshift of the \ion{Fe}{I}\,6301.5\,\AA\ (panel a) and \ion{Fe}{I}\,6302.5\,\AA\ (panel b) line. Each data point represents the mean Doppler velocity of the lower 5\,\% of the bisector of the temporally averaged observation sequence. Error bars indicate the mean error. Radial axes are indicated by colors and symbols. The black solid line and gray shaded area display the average center-to-limb variation and its standard deviation.}
\label{fig_A11}
\end{figure}

\begin{figure}[htbp]
\vspace{0.87cm}
\textbf{a)}\\[-0.3cm]  \includegraphics[width=0.937\columnwidth]{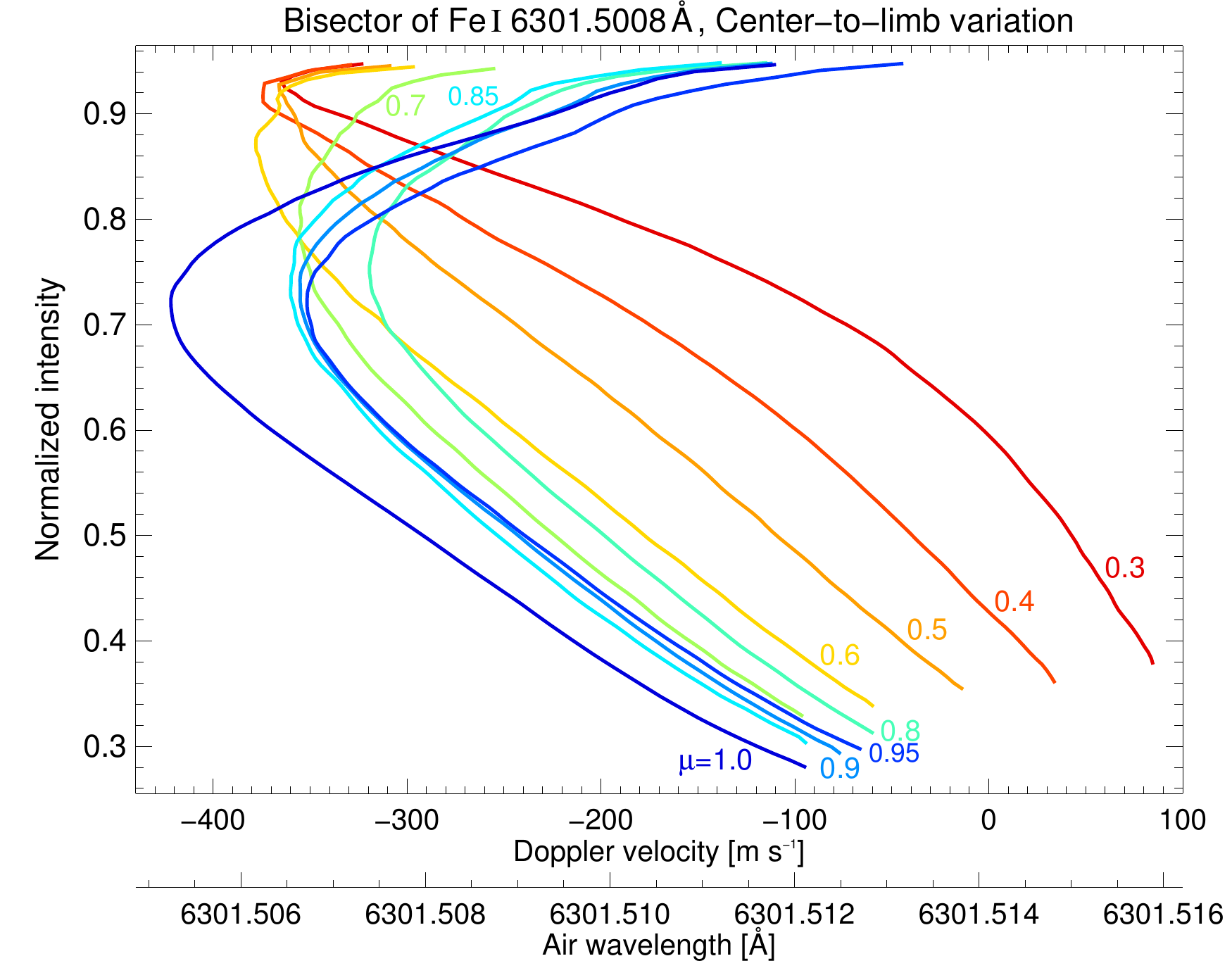}\\[0.1cm]
\textbf{b)}\\[-0.3cm]  \includegraphics[width=0.937\columnwidth]{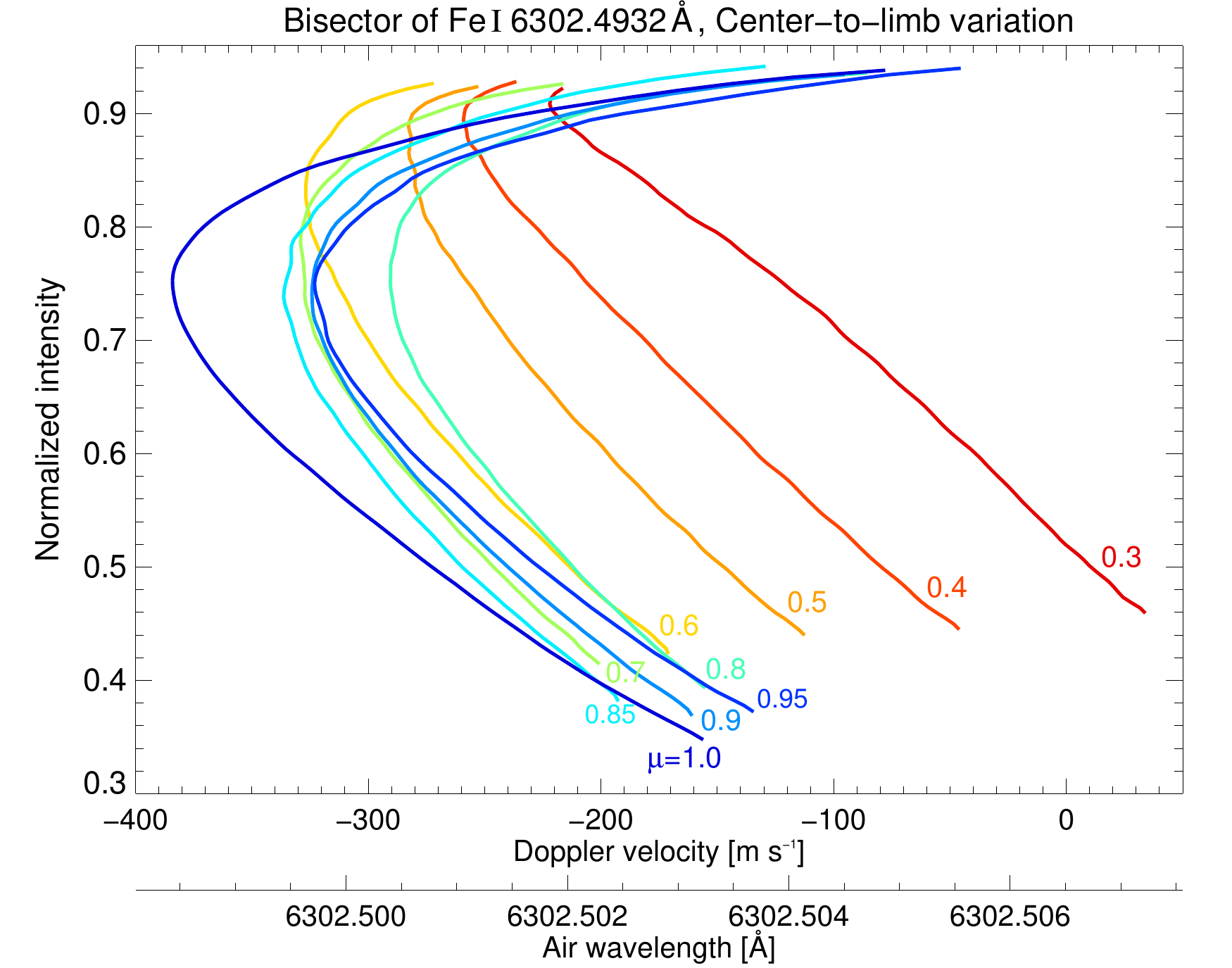}
\caption{Center-to-limb variation of the \ion{Fe}{I}\,6301.5\,\AA\ (panel a) and \ion{Fe}{I}\,6302.5\,\AA\ (panel b) line bisector, from the solar disk center ($\mu=1.0$, blue curve) toward the limb ($\mu=0.3$, red curve). The normalized intensity is displayed against the absolute air wavelength and Doppler velocity. Each curve represents the average bisector for all measurements at the respective heliocentric position.}
\label{fig_A12}
\end{figure}

\end{appendix}

\end{document}